# NEW MATERIALS PHYSICS

Paul C. Canfield
Ames Laboratory and Department of Physics
Iowa State University, Ames, Iowa 50011

**Abstract**

This review presents a survey of, and guide to, New Materials Physics research. It begins with an overview of the goals of New Materials Physics and then presents important ideas and techniques for the design and growth of new materials. An emphasis is placed on the use of compositional phase diagrams to inform and motivate solution growth of single crystals. The second half of this review focuses on the vital process of generating actionable ideas for the growth and discovery of new materials and ground states. Motivations ranging from (1) wanting a specific compound, to (2) wanting a specific ground state to (3) wanting to explore for known and unknown unknowns, will be discussed and illustrated with abundant examples. The goal of this review is to inform, inspire, an even entertain, as many practitioners of this field as possible.

# INTRODUCTION TO NEW MATERIALS PHYSICS

"No one is born a great cook, one learns by doing" Julia Child

Condensed matter physics (CMP) offers an intoxicating combination of fundamental and applied physics that allows a researcher to address some of the most basic questions of symmetry and mechanism and yet have the potential, and even promise, of direct applications that can jump from the lab to market in the time span of less than a decade. CMP poses some of the most fundamental questions about spontaneous symmetry breaking, energy, time scales and complexity. Amazingly (and with great esthetic and social satisfaction) out of these quests for fundamental answers come states like high temperature superconductors, emerging from quantum critical fluctuations like phoenixes from the ashes of dying magnetism. Whereas historically many of the early triumphs of CMP were associated with the description and understanding of known elements, minerals and simple compounds, increasingly the wealth of systems and phenomena treated in CMP comes compounds identified or discovered by research efforts focused on New Materials Physics.

New Materials Physics (NMP) draws on ideas and tools that were historically, often found in separate academic disciplines and departments. NMP researchers use binary and ternary phase diagrams with the shameless ease often associated with Geology or Materials Science efforts. NMP researchers will often browse compilations of crystal structures and be conversant about group and point symmetries in a manner more often associated with Solid State Chemistry or scattering groups. NMP researchers ultimately become, to varying degrees, bilingual in the languages of bonds and bands. NMP researchers often utilize a multiplicity of thermodynamic and transport measurements to characterize their compounds, measurements most often associated with CMP efforts. Depending upon the states of specific interest, measurements can include temperature dependent specific heat, magnetization, thermal expansion, electrical resistivity, thermal conductivity, and/or thermoelectric power, often as a function of applied field and/or pressure. In essence, a practitioner on NMP must be a "conversational charlatan" in many fields and sub-fields of CMP, Solid State Chemistry, and Material Science. At a picturesque level, NMP can be considered as the illicit love-child of solid state physics and solid state chemistry with materials science acting as a foster parent. NMP sits on the physics side of the middle regions of a ternary phase diagram that has each of these disciplines on one corner.

The power of NMP is that it offers the possibility of mapping the terms in a Hamiltonian onto the periodic table. Depending on the problem of interest, of the ground state at hand, this mapping ranges from the conceptual (or poetic) to very concrete and explicit. Figure 1 illustrates a schematic version of this mapping. The pre-transition metal elements (2s2p and 3s3p shells) offer low masses and tight bonding, both of which can lead to high characteristic vibrational frequencies; the 3d- and 4d-transition metals present tunable band-filling that offers a Janus-like potential for large density of states and/or magnetism that can vary from fragile to very robust. The nitrogen and oxygen columns (the pnictides and chalcogenides respectively) offer the ability to increase or decrease the bandwidth by moving down or up whereas the Li and Be columns (alkali and alkali-earth metal elements respectively) offer monovalent and divalent substitutions that can be used to tune band filling and matching (or mismatching) ionic size. This small description of basic trends encompasses many of the ideas and themes running through both low-$T_c$ and high-$T_c$ superconductivity research. For old-school, electron-phonon mediated, low-$T_c$ superconductivity, the simplest expression for transition temperature is $T_c \sim \omega_D \exp(-1/[N(E_F) V])$ where $\omega_D$ is the Debye temperature which is proportional to the average phonon frequency, $N(E_F)$ is the Density of States at the Fermi level and V is the electron phonon coupling. [1] Transition metals were thought to be important to increase the DOS term and compounds bearing light elements were often sought out in the hope of increasing $\omega_D$. Indeed, with the discovery of $MgB_2$ in 2001 [2] [3] [4]and the more recent pressure induced, very high $T_c$ values associated with $H_3S$ [5] [6], the study of compounds very rich, or purely made from very light elements is enjoying a renaissance. For high-$T_c$ compounds, a proximity to transition metal based magnetic order that can be fully suppressed by pressure or substitution is emerging as a generic feature in CuO- and FeAs-based superconductors as well as the earlier, heavy Fermion, superconductors (compounds with very low $T_c$-values that are actually relatively high when compared to their bandwidths). [7]Searches for other examples of high-$T_c$ compounds have focused on tuning / controlling transition-metal-based-magnetism by tuning of band filling via substitution of transition metals and/or alkali / alkali-earth metals and control of band width via substitution up and down the pnictide and chalcogenide columns.

The 4f-shell offer similar mappings and opportunities. At the grossest level, Hund's rules allow for the creation of Table 1 and the values of $\mu_{sat}$, $\mu_{eff}$ and the de Gennes parameter, dG, give a measure the moments as well as the ordering temperature for RKKY mediated ordering. More refined study identifies Ce, Eu and Yb (and to a lesser extent Pr and Tm) as being prone to ambivalence / hybridization, making them happy hunting grounds for Kondo- and related physics. Crystal electric field (CEF) splitting and Kramers / non-Kramers variation across the series provides an alternating series of compounds with protected and non-protected low temperature entropy that can be removed by a variety of mechanisms and transitions. CEF splitting (based on the ion's point symmetry) also allows the rare earth ions to assume extreme examples of Heisenberg, Ising, and even n-state clock models (n = 6, 4, etc.). Such anisotropies can be used to create compounds ranging from ones that allow for the study the effects of a transverse magnetic field on an Ising spin-glass to ones that assume geo-political importance due to their key role in high energy permanent magnets such as $Nd_2Fe_{14}B$.

NMP is the exploration and navigation of the ideas - structure - composition phase space that is delineated by the periodic table, a century of solid state physics and chemistry, and the limits of our imagination. The mantra of NMP is "think, make, measure, think". On good days this cycle can be completed in full; a growth can come out in the morning, a measurements can be done by mid-afternoon, and, based on analysis and gut feelings a new growth can be in before dinner. In this article the basic tools, mind set and goals of NMP will be outlined and examples of its implementation will be reviewed. In the final analysis, facility with the mind set and tools of NMP can enable a researcher to try to design or discover the compound that will allow her / him to study the physics or ground state of interest.

To start us on our way, let me end the introduction with an anecdote. At one point, early in my career, I had a colleague ask me why I tried to make so many different, and difficult, materials. He very sincerely said that since all physics can be found in silicon, I seemed to be making things harder then they needed to be. This comment reminded me of the old joke about a cold war meeting between generals from NATO and the USSR. The NATO general commented that, on average, their soldiers consumed 3,200 calories a day. “NO!” responded the Soviet general. “Perhaps as little as 2,800 calories,” replied the NATO general, “but not less.” Again the Soviet general responded “NO!”, followed by, “No man can eat that many potatoes in one day.” No doubt the meeting of generals is apocryphal, and serves as a joke, but, as far as I am concerned, “finding all physics in silicon” is more or less as pleasant or fruitful as eating only potatoes. It might be possible, but it is not the best way forward.

The goal of this paper is to introduce the ideas and tools associated with NMP to new comers to the field and to elaborate on key points and neat points of interest for more experienced readers. To this end, I will start by reviewing how to read, use and abuse composition – temperature phase diagrams. These are the maps we use and create to navigate around our N-dimensional phase space and, hopefully, grow crystals of known and unknown compounds. After phase diagrams, I will provide a cursory overview of a variety of crystal growth methods, followed by a detailed review of solution growth. With phase diagrams and solution growth covered, I will turn to the most important part of this manuscript: trying to answer the often asked question, “how / why did you decide to grow that?” This section tries to explain how ideas can be generated, how inspiration can be cultivated. This is rarely done, and I sincerely hope it will be of use or comfort to others starting in this field.

On with the show.

## MAPS AND PHASE DIAGRAMS

### Hic sunt Dracones

Ancient maps sometimes bear terms such as “terra incognita”, “hic sunt leones” or “hic sunt dracones”. The designations of “unknown land”, “here are lions” or “here are dragons” kindle an existentially human desire to explore, to find the lions, to vanquish or control the dragons. Although the golden age of global exploration is long past, such designations still exist on many other maps; they produce the siren call that beckons physicists, chemists and material scientists to quest for new phases and ground states; they are the very essence of NMP; dragons and lions can still be found in the terra incognita regions of phase diagrams.

Most condensed matter physicists think of a phase diagram in terms of perturbing a phase transition such as superconductivity, magnetic order, etc. Figure 2 shows a common example; [8] CeSb can assume many different magnetically ordered states depending upon the temperature and applied magnetic field. Indeed the richness of this particular phase diagram inspired the ANNNI (anisotropic next nearest neighbor interaction) model [9] as well as the idea of the "devils staircase" cascade of phase transitions. [10] Figure 3 provides another example of the genre: the T-x phase diagram for $Ba(Fe_{1-x}Co_x)_2As_2$, [11] [12] [13] [14] the archetypical phase diagram of the FeAs-based superconductors. It was this phase diagram that codified the relation between the magnetic, structural and superconducting phases and clearly suggested that quantum critical fluctuations may well be key in understanding this system.

The creation of such phase diagrams is key to quantifying, understanding, and ultimately controlling the physics of these systems. As such they are an end point, a goal, of intense experimental efforts. What is key to note, and what many condensed matter physicists do not often appreciate, is that before crystals of these compounds can be discovered or made, other types of phase diagrams play a key role: binary, ternary and quaternary, composition - temperature phase diagrams. [15] [16] These are the phase diagrams that are the bread and butter of materials scientists, geologists and students of NMP. Let's examine the binary and ternary phase diagrams associated with the example compounds mentioned above. This will help develop some of the ideas and vocabulary needed in the following sections. For alternate discussions of these and related phase diagrams, specifically in relation to NMP and crystal growth, references [17] [18] [19] [20] [21] [22] [23] are recommended as background or supplemental reading.

Figure 4a presents our (Humanity's) current understanding of the Ce-Sb binary phase diagram. Before going into some of the detail of this diagram, it is important to note that any phase diagram is merely a suggestion of reality. Figures 2 and 3 as well as Figure 4 and others to follow simply represent our current understanding of the "system". Whereas this may seem obvious for physicists when discussing Figures 2 and 3 it is important to emphasize that binary and ternary compositional phase diagrams are still being refined and new phases, even in binary compounds are being discovered on a regular basis [24] [25]. When using phase diagrams it is important to appreciate that they are not, "set in stone" and can often be less than accurate. To illustrate this point, Fig. 4b presents a phase diagram, for the same Ce-Sb binary system, from two decades earlier. It is significantly different, not only in the location of key temperatures and decomposition lines, but even in the composition of reported compounds. In recent years, there has been a proliferation of computationally generated phase diagrams; the "caveat emptor" warning is of even greater importance with these.

With all of that said, back to Fig. 4a; binary phase diagrams (like geographic maps) show quickly where there is liquid and where there is solid (water or land). Again, like a geographic map, in the solid regions binary phase diagrams indicate which state, or states, are possible and delineate borders. For the Ce-Sb binary system, below 625 °C there are only solid phases. At higher temperatures, a single phase liquid state is possible and for temperatures as low as 770 °C (on the Ce-rich side) or 625 °C (on the Sb-rich side), or for all compositions above a purported 1820 °C. The dome-like separation between the single-phase, liquid region and the lower temperature parts of the phase diagram is composed of the liquidus lines for the various reported compounds: Ce, $Ce_2Sb$, $Ce_4Sb_3$, CeSb, $CeSb_2$, and Sb. Ultimately, for solution growth of elements or compounds, we will be keenly interested in the location and accessibility of these liquidus lines. For now, though, let's identify a few other features of this phase diagram.

Elemental Ce, Sb and the one of the binary compounds, CeSb, melt congruently. This means that upon heating they transform from a single phase solid to a single phase liquid of the same composition. Perhaps more importantly, a single phase solution of CeSb, when cooled, will solidify into a single phase solid of CeSb at its melting point (near 1820 C). This is "what we are used to" based on our everyday experience with water and some elements. Many, if not most, compounds do not behave so simply though. $Ce_2Sb$, $Ce_4Sb_3$ and $CeSb_2$ melt incongruently or, to use another term, decompose peritectically. This means that when $CeSb_2$ is heated it warms to 1130 °C as a solid phase, and then decomposes into CeSb and an Sb-rich liquid. Only at a temperature of roughly 1600 °C, when the initial composition of $Ce_{0.333}Sb_{0.667}$ intersects the liquidus line, does the system become a single phased liquid. In the case of $Ce_2Sb$, upon heating, two peritectic lines are crossed: first one for the decomposition of $Ce_2Sb$ into $Ce_4Sb_3$ and liquid and then, at higher temperatures, a second one associated with the decomposition of $Ce_4Sb_3$ into CeSb and liquid. For temperatures above roughly 1600 °C a single phase liquid of $Ce_{0.667}Sb_{0.333}$ is found.

Whereas, when discussing heating of the system, these peritectic decompositions may sound like a somewhat complicated detail, they become profoundly important when you think about cooling the system from the single phase, liquid state. Continuing on from the last example, we can consider cooling a liquid of composition $Ce_{0.667}Sb_{0.333}$ from 1700 °C to lower temperatures. Below ~1600 °C crystals of CeSb will start to grow. This will continue until, below 1510 °C the system cools below the peritectic line for $Ce_4Sb_3$; between 1510 °C and 1330 °C $Ce_4Sb_3$ will form. As the system cools below 1330 °C, finally, $Ce_2Sb$ will grow. During this whole cooling process, as crystals of first CeSb, then $Ce_4Sb_3$ and finally $Ce_2Sb$ grow the remaining liquid becomes increasingly Ce-rich. The liquidus line defines the composition of the remaining liquid phase at a given temperature. Finally, as the system is cooled to the eutectic temperature of 770 °C, the remaining liquid solidifies. Under the assumption that there is little re-consumption of crystallized phases on cooling (i.e. this experiment is not happening on the time scale of millennia but rather hours or days) at the end of this process of cooling a liquid of composition $Ce_2Sb$ we end up with a mixture of solids: CeSb, $Ce_4Sb_3$, $Ce_2Sb$ and Ce. *What a mess!*

Fortunately, growing single phase, single crystals of incongruently melting compounds is far from impossible. Lovely single crystals of $Ce_2Sb$ or $CeSb_2$ can be readily grown from excess Ce or Sb respectively by intersecting the liquidus line below their respective peritectics. [18] [20] [26] [27] For example, $Ce_2Sb$ single crystals can be grown by cooling a $Ce_{0.9}Sb_{0.1}$ melt from 1100 °C down to 800 °C; $CeSb_2$ single crystals can be grown by cooling a $Ce_{0.06}Sb_{0.94}$ melt from 1000 °C down to 650 °C. In each case once the lower temperature is reached, it behooves the researcher to remove the remaining solution from around the crystals. More about this soon.

We must remember that Fig. 4 was introduced under the pretense of discussing CeSb and its rich and complex H-T phase diagram. The growth of CeSb is nominally possible from a binary melt, but it is not simple. The experimental reality of heating a 50-50 mixture of Ce and Sb to above 1820 °C is daunting. First you need to be able to generate that high of a temperature, ideally in a uniform and controllable manner, i.e. you want to be able to control heating and cooling rates over many hours and ideally days or weeks. Secondly you need to be able to contain this mixture at these temperatures. For Ce-Sb, at temperatures in excess of 1800 °C, this is not simple. This is both a reactive (Ce) and volatile (Sb) mix that requires chemical and structural stability from its container. Single crystals of CeSb could be made by sealing the elements in a thick walled metal crucible and heating, relatively briefly, to temperatures in excess of 1820 °C. This process was, not surprisingly, described as “mineralization”. [28] The equipment needed to perform such “mineralization” growths is not trivial and the resultant crystal were not always well formed, or low in defects.

Fortuantely, CeSb can be grown in a much simpler fashion. Much like the protagonist, A. Square, in the book “Flatland”, [29] we can appreciate the wonders and use of an extra dimension. Instead of being limit to two dimensions for our phase diagram, we can add an extra element (or more) and enter into a three (or higher) dimensional phase diagram. In essence we can look for some lower temperature liquid that we can dissolve, and recrystallize, CeSb out of. Again, this should not be a new or confusing idea since virtually every reader of these words will be familiar with at least the idea that common salt crystals form out of water. NaCl melts near 800 °C, and yet it crystals of NaCl can be grown out of $H_2O$ well below 100 °C. The key question for CeSb growth is what will play the role of $H_2O$?

One possible way of answering this question would be to check existing data on Ce-Sb-X ternary phase diagrams. Unfortunately, even today, instead of there being scores of such ternary phase diagrams for all possible X, there are single temperature cuts for X = Si, Mn, Ni and Ag and only partial cuts for X = Al, Fe and Se. [16] When faced with this problem roughly 25 years ago the “Casa Blanca” [30] approach was the most expedient way forward: round up the usual suspects. In this case the “usual suspects” were low melting metals that Ce and Sb could be diluted into: X = Ga, In, Sn and Pb. Sn, it turns out worked best. Figure 5 presents the Ce – Sn and Sb – Sn binary phase diagrams; the Sb – Sn binary phase diagram is particularly promising in that the liquidus line monotonically decreases from elemental Sb at 631 °C down to elemental Sn at 232 °C. This means that there are not any high melting Sb-Sn binary compounds that might form instead of CeSb. The Ce – Sn binary phase diagram is a little less promising in that there are eight reported Ce-Sn binary compounds; the liquidus line reaches up to above 1500 °C for $Ce_5Sn_3$ and $Ce_5Sn_4$; and even the most Sn-rich compound, $CeSn_3$, melts congruently at 1170 °C. All of this means that as Ce is added to Sn, attention must be paid to the possible formation of $CeSn_3$. [8] [18] With all of this said, the relatively dilute $Ce_{0.05}Sb_{0.05}Sn_{0.90}$ solution of CeSb in Sn, when cooled from 1050 to 800 °C produces exquisite single crystals of well faceted and essentially free standing CeSb. [8] The fact that the Ce is very dilute in Sn means that lower cost and easy to work with $Al_2O_3$ crucibles can be used to contain the melt. [18] [23] The relatively low temperature range of the growth means that silica tubing can be used to contain the growth crucibles and that there is relatively little or no partial pressure of Sb or Sn over the melt. In addition, the lower growth temperatures and ability to cool over hundreds of hours using a simple furnace and temperature control equipment lead to samples that have very few defects that scatter electrons, giving rise to residual resistivity ratios, RRR = $\rho$(300 K) / $\rho$(2 K), approaching 1000. With samples such as these the H – T phase diagram shown in Fig. 2 can be assembled from $\rho$(H,T) and M(H,T) data sets. [8]

As was just demonstrated, although there is no reported Ce-Sb-Sn ternary phase diagram data, we can get some idea of what the outer parts of the ternary diagram may look like by assembling the three binary phase diagrams along the edge of the compositional base. This is done in Fig. 6a. Figure 6b presents the pseudo-binary cut that we are accessing during the growth of CeSb out of Sn. Conceptually, the pseudo binary cut is very helpful since it maps directly onto what we have learned from binary phase diagrams. We cool from a single phase liquid until we hit the liquidus line for the formation of CeSb. The goal of a solution growth is to hit the liquiudus line of the desired compound and grow well formed, single crystals of said compound as the system is cooled. For the case of CeSb, the desired compound was a known one. For many growths the compound is known, at least crystallographically. For some growths, though, the idea is to use the crystal growth process to explore the liquidus surface and see what new compounds may be discovered.

It is not by accident that Sn was, “one of the usual suspects” when a variety of third elements were tried for the growth of CeSb. Sn is one of the lower melting elements, is not toxic, and can dissolve many of the other elements, to varying degrees, into Sn rich melts. For this reason Sn if often tried when there is a dearth of phase diagram information. Given its rough, compositional similarity to CeSb, it should not be surprising that $Yb_{14}MnSb_{11}$ was readily grown out of Sn (being basically RSb with a little Mn thrown in for luck). [31] Over the years a variety of $RT_2X_2$ (R = rare earth, T = transition metal, X = Si, Ge) compounds have been grown out of Sn (and In too). [32] [33] Sn has also been used for growths of P-bearing compounds. [34] [35] Not only does Sn incorporate a lot of P into Sn-rich solutions, but it also dramatically reduces the partial pressure associated with elemental P. Whereas phosphorous boils below 500 °C (in the red form) melts of dilute P-Sn melts can be taken up to 1200 °C or higher without encountering significant P-partial pressures. (This can be considered to be an extreme violation of Raoult’s law of simple liquids with there clearly being strong interactions between the Sn and P in the melt keeping the vapor pressure down. Later in this article we will review related Fe-P, Mn-P and even Li-N melts which are even clearer deviations.)

Figure 3 presented the archetypical phase diagram for the Fe-base superconductors. $BaFe_2As_2$ is the patriarch of the second large family of Fe-based superconductors (known as 122’s), with LaFeAsO being the first member of the so-called 1111’s. [13] The 122’s became the much more extensively studied FeAs-based superconductors due to the fact that they could be grown and studied in single crystal form, not only as pure compounds but also extensively substituted on each of the sites.  Not surprisingly, the first single crystals of $BaFe_2As_2$, as well as $(Ba_{1-x}K_x)Fe_2As_2$, were grown out of a quaternary melt rich in Sn.  There were several reasons for trying Sn as a solvent:  as show in Fig. 7, Ba, Fe, and As have fair solubility in Sn; Sn allows for growth over a wide temperature range, extending to relatively low temperatures; and As has a substantial vapor pressure (it does not melt at ambient pressure it sublimes) and Sn greatly reduce this.  Although even trying to imagine what a true quaternary phase diagram looks like is difficult, growth of ternary compounds out of quaternary melts is not unusual.  At the simplest level we can again use the idea of a pseudo-binary cut through the multi-dimensional space and simply imagine Fig. 6b with "CeSb" replaced by "$BaFe_2As_2$".

We will go over more details about Fe-based research below but, for our current discussion about phase diagrams, it is worth noting that, although we could grow $BaFe_2As_2$ out of Sn, there were difficulties too.  There was partial substitution of Sn for As.  The crystals, while well-formed, were on the small side, especially for neutron scattering experiments. Growth out of Sn did not allow for ready transition metal substitution for Fe.  For all of these reasons it became necessary to grow $BaFe_2As_2$ in an alternate manner.  What is truly remarkable is that $BaFe_2As_2$ can actually be grown out of a ternary melt.

Figure 8a shows the Fe-As binary phase diagram and Fig. 8b presents the Ba-Fe-As ternary compositional phase space.  $BaFe_2As_2$ is currently the only reported ternary compound and, as can be seen in Fig. 8a the binary, FeAs melts congruently at 1030 °C.  Remarkably, FeAs does not have substantial partial As-pressure over the melt, even at temperatures approaching 1200 °C.  This in NOT OBVIOUS and had to be established, *carefully*, experimentally.  But, once this was appreciated, then the growth of $BaFe_2As_2$ out of excess FeAs was straight forward.  This is illustrated by the pseudo-binary cut shown in Fig. 8c.  This is a different type of pseudo-binary cut from the Sn-growths we discussed above in that $BaFe_2As_2$ is not grown out of excess of a single, low melting element.  In this case, it is grown out of excess of a binary melt, in this case FeAs.  With growth of $BaFe_2As_2$ and $Ba(Fe_{1-x}T_x)_2As_2$ (T = transition metal) samples possible out of excess FeAs, the physical measurements needed to create the phase diagram shown in Fig. 3 could start.

NMP uses and creates phase diagrams.  It tries to identify compelling areas on these maps of human knowledge that may hide unprecedented flora and fauna or even mythical beasts.  NMP uses existing phase diagrams to serve as the spring board for dives into the unmarked portions of these compositional – structural – properties waters.  In the next section we will broadly review a variety of growth methods and focus, in detail, on solution growth.  After that we will start examining various modes of operation for NMP research.

## FROM IDEA TO REALITY:  MATERIALS SYNTHESIS AND CRYSTAL GROWTH

"The future belongs to those who believe in the beauty of their dreams." Eleanor Roosevelt

Crystalline materials can be made in single crystalline or polycrystalline form. At some level the distinction between what is a single crystal and what is a small grain of a polycrystalline sample is in the eyes of the beholder, or, more formally defined by the measurement at hand. Single crystal X-ray diffraction measurements can be performed on samples that are virtually invisible to the naked eye; transmission electron microscopy (TEM), scanning tunneling microscopy (STM) and even angle resolved photoemission spectroscopy (ARPES) measurements can be made on crystals with dimensions of a few microns (although in each case mounting and finding the sample makes this difficult). On the other end of the basic research spectrum, inelastic neutron scattering measurements still need, or certainly benefit from, single crystals with dimensions measured in cubic centimeters. In general, many physical measurement require samples with dimensions of several mm in one or more dimensions. How readily a compound can be made in phase pure form, in single or polycrystalline form, depends greatly on the details of the compound's phase diagram.

Having introduced and discussed the vocabulary of phase diagrams, we can now divide compounds into the categories of congruently and incongruently melting. Figure 9a presents a generic binary, A – B phase diagram with two compounds: AB that melts congruently and $A_2B$ that melts incongruently (or decomposes peritectically). The materials synthesis and/or growth of these two types of compounds can be very different. If a compound melts congruently, as AB does, there is the possibility (modulo experimental banalities such as accessible temperatures, vapor pressures, reactivity, etc.) of cooling a stoichiometric melt of the compound and having it form a stoichiometric solid of the same composition at a single, well defined, temperature. If the melt is cooled slowly, and with controlled nucleation, single crystals can be grown. This is the operational principle behind crystal growth techniques such as Verneuil, Czochralski, Bridgman, and zone refining. [36] [37] On a much faster time scale, this is also the operational principle behind the production of polycrystalline ingots via arc-melting.

An arc-melter is a device used to rapidly heat/cool compounds via controlled electrical discharge, often in an inert atmosphere. Arc-melting is a common and versatile method of producing polycrystalline samples of wide classes of materials, most often used on intermetallic compounds and extending to various borides, oxides, etc. If we want to make polycrystalline AB and neither A nor B have significant vapor pressures for $T > T_{AB}$, then rapidly heating A and B to form liquid AB followed by a rapid cooling through $T_{AB}$ (over the course of a 0.1 to 10 seconds) can produce a polycrystalline mass of AB.

If single crystals are needed (or desired) then slower cooling and/or some attempt at control of nucleation are needed. Between 1880 and 1930 Verneuil, Czochralski and Bridgman devised different methods of, effectively, establishing a temperature gradient across a sample that put solid and liquid in contact with each other. In terms of Fig. 9a, solid AB and molten AB exist, in contact with each other in a region that has temperatures spanning $T_{AB}$. Whereas the Verneuil and Czochralski processes produce crystals outside of a crucible, the Bridgman process has the crystal confined in a crucible with a molten zone moving along the crucible length as the crystal grows. A subsequent modification to these techniques was float-zone refining, which in the recent two decades has experienced explosive growth in application with optically heated floating zone units.

What has to be borne in mind is that, for any of these process to work in the simple manner described, the compound has to be congruently melting. There has to be a co-existence of liquid and solid compound (AB). If the material is not congruently melting, then we return to the "what a mess" situation described earlier for $Ce_2Sb$. If, for example a melt of stoichiometry $A_2B$ was cooled it would lead to the formation of phases AB and $A_2B$ as the temperature decreased from above $T_{A2B}$ down to $T_{Eu1}$ and then $A_2B$ and A when the remaining eutectic composition solidifies. In general, the synthesis and growth of incongruently melting compounds can be more complex, unless solution growth is used.

If polycrystalline AB can be made via arc-melting, then it is possible that single phase $A_2B$ can also be made, but with an additional step. Once the liquid of $A_2B$ is quenched (rapidly solidified) it will contain multiple phases that are relatively fine grain and in intimate contact with each other. If the ingot is reheated to a temperature just below $T_{A2B}$ and held there for a long time (often this means hundreds or thousands of hours) an increasingly single phase ingot of $A_2B$ can be achieved via what is referred to as a solid state reaction, basically a diffusion mediated reaction. Depending on the initial grain sizes, grinding, pressing and re-reacting may be necessary.

Solid state reactions are also possible and even convenient for the synthesis of congruently melting materials in polycrystalline form if their formation requires very high temperatures or temperatures that are inconsistent with the vapor pressure of the elements used. If, for example, $T_{AB}$ was at temperature above that reachable by the furnace at hand, a solid state reaction of a pressed pellet of A and B or of $A_2B$ and B could produce polycrystalline AB. Solid state reactions were a common way of making the initial high $T_c$ cuprate superconductors. At the time it was referred to, in some circles, as shake-and-bake processing since often the mixture of different oxide powders was involved. It should be noted, though, that for many solid state chemists or new material physicists this is considered to be a pejorative description.

A technique that can produce very nice single crystals of both congruently and incongruently melting materials is vapor transport. [38] This technique involves the motion of constituent parts of the compound across a temperature gradient in vapor phase. In many cases a small amount of transport agent (often iodine, chlorine or related volatile element or compound) is used to act as a chemical shuttle bus, moving constituents to the growth side of the reaction ampoule. Many crystals, [38] including, recently, $Bi_2Se_3$, [39] can be grown using iodine as a transport agent; various vanadium and molybdenum oxides are grown (using $TeCl_3$ or $VCl_3$ as transport agents), [40] more recently FeSe has been grown using a KCl / $AlCl_3$ salt mixture that acts, in part, as a transport agent. [41] The difficulty with implementation of vapor transport growth is identifying a transport agent (if any, in many cases transport is simply impossible) and also identifying the gradient. Even if/when these are identified, chemical transport growth can require weeks or months to produce crystals of the necessary size.

The most versatile crystal growth method, the one that works for the largest number of inorganic compounds and, as a bonus, is relatively cheap in term of equipment, materials and time, is solution growth. [17] [18] [19] [20] [21] [22] [23] [34] [42] The viability of solution growth is not determined by the target compound being congruently or incongruently melting, rather it is determined by the accessibility of the liquidus line (surface). Using Fig. 9b, we can see that $A_2B$ can be grown out of excess A between the temperatures of $T_{A2B}$ and $T_{Eu1}$ (shown schematically as temperature – composition trajectory of the liqiudus composition). AB can be grown out of excess A or excess B. For the case of excess A, the temperature range is limited on the low temperature side by the peritectic line for $A_2B$. Cooling below $T_{A2B}$ would lead to the formation of $A_2B$ instead of AB. Figure 9b shows a composition trajectory for a hypothetical growth of AB out of excess A. For the case of growing AB out of excess B a wider temperature range for cooling exists because there is no, more B-rich binary compound between AB and the B-rich eutectic. Figure 9b shows a composition trajectory for a hypothetical growth of AB out of excess B. For both growths of AB an initial composition that is well removed from AB is chosen so that there will be an opportunity for growth of isolated crystals of AB in the cooling solution. It should be noted that even pure A and pure B can be grown out of solution for compositions between the pure element and their respective eutectics. Although this seems a bit tight or needlessly complicated in Fig. 9, there have been cases when we have taken advantage of this possibility, for example growing single crystals of pure Co out of a Co-S eutectic at temperatures below cobalt's Curie temperature [43] [44].

A growth method that is strictly a variant of both solution growth or vapor transport is hydrothermal growth. This is an exceptionally powerful growth technique for oxides and related materials. As the name suggests, water is heated, often to temperatures and pressures well above its critical point (i.e. to temperatures above 400 °C and pressure above 200 atmospheres). At these temperatures and pressures water is an excellent solvent. Hydrothermal growth is most famous for the growth of large, high quality quartz. It is also used for many other compounds, including many non-linear optical compounds [45].

For any given compound there may be a multiplicity of synthetic methods available or, in some cases, only one or even none. The choice of how to try to grow a compound is dictated by a variety of considerations, not the least of which are, (1) what are the requirements of the measurement you want to make on the samples and (2) what is convenient or possible in your lab. Given its versatility both for growth of specific, known, compounds as well as for exploration for new compounds, solution growth is a powerful tool for NMP research and will be the focus of the next section.

PRIMER TO SOLUTION GROWTH OF SINGLE CRYSTALS

The time has come, the Walrus said, to talk of many things…. Lewis Caroll

Solution growth of single crystals requires, among other things, an idea, an initial composition, a crucible, a temperature profile, and a mechanism for separating the grown crystals from the remaining solution. We will spend the subsequent sections of this review discussing where ideas come from and how to encourage them to visit more often. In this section we will discuss the experimental details, guidelines, rules of thumb, etc., of growth of single crystals from high temperature solutions. There are several excellent text dedicated in full or in part to details of solution growth of single crystals [36] [42]. For the purpose of NMP, more recent review and research articles such as: [17] [18] [19] [20] [21] [22] [23] and/or [34] are also very helpful.

If, as described in the section above, we decide to grow either AB or $AB_2$ out of solution, (see Fig. 9), or if we refer back to Fig. 4a and grow $CeSb_2$ out of excess Sb, we will want to place the desired ratio of elements into a "growth" crucible, place a "catch" crucible with some form of filtration on top of the growth crucible and seal this assemblage into a silica tube with some silica wool to act as cushioning. This is shown schematically in Fig. 10. In Figs. 10a and 10d the growth crucible is shown packed to a level of roughly 2/3 full with the solid lumps of A and B, the only concern is that there not be any chance of the melting materials touching/adhering to the frit or filtering material and thereby altering the chosen stoichiometry of the melt in an uncontrolled manner. As shown in Figs. 10a and 10b, after packing the growth crucible the frit can be fitted on top and the catch crucible is placed on top of the frit. (The use of fritted crucibles - also known as Canfield Crucible Sets or CCS - is reviewed in detail in ref. [23].) The assembled growth and catch crucibles are then placed into a silica tube (Fig. 10a). Silica wool is used as cushioning material on top of the catch crucible (Fig. 10e). This is vital if a centrifuge is to be used to provide rotationally enhanced acceleration during the decanting step. In some cases a plug of silica wool is also used below the growth crucible to provide support, cushioning, and/or a bit of spring constant to help hold the crucibles together. Figure 10d shows an alternate assembly of the growth/catch crucibles. Before the development and ready use of the fritted CCS, another plug of silica wool was placed in the catch crucible to help separate the remaining liquid from the grown crystals. Although this is certainly cheap and convenient, it means that there will be silica fibers in the melt possibly acting as contamination and/or nucleation sites. A perhaps even worse problem with use of silica wool in the catch crucible is that the decanted liquid is hopelessly contaminated by it and cannot readily be reused. The advent of relatively cheap, fritted CCS has led to their default use in our labs.

Figure 10 shows the growth assembly for a growth that is compatible with alumina ($Al_2O_3$) crucibles. This is often the case for growths rich in less reactive elements such as Ge, Al, Ga, In, Sn, Sb, Pb or Bi. On the other hand, growths rich in more reactive metals such as Li, Na, Mg, or rare earth metals are not compatible with the use of alumina since their oxides are more stable than alumina, leading to thermite-type reactions at higher temperatures. In order to hold melts such as these, a different crucible material is needed.

For many such growths Nb or Ta tubing can be used to make crucibles. Figure 11 shows the parts associated with the assembly of what we refer to as a "three-cap-crucible" [19] [46]. This name derives from the fact that we make the crucible from Ta tubing and caps made from Ta sheet. The bottom cap is pressed into the Ta tube and welded into place using an arc-melter and the jig shown in Fig. 11b. Next, the load is place in the crucible and, if there were no desire to decant, a top cap could be pressed into the top of the tube and welded into place, again using the arc-melter and jig. On the other hand, separating the crystals from the remaining liquid, i.e. decanting, is very useful. To accomplish this we take a third Ta cap and drill holes into it. This Ta-frit, or strainer, is press fit into the tube, just above the load. Once this is done the top cap can be welded in place. The whole 3-cap assembly can now be sealed into a silica ampoule in a manner similar to what was done for the alumina crucibles, as shown in Fig. 10.

Other crucible materials can be used. The choice depends a lot on and optimization between non-physical parameters such as convenience and cost with physical concerns such as chemical and mechanical stability. Other possible materials range from graphite, silica, MgO, $Y_2O_3$ and BN to Fe, Re or Pt. At some level, you use what you can get away with….When we grew copper nitrate crystals (for inelastic neutron scattering) we used glass jars similar to what some of us use to preserve our summer jams in.

After heating and slow cooling (the details of which will be discussed below) it behooves us to separate the grown crystals from the excess liquid. Recall, we are desperately trying to end up with a two phase system: crystals of the phase we want (or some wonderful new phase we are about to discover) and excess liquid. In this scenario we now would love to quickly and efficiently remove the crystals from the liquid. Happily, we can do this, at least in many cases, with the help of a lab centrifuge, Fig. 12a. The ampoule is removed from the furnace and promptly placed into one of the centrifuge cups, Fig. 12b. (This is most often done by "pouring" the ampoule out of a 50 ml $Al_2O_3$ crucible into one of the centrifuge cups.) The lid is closed and the centrifuge turned on. After a 5-10 seconds of angular acceleration the centrifuge is turned off and, once the rotation has stopped, the lid opened and the ampoule removed. This is, shall we say, a dynamic process; the time involved in removing the ampoule from the furnace and decanting is less than the time it took to type, or read, these sentences.

It is important to emphasize that the centrifuge is just a standard lab centrifuge with a brass rotor and cups. It is not heated in any way. The cups are at room temperature when the ampoule is put in (they do heat up as a result of the decanting, so be careful handling them….). There clearly is some decrease of melt temperature as the ampoule is going from furnace to centrifuge, but it is not too large. We have been able to decant liquid that is only a few 10's of degrees above the reported eutectic temperature. This is possible due to (i) the overall small amount of time required to decant (5-10 seconds), (ii) the partial thermal isolation of the melt (separated from room temperature by silica, partial vacuum and the crucible) and (iii) the thermal inertial of the melt and ampoule. It is important to point out something that the biologists are keenly aware of; a centrifuge that can operate at 2000 rpm and has the sample rotating roughly 15 cm from the axis of rotation can produce accelerations in excess of 500 times Earth's gravitational acceleration. Even if we say that in the first 5-10 seconds we only reach 1/10 of this value, 50 x g is a very large acceleration. It is this sudden and large acceleration that allows for such clean and, in many cases, complete separation of crystal from melt.

Figures 10a and 10e show the crucible sealed into a silica ampoule. This is generally necessary to protect the melt from reacting with air. For the case of a Ta crucible, it is necessary to prevent the Ta from oxidizing and falling apart (and then having the melt react with air). Although silica is exceptionally versatile, i.e. thermal shock resistant and also workable with a $H_2$ / $O_2$ blow torch, it does come with a limit attached to it: silica softens above 1200 °C. This softening temperature is rather well defined, silica can be used reliably for temperatures less than 1190 °C and for temperatures greater than 1210 °C the risk of either collapse or inflation, depending on the internal pressure, grows rapidly with each additional 10 °C. If a growth needs to be at temperatures well above 1200 °C alternative environmental control is needed. One of the simplest ways of protecting a growth for temperatures up to 1500 °C is to use a vertical tube furnace with high purity Ar slowly flowing through the tube. The 1500 °C limit is now set by the temperature that SiC heater elements can readily reach. For higher temperatures different heater elements and, very soon, different refractory materials for furnace construction will be needed.

Figures 13a and 13b show a schematic diagram and a photograph of a vertical tube configuration. In our lab, mullite is the actual vertical tube. It is capped on each end with metal flanges that are sealed with neoprene gaskets and cooled by muffin fans (Fig. 13c). High purity argon is supplied on one end and a positive pressure is maintained by a slow bubbling of the Ar through an impedance on the other end of the tube. The physical manifestation of this one-way impedance is shown in fig. 13d. This "bubbler" is filled with diffusion pump oil and the Ar flow is set so that one bubble every few seconds passes through. This is an exceptionally simple, inexpensive and effective way of providing a protective environment. We have been able to grow rare earth bearing intermetallic compounds out of metal based solutions at temperatures up to 1490 °C without any significant oxidation problems. A clear difficulty associated with this higher temperature growth configuration is that it does not allow for rapid decanting. This can be worked around by either using post growth mechanical, physical or chemical separation or, if the growth ended at a temperature below 1200 °C, by resealing the growth crucible in a silica ampoule, reheating it to some temperature below 1200 °C and then decanting. An example of this can be found in our $RNi_2B_2C$ growths. [19] Crystal growth took place between 1480 °C and 1180 °C in the vertical tube under bubbling Ar. After reaching 1180, the furnace was turned off and cooled over roughly 10 hours. The growth crucible was then sealed in a silica tube (with a catch crucible filled with silica wool), heated to 1180 °C and held there for 5-10 hours. At this point the excess solution could be decanted using a centrifuge, as described above.

Although decanting is very efficient at separating grown crystals from the remaining liquid, there are cases when it is either not practical or not possible. In addition, sometimes decanting does not remove all of, or enough of, the remaining liquid; the viscosity and surface tension of liquid metals vary with temperature and sometimes prove problematic.  In these cases there are alternate, or additional methods for removal of solidified melt from the grown crystals.  Mechanical methods can range from simply breaking pieces of crystal out of the crystal/solidified-melt composite to sanding and polishing.  Chemical etchants are useful as long as they attack the solidified melt more rapidly than the desired crystalline phase.  These can be either acid or base, depending on the composition of the solidified melt.  For example, NaOH solutions work well on excess Al and HCl works well on excess Sn.

Distillation is possible when the excess melt is volatile.  We can distill Mg off of $MgB_2$ growths and have also used distillation to remove excess Zn, Te, and Se.  Distillation can often be accomplished by sealing the crystals into a silica tube and heating the end with the crystals with the other end staying near room temperature.  Depending on the temperature gradient the distillation can be completed in a little as a few hours.

Having discussed choice of crucible, as well as the temperature constraints placed on growth by the use of silica ampoules and/or silica carbide heating elements.  We should now go over the coupled choices of initial composition and temperature profile.  There are three basic parts to a temperature profile for a solution growth:  preparing the melt, the slow cooling associated with nucleation and growth and, finally, decanting.  We have already discussed decanting and we will postpone for a few paragraphs the details associated with preparing the melt.  If we return again to Fig. 9 and consider the growth of AB out of excess B we can see that there is an exposed liquidus line for AB ranging from 50% B all the way over to the eutectic composition on the B-rich side of the diagram.  Any initial composition in this wide range would intercept the liquidus line for AB as it is cooled from above $T_{AB}$ down to $T_{EU2}$.  In general we want to start the slow cooling associated with nucleation and growth of crystalline AB from the single phase, liquid, region of the phase diagram.  No crystal growth takes place while cooling through the liquid region above the liquidus line, so, once we have a homogeneous liquid we want to be cooling into a two phase liquid sooner rather than later.  This simply means that if the initial melt stoichiometry is close to AB, then the slow cooling will have to start at temperatures close to $T_{AB}$; if the initial melt stoichiometry is closer to the eutectic composition, then the slow cooling can start from a much lower temperature (still above $T_{EU2}$ of course).

Greater constraints on cooling temperature ranges are found on the A-rich side of the phase diagram, associated with the peritectic decomposition of $A_2B$.  If we choose to grow AB out of excess A, we can only do this if we intercept the liquidus line for AB formation.  This will only happen for initial temperatures above $T_{A2B}$ and initial compositions between AB and liquidus line composition at $T_{A2B}$.  If we cool the melt below $T_{A2B}$ we will start growing $A_2B$ instead of AB.  Similarly, if we want to grow $A_2B$ out of excess A, then we need an initial composition between the A-rich eutectic composition and the composition of the liquidus line at $T_{A2B}$ and the slow cooling will only need to begin a little above $T_{A2B}$.

The above paragraphs are basically restatements of how to read a binary phase diagram.  As my Russian-speaking colleagues say, paraphrasing the Latin, “Repetition is the mother of learning." Hopefully such repetition reinforces the importance and power of these maps. Unfortunately, choosing the best initial stoichiometry and temperature profile for a growth is a bit more complicated and it often involves intuition, experience and a few trials. If I knew nothing about AB growths, and if there were no vapor pressure, reactivity, expense issues, I would probably try an initial growth at the B-rich-composition shown in Fig. 9b and start slow cooling about 75 °C above the reported liquidus temperature and cool down to a temperature about 50 °C above the reported value of $T_{EU2}$. The reason for the "75 °C above liquidus" and "50 °C above $T_{EU2}$" is the acknowledgement of the inherent inaccuracy of any given phase diagram combined with the inherent inaccuracy of knowing the exact temperature of the melt (i.e. the lack perfectly calibrated thermometers as well as lack of perfect isothermal conditions) in any given furnace. Another reason for the "50 °C above $T_{EU2}$" is an acknowledgement that even though the decanting process is rapid, there may be some cooling of the melt as the sample is moved to the centrifuge and angular acceleration is building. Both of these numbers can be diminished, by even factors of two or three, if needed and supported by previous growths.

One of the beauties of solution growth is the fact that the specific choice of B-rich-composition for a first attempt to grow AB is not all that critical. In general you want to balance not being too close to the compound you are trying to grow (i.e. hopefully allowing individual crystals to grow with plenty of solution around them) and still having the system cool in the two phase region (e.g. intersect the liquidus line) enough to produce adequate hypothetical yield of crystalline material. Figure 9b shows my initial default first growth choices for this hypothetical/schematic system.

Now comes a clear bifurcation: if the growth yields a crystal of AB that is adequate for the desired measurement (remember we are growing this sample so as to make some sort of measurement), then we can cycle from the "make" step to the "measure" step, soon to be followed by "think". If, on the other hand, the initial attempt at AB growth from the B-rich-composition shown in Fig. 9b did not work (either well or at all) then there needs to be a post-mortem of the growth; in what way did the growth not work? There can be many answers and many possible routes to a better result. If all of the material decanted to the catch side (referred to in our group as a "total spin"), then either the system needs to be cooled to a lower temperature before decanting or the next growth needs to have an initial composition closer to AB. If there is no material decanted to the catch side (referred to as a "no spin"), then a higher temperature should be used for decanting. In both of these cases the result implies that, as long as no mistake was made in either the initial composition or the temperature profile, then the reported phase diagram is incorrect.

If there are crystals on the growth side, and decant on the catch side, there can still be problems: the crystals may be too inter-grown, too dendritic and/or too abundant and small. Each of these problems may possibly be remedied by increasing the amount of time taken during cooling. Usually factors of two are used, i.e. if the cooling time was 30 hours, change to 60 or 120 hours may make a qualitative difference. Other possible changes in subsequent growths can be trying a more dilute growth (i.e. moving the initial composition further away from AB) and cooling over a more limited temperature range, or, in some cases, trying to create some cold spot on the growth crucible to act as a preferential nucleation site.

For the case of trying to grow AB, we also have the option to try growing out of excess A instead of excess B. Sometimes the details of nucleation and growth can be quite different for growth out of excess A as opposed to excess B. The primary reason for not starting with attempts out of excess A is the limitation imposed by $A_2B$ and its peritectic decomposition. Refining subsequent growth attempts of AB out of excess A and even $A_2B$ out of excess A would progress along the same lines as outlined above for the attempts at AB out of excess B.

Clearly, when A and B leave the realm of the abstract and become real elements with very specific properties such as vapor pressures, reactivity with crucible, toxicity, cost, etc., then other concerns enter into the choice of initial and subsequent growths. If for example element B was much more expensive, volatile, or reactive than A, it may be better to try growing AB out of excess A even though the phase space is more limited.

Now, as mentioned a few paragraphs ago, we should discuss the first part of the temperature profile: preparing the melt -- basically melting all of the starting materials so as to produce a single phase liquid that can then be slowly cooled for crystal growth. This can be as simple as heating the growth ampoule to the "75 °C above the reported liquidus temperature" mentioned above, but even this has some implicit assumptions that we should spell out. For example, if the target temperature for the start of slow cooling was 1000 °C, then simply placing the growth ampoule into a furnace at 1000 °C and immediately starting to cool would not insure that a homogeneous melt was being cooled from 1000 °C. Indeed, in this case, depending on how long the growth is given to cool through the first 75 °C, it is possible that a single phase, homogeneous liquid is never attained, leading to possible problems with stoichiometry of melt, nucleation of crystals, and, very fundamentally, reproducibility. For a simple growth, let's say a 2 ml size growth crucible, if the desire is to start slow cooling from 1000 °C down, the ampoule should be warmed to 1000 °C over roughly 3-5 hours and dwell at 1000 °C for another 3-5 hours before cooling. Alternatively, if there are no concerns (see below) about very rapidly heating the ampoule to 1000 °C, then having the ampoule placed into a 1000 °C furnace and sitting there for 4-6 hours before cooling would probably be fine too. Either way, you want enough time at maximum temperature to allow for a homogeneous liquid to form.

Now, there are many reasons that life might not be as simple as outlined above. The most obvious one is associated with vapor pressure. For example, when Sn is used to incorporate As into the Sn rich melt (for example, for the growth of $BaFe_2As_2$, $SrFe_2As_2$ or $CaFe_2As_2$ out of Sn [13]) it is important to heat to the melting point of Sn and then slowly heat to higher temperature so as to incorporated the high vapor pressure As into the Sn and creating a low vapor pressure melt. A somewhat more subtle version of this same idea can be found in some of the transition metal (TM) -chalcogen (Ch) or – pnictogen (Pn) binary melts. Again, a clear example can be taken from recent work on the FeAs based superconductors, in this case, the preparation of FeAs (Fig. 8a). FeAs is reported to melt congruently at 1030 °C. Indeed, a solution rich in FeAs is used to growth $BaFe_2As_2$, $SrFe_2As_2$ and $CaFe_2As_2$ based compounds [11] [13] [47] [48]. This being said, it is a grave and dire mistake to blindly put Fe and As into a sealed ampoule and heat to 1030 °C; a lot can go wrong, and has in different labs around the world. If As is heated above its sublimation temperature, and if there is not enough Fe available to react with it, then an As partial pressure will build up to many atmospheres and cause the reaction ampoule to explode. Not only does "enough" Fe need to be present in the ampoule, but it also has to be accessible to the As vapor. This means that finely divided Fe is far better than a single lump or rod of Fe since not all of the lump is "readily available" to react with the highly mobile As vapor. For FeAs, as well as other TM – Pn or –Ch melts it is important to prepare the melt by very slowly heating a mixture of finely divided TM with Pn or Ch so as keep vapor pressures low. Recent examples of some our own TM-S, TM-Se, TM-Te and TM-P growths can be found in refs. [43] [44] [46] [49] [50] [51] [52] [53] [54].

At this point, it is probably most useful for the reader to find more detailed examples in some of the many references that have been provided so far (or are yet to come), or find specific compounds of interest that have been grown in single crystal form and read about the details of the growths made in the past….***or not***. At this point, for the end of these sections about growth, I fear I need to get up on my soapbox for a paragraph or two and address this "or not". Ideally, specific details can be found in research papers about specific systems. In our own group, as a rule, we try to provide adequate detail about initial composition, temperature profile, crucibles, etc. This is really important for the implementation of what is traditionally know as, "the scientific method". Unfortunately many research groups feel that there is a distinct advantage to keeping details of materials synthesis secret and, for example, simply state that a given compound was grown using solution growth techniques, or was grown out of excess X with no further details. This really should be strongly discouraged by both the journals as well as the referees. It hinders the advancement of the field and leads to either a disincentive to try to make the compound or substantial lost/wasted time, or, for potentially dangerous growths, puts subsequent researchers at risk by not providing useful guidance.

The corollary of this really should be, though, that if you use a similar growth technique to an earlier published paper, then you should refer to that paper, in the way you refer to earlier measurements of a physical property. Over the past decades my own group has "taught the world" how to grow many types and classes of materials. Sadly, in some of these cases, having laid out full and detailed growth information, often in our initial, or discovery paper, there have been a large number of subsequent papers that clearly are making the samples in exactly the same manner (even down to times and decanting temperatures we chose in a somewhat arbitrary manner) as we did without any reference to our earlier work. If "we", as the New Materials Physics community, feel that having full and detailed growth details published, even with a discovery paper, is important, then "we" as a community need to acknowledge this with proper references and citations. This will be important if NMP is to really grow as a thriving field.

OK, off of the soap box now….Carry on.

## MODES OF RESEARCH

"The best way to have a good idea is to have lots of ideas" Linus Pauling

Over the past sections we have reviewed the basics of phase diagrams and materials synthesis as well as some of the specifics of solution growth of single crystals. Now, as John Cleese used to say, it's time for something completely different [55]: I want to outline how NMP can be deployed to enable various modes of research. I want to also, hopefully, give some idea of where ideas come from and how, so that, like small timid woodland creatures, they can be lured out into the open and studied and enjoyed.

At some level, we are now starting to discuss the "think" step in the "think – make – measure – think" mantra. The question at hand is how do you decide what to try to grow? Pearson's Crystal Structure Database for Inorganic Compounds [56] has over 250,000 entries; how do we decide which of these compounds might be of interest? How do we decide that there may be new, yet to be discovered compounds lurking "here" rather than "there"; which poorly explored ternary phase diagram do we choose to study, and why? The answers to these questions are often complex and can be part of a larger family tree of ideas and innovations. We will return to the genealogy of ideas near the end of this review. First, though, I outline the three basic motivations for attempting a given growth: (1) you want a specific compound, (2) you want a specific ground state or (3) you want to explore, searching for known unknowns as well as unknown unknowns. Although these motivations describe overlapping Venn diagrams they do offer some useful distinctions and implications. Each motivation, to differing extents, narrows the phase space for selecting specific composition for a given growth and provides the rational for taking the next step.

## YOU WANT A SPECIFIC COMPOUND

"A horse, a horse, my kingdom for a horse!" Richard III

Every now and then there is an exceptionally strong reason to want a specific sample; research priorities sometimes narrow dramatically and what had been a rather wide field of view can focus into a pin-point of current interest. Usually "wanting a specific sample" is based on the discovery of exceptionally compelling phase transitions or states in a particular compound. This has happened several times over the past decades and when it does it means that "other discoveries" such a second phases or differing ground states get put aside for a while (hopefully to be returned to at a later time). In our group the "specific sample" wanted can come from outside discoveries or as part of our own internal efforts. In the latter case it often evolves more naturally out of ongoing projects focused on searching for specific ground states or explorations. We will discuss those examples in the paragraphs below. For now, though, let me illustrate how "*want this*

*NOW*” can bring new ideas and projects to an effort and just how quickly it smoothly flows into and feeds other motivations for growth as well. At some level, in a healthy research environment, “want this NOW” is like a big rock hitting a pond. There will be a large, initial disruption, lots of unexpected activity, but soon the waves turn into ripples and ultimately the new ideas and techniques get incorporated into the gestalt of the group.

The discovery of superconductivity in the Y-Ni-B-C quaternary system, specifically in the polycrystalline samples of $RNi_2B_2C$ compounds, [57] [58] [59] led to a strong desire to have single crystals of these quaternary materials. These held very clear promise of being a fascinating set of magnetic superconductors and measurement of anisotropic properties on single phase crystals was desirable. The problem with growing these compounds was multifold, though. Not only were they quaternary, but with B and C they were refractory. Initial attempts were based on seeing if any fifth element, ideally with a relatively low melting point could serve as a flux. Sadly, in this case, none of the “usual suspects” worked out. Fortunately initial studies showed that the $RNi_2B_2C$ could be found in non-stoichiometric arc-melted pellets rich in excess $Ni_2B$. [58] [59] Indeed, annealing of such arc-melted pellets at temperatures approaching 1500 °C led to the growth of small single crystals. These hints led to the realization that $RNi_2B_2C$ could be grown out of excess $Ni_2B$. [19] [60] Figure 14 shows the binary Ni-B phase diagram and illustrates the series of eutectics that create a fairly wide region of sub-1200 °C liquid.

The very strong desire to grow single crystals of $RNi_2B_2C$ pushed our decanting temperature to the limit that silica offers. We developed a two step cooling process, the first using the vertical tube furnace and bubbler arrangement, shown in Fig. 13 and described above, to slow cool from 1480 to 1190 °C. The growth was then cooled to room temperature over ~ 12 hours, sealed into amorphous silica, and then heated to 1200 °C in a more standard box furnace, held there for ~6 hours to re-melt the excess $Ni_2B$ and, finally, decanted. Single crystal plates, with dimensions sometimes only limited by the inner diameter of the growth crucible, [19] resulted from these growths, see inset to Fig. 14.

Whereas the initial motivation was wanting single crystals of $RNi_2B_2C$ for anisotropic $H_{c2}(T)$ measurements, developing the ability to grow these compounds marked a transition of its own: we transitioned from the “can we make it?” stage to the much less uncertain and much more fun, “now that we own it, what can I do with it?” stage. We could, for example, study details of the local moment magnetism ranging from zero field studies of ordering wave vector based on neutron diffraction measurements on isotopically enriched $^{11}B$ based samples, [61] [62] to detailed angular dependent studies of metamagnetism which led to our discover of 4-state clock model moments for $HoNi_2B_2C$. [63] Again using B-isotopes we could study the isotope shift (relatively small, but a great learning exercise), [64] and by extending our studies of rare earths to R = Yb, we discovered a rare, at that time, example of an Yb-based heavy Fermion in $YbNi_2B_2C$. [65] As we collaborated more and more widely we discovered a long sought after example of a hexagonal to square flux line lattice (FLL) transition [66] and, based on very tight collaborations with theorists, we embarked on a systematic substitution of Co onto the Ni site so as to change the electronic mean free path and tune the critical field for the FLL hex-to-square phase transitions. [67]

Another example of "We really want to grow a single crystal of *this* specific compound" was the icosahedral R-Mg-Zn quasicrystal (QC) family. In 1997 we already had a deep appreciation for the beauty of using the rare earths (R) as tuning mechanism for RXY series,[25,32] and the discovery of what appeared to be a stable, rare earth bearing quasicrystalline phase was unignorable, [68] especially when there were initial claims of long range local moment ordering added to the mix. [69] We *really* wanted single grain, single phase samples of these materials to study their magnetic (and other) properties. Fortunately, at about this time an initial, pseudo-binary phase diagram had been published [70] with a very clear indication that (i) the desired QC phase was most likely incongruently melting and (ii) it appeared to have a fairly exposed liquidus surface. At this point in the article, I hope that it is clear that such a combination of desire and phase diagram made this irresistible.

In order to hold the Mg-rich melt and, simultaneously, be able to separate the crystals (quasi- or not) from the remaining liquid when done with growth, we developed the 3-cap tantalum crucibles [18] shown in Fig. 11. With these crucibles we were able to readily fine tune the growth binary phase diagram (see Fig. 15) so as to produce some of the first, large grain, icosahedral quasicrystals [71] [72] with natural growth habit, e.g. pentagonal dodecahedra (inset to Fig. 15 and Figs. 16 a-c). Although these samples were grown 16 years after the discovery of quasicrystalline order, [73] there was still a lingering doubt in the wider physics community about whether these were truly macroscopic solids with five-fold symmetry. Images such as those shown in Figs. 15 and 16 led to two divergent types of interesting questions / discussions at meetings: (i) "How long did it take you to cut and polish the sample into that shape?" (A: Only as long as it took to grow….but no polishing was needed; this is the natural habit of this compound.); (ii) "My goodness, is that the real scale and morphology? I finally believe that these are stable and five-fold." (A: Thank you for the kind comment; isn't solution growth wonderful?)

As in the case of the $RNi_2B_2C$ materials, having access to, and control of, single grain i-RMgZn samples for R = Y, Gd – Tm allowed us to again transition from "can we grow it?" to "how can we use it as a model system that lets us ask questions?". Measurements on single phase, single grain i-RMgZn very quickly demonstrated that there was no long range magnetic order, [74] but instead the i-RMgZn compounds presented a model spin-glass system with large local moments on and ordered, albeit aperiodic sublattice. [75] By taking advantage of the fact that we could populate this sublattice with local moments that were Heisenberg-like (Gd), non-Heisenberg-like (Tb-Tm) and non-magnetic (Y) we were able to study the differences and evolution between Heisenberg and non-Heisenberg spin-glass systems. [76] In addition, since with i-RMgZn we appreciated that solution growth was a synthetic tool that had not been well applied to quasicrystalline systems, we expanded our repertory to non-moment bearing systems such as i-AlPdMn, i-AlPdRe, and decagonal AlNiCo (examples shown in Fig. 16). [77] [78] [79]

Whereas for $RNi_2B_2C$ and i-RMgZn we had to create new solvents, extend growth techniques to extremes of temperature, devise new crucibles to handle reactive and volatile elements, etc., sometimes nature is actually rather kind and the sample you "really want" can be grown more or less readily and easily. $Yb_{14}MnSb_{11}$ and $BaFe_2As_2$ provide two examples of this. $Yb_{14}MnSb_{11}$ had been studied in polycrystalline form and there were, at the time, concerns that Yb may be moment bearing. [80] In order to address this, clean, single phase, single crystals were needed. What was particularly attractive about this material was the fact that it had roughly the same rare earth to antimony ratio as CeSb, one of the compounds that we discussed in detail when reviewing the basics of solution growth. For our first attempts at growing single crystal $Yb_{14}MnSb_{11}$ we tried Sn as the flux and immediately found that we could grow this ternary compound out of a quaternary melt. [31] Every now and then "rounding up the usual suspects" does work. Yb was not moment bearing in this material, but single crystals of $Yb_{14}MnSb_{11}$ and the related $Yb_{14}ZnSb_{11}$ [81] [82] proved very important for the development of thermoelectric materials.

When superconductivity was discovered in polycrystalline samples of K-doped $BaFe_2As_2$ [83] we really wanted to grow single crystalline samples as quickly as we could. Recalling that As is just above Sb in the pnictogen column and based on our past experience with CeSb and $Yb_{14}MnSb_{11}$, Sn was a logical flux for initial growth. After a few days of initial growth and optimization we has the world's first single crystals of superconducting, K-doped $BaFe_2As_2$. [84] Again, a combination of experience and good fortune allowed for rapid progression from "wanting it" to creating a massive research agenda based on growth and control of these compounds. [13] We will discuss the Fe-based superconductors more in the next sections.

It should be noted that not every expedition comes back with single crystalline trophies. There are many compounds that "we really want" that are still eluding us. In some cases, though, it is important to realize what you ***can*** achieve, and figure out how (or if) you can work with what you can make. $MgB_2$ is an excellent example of this. In early 2001 we heard rumors that $MgB_2$ had been found to superconduct near 40 K [85]. Given that this was only rumor, we immediately wanted to determine if it was true or if this was just another USO (unidentified superconducting object) sighting. On one hand, as part of our initial attempts to make $MgB_2$ out of Mg rich melts we quickly found out that we could make very high purity, polycrystalline $MgB_2$ by reacting high purity B with Mg vapor in a sealed Ta tube. [86] [87] [88] On the other hand, despite trying scores of ternary and quarternary growths we were not able to finesse single crystals of $MgB_2$ out of any ambient pressure growths. Fortunately we were able to perform extensive, basic characterization on polycrystalline $MgB_2$, [3] [4] [85] [86] [87] [88] ranging from the boron isotope effect to even developing a way of inferring anisotropic upper superconducting critical field, $H_{c2}(T)$, data from polycrystalline measurements. [89] As new material physicists we use the samples we can make and extract as much information from them as possible. To quote Mic and Keith, "You can't always get what you want but, if you try, some time, you find, you get what you need." [90] [91]

In each of these cases of "wanting a specific sample" I have highlighted what could be called "a disruptive event", a new sample type with specific, highly desirable, properties. There are, of course, thousands of examples of growths we make where a specific sample is hoped for or even anticipated. But these growths are more readily associated with one of the other two motivations for trying to grow something.

YOU WANT A SPECIFIC GROUND STATE

"I want you, I want you so bad, it's driving me mad, it's driving me mad."
The Beatles

Often the backbone of NMP research is based on the design, discovery, growth, characterization and manipulation of specific ground states or phase transitions. Whereas such programs are not necessarily focused on a single material, there is often a narrowing of phase space associated with such efforts and goals. This narrowing can be considered as a negotiation with Nature (as in "Mother") in which the NMP researcher tries to identify specific physical requirements that can be used to limit the materials phase space that needs to be searched. Often such limitations can often result in (i) the use (or non-use) of specific elements, (ii) a focus on specific lattice or atomic point symmetries, or (iii) studies of specific crystal structures.

A conspicuous example of using specific elements can be found in the search for non-radioactive, heavy Fermion compounds. A first cut at such a search would be limited to intermetallic compounds containing Ce and Yb. These are both elements with well documented ambivalency and compounds bearing them define rich hunting grounds for strongly correlated electron physics. If the research focus is to find model / toy / simple systems, then a crystal structure with a single unique crystallographic site for the Ce or Yb is desirable since multiple sites would allow for a comparable increase of salient energy scales and CEF splittings. In a similar manner, a crystal structure that does not report a width of formation would be desirable as well. If there is a known width of formation for either of these rare earths or for their ligands, it can lead to poorly controlled scattering, as well as a distribution of local environments and energy scales (e.g. possible distribution of Kondo temperatures).

As a general rule, my own group often tries to examine Ce and Yb compounds in the context of the wider rare earth series that they are parts of. This allows us to place their ordering and/or anisotropies in context of (i) more local moment members and (ii) non-magnetic members of the same series. In many cases the properties of Ce and Yb compounds are discovered as an additional part of larger studies of rare earth series; YbBiPt, $YbNi_2B_2C$, $YbNi_2Ge_2$, and YbAgGe were found as part of such progressions. [33] [60] [65] [92] [93]

Perhaps a less obvious, and much more focused, search for a specific ground state can be illustrated by the design and discovery of heavy Fermion behavior in $PrAg_2In$. Trivalent Yb and Ce are Kramer's ions and can have their Hund's rule ground-state multiplet J split by crystalline electric fields (CEFs) only as far as a doublet with a guaranteed Rln2 worth of entropy that has to be removed at low temperature. Trivalent Pr, on the other hand, is a non-Kramer's state and can have a CEF split ground state that is a singlet with a disappointingly small Rln1 = 0 worth of entropy. The reason that this is of note is that praseodymium is also ambivalent, $Pr_4O_{11}$ being an oxide with mixed valency. This means that Pr ions can, potentially, hybridize at low temperatures, forming a potential Kondo state if there is still some entropy to play with. Fortunately, if the $Pr^{3+}$ ion is in a cubic point symmetry, then it can have a non-magnetic, $\Gamma_3$ doublet as a ground state. [94]

The statements made above provide a clear example of how a "negotiation with Nature" can work. If we want to try to find an example of a Pr-based heavy Fermion, then we need to preserve some entropy to low enough temperatures for the hybridization to lead to a correlated electron state below the Kondo temperature. In order to preserve entropy in a non-Kramer's ion we look for Pr in a cubic point symmetry. Cubic point symmetry only occurs in cubic unit cells. So, the marching orders were to look for Pr-based intermetallic compounds with cubic unit cells and a single Pr site that is in a cubic point symmetry. This actually narrowed the phase space dramatically. $PrAg_2In$ rapidly emerged as a likely candidate and with it we were able to establish Pr-based heavy Fermions as a reality. [7] [95]

We do not have to limit our examples to heavy Fermion physics; another example of how the unit cell and point symmetry can be vital parts of the search for a specific ground state is the search for local moment systems that manifest extreme single ion anisotropy. During our studies of the interaction between local moment magnetism and superconductivity in the $RNi_2B_2C$ materials we discovered exceptionally planar anisotropies in $HoNi_2B_2C$. [63] Not only are the $Ho^{3+}$ moments confined to the basal plane at low temperatures, but, instead of behaving like the idealized xy-moment that is considered to be isotropic in the plane it is confined to, in $HoNi_2B_2C$ the $Ho^{3+}$ moments become confined to the four [110] directions, basically behaving like a 4-state clock model. This profound, in-plane anisotropy manifests itself in exceptionally clear angular dependencies of the meta-magnetic phase transitions at low temperature. Indeed, by analysis of these angular dependencies the net distribution of ordered moments along the [110] directions could be inferred for each field stabilized state. [63] At the time, these very sharp meta-magnetic phase transitions, that manifest such clear and tractable angular dependencies, were considered remarkable and we wanted to confirm that it was not somehow pathologic to the $RNi_2B_2C$ family. Our desired state, then, was extreme planar local moment antiferromagnets.

This round of negotiations with Nature led to rare earth compounds (for well-defined local moments) with a single unique crystallographic site for the rare earth (for simplicity) with a tetragonal point symmetry (for the planar anisotropy that could then manifest 4-fold in-plane variation). This final requirement brought an extra constraint of looking at systems with tetragonal unit cells. This cascade of requirements led to the creation of what we called "the tet-list", a compilation of rare earth series (RX binaries or RXY ternaries) that survived the negotiations. Since prediction of the sign and size of the CEF splitting was (and to a large extent still is) beyond the grasp of computational methods, we then needed to grow representative members of each series to determine whether the extreme anisotropy we were interested in was present. Whereas extreme axial anisotropy is readily found in the R=Tb member of the $RNi_2Ge_2$ series (which orders antiferromagnetically below 17 K), planar anisotropy became extreme for the R=Er and Tm members for which the ordering temperatures had dropped to or below a conveniently reached 2 K. [33] The $RAgSb_2$ series, on the other hand, manifest extreme planar anisotropy for R = Dy with AF order occurring just below 10 K. [96] Again the angular dependence of the single ion anisotropy as well as the metamagnetic phase transitions were consistent with local moments constrained to the four [110] directions [97] and, as for $HoNi_2B_2C$ before it, the net distribution of moments along each direction could be inferred.

Having Heisenberg ($Gd^{3+}$ or $Eu^{2+}$) moments, Ising moments ($Tb^{3+}$ in $TbNi_2Ge_2$), 4-state clock model moments ($Ho^{3+}$ in $HoNi_2B_2C$ and $Dy^{3+}$ in $DyAgSb_2$) we wanted to further study the potential for extreme in-plane anisotropy. An initial attempt to find a 6-state clock model system (via the creation and use of a "hex-list") failed, but instead we were able to find related behavior in the trigonal RAgGe structure with the rare earth in an orthorhombic point symmetry instead of a hexagonal one. [93] This lead to the study of the metamagnetism of TmAgGe with three crossed Ising moments, all lying in the basal plane. [98] We are, actually, still negotiating with Nature about a RXY compound that has R in a hexagonal point symmetry and manifests extreme planar anisotropy. Since it has been a decade since our last foray into this phase space there have been some new structures discovered that offer new possibilities. [99]

Sometimes the search can be for manifestations of a specific structure. Whereas there are many examples of crystalline solids with cubic, tetragonal, hexagonal, etc. unit cells, the number of stable quasicrystalline compounds is still relatively small, especially when you limit the search to binary compounds. Having more or less created the chance for the accidentally discovered a Sc-Zn, binary quasicrystal [24] (as will be discussed below when we turn to the third motivation for growth: exploration), we speculated that other, undiscovered, binary quasicrystals may be found in proximity to crystalline binary compounds, with some of the same structural motifs, called crystalline approximants. [24] [100] [101] Specifically we thought that such undiscovered QC-phases may well be compounds that decomposed peritectically at relatively low temperatures, with rather small exposed liquidus lines. We then needed to find binary crystalline approximants that, ideally, have large, exposed liquidus lines on one or both sides. The $RCd_6$ system was ideal, [100] [101] it had been identified as an approximant and existing binary phase diagram data, both experimental and computational, showed $RCd_6$ as the most Cd-rich binary with an extensive liquidus line extending from roughly 720 °C near $Gd_{10}Cd_{90}$ down to roughly 320 °C near pure Cd.

Our first test of the R-Cd system was to cool a melt of $Gd_{07}Cd_{93}$ from 700 °C down to 340 °C and decant. If indeed a more Cd-rich phase existed we should grow both the $GdCd_6$ as well as the subsequent, lower temperature new phase. Indeed, upon decanting, cooling and opening we found clear pentagonal dodecahedra growing both on the large, well faceted $RCd_6$ crystals as well as independently nucleating off of the crucible wall (inset to Fig. 17). This was a cherished moment of "***Ha!*** Exactly as we predicted!" followed by a large amount of work. [25] Given this new Gd-Cd compound, we re-evaluated the Cd rich part of the Gd-Cd phase diagram, determining the peritectic temperatures as well as the extent of the liquidus line for the formation of i-GdCd (Fig. 17). [25] We grew single phase samples of i-RCd for R = Y, Gd – Tm allowing for non-magnetic, Heisenberg and non-Heisenberg behavior. [25] [102] We also embarked on a painstaking set of growths using isotopically enriched $^{114}$Cd that would allow for elastic as well as inelastic neutron scattering measurements. [103] [104] In order not to waste large amounts of isotopically enriched Cd, we took advantage of the fritted crucibles (CCS) described above. [23] This allowed us to re-use the nearly pure Cd decant, add more rare earth and perform subsequent growths. Whereas research into the details of the low temperature state of the i-RCd system is ongoing (it is spinglass-like in nature), with both i-ScZn and i-RCd appearing as low temperature, peritectically decomposing phases adjacent to a crystalline approximant phase, we feel that we have established a viable route to finding other quasicrystalline phases.

A final example of wanting a specific ground state is our ongoing, long-term quest for manifestations of what we call fragile magnetism as a potential route to new classes of high temperature superconductors. This is presented and discussed in detail in ref. [7], but, in short, we believe that given the proximity of copper- and iron based high-Tc states to suppressed antiferromagnetic order, combined with a similar phase diagrams and exotic superconductivity in heavy-fermion systems, there is a compelling argument that new classes of high-temperature superconductivity may well be found close to other, fragile, transition metal based antiferromagnetic states. The rate-limiting step here, the difficulty, is identifying potential systems for testing with either dilution or pressure. This is not a completed project, but rather one that is ongoing and hopefully might lead to the discovery of high temperature superconductivity based not on CuO or FeAs, but some other transition metal and structural motif. Time will tell.

## YOU WANT TO SEARCH FOR KNOW AND UNKNOW UNKNOWNS

> "Exploration is in our nature. We began as wanderers, and we are wanderers still." Carl Sagan
>
> "Remember, no matter where you go…there you are."
> Buckaroo Banzai

Another motivation for the growth of a new material is the desire to explore new territories. There are actually two versions or variants of such exploration; in some cases there is the desire to start by the campfire and walk out, exploring the dark edges or peripheries. This is the exploration of known unknowns. The other variant is to parachute deep into unexplored territory. This is exploring for unknown unknowns. We will start with known unknowns and the work our way out form there.

Explorations of known unknowns are the logical consequence of succeeding in an earlier growth motivated by reasons 1 and 2 above, i.e. you got the compound or the ground state you wanted. An obvious known unknown experiment is to start from a given member of a RXY series and "run across the series", i.e. grow the series for as many different R = rare earth members as is possible. [25] [26] [32] [33] [52] [60] [75] [92] [93] This is a known unknown because if, as is often the case, other RXY members are *known* to form, but have not been measured beyond crystal structure, then their physical properties are *unknown.* It is known that there will be magnetic properties with varying anisotropies as well as possible hybridizations for R = Ce and Yb members. What is not known is how extreme these may be (both anisotropies as well as hybridizations). In addition, sometimes there are surprises, even associated with the non-moment bearing R = Y, La or Lu members. For $LaAgSb_2$ we discovered multiple charge density wave transitions, [96] [105] for $LuFe_2Ge_2$ we discovered a density wave like feature that has been associated with a much rarer spin density wave transition, and possible fragile magnetism. [32] [106]

More specific, or focused, versions of exploration of known unknowns can be made as well. Once we knew how to grow $RNi_2B_2C$ materials and understood the basic properties of the pure compounds we wanted to see what happened as we progressed from $HoNi_2B_2C$ with $T_c > T_N$ to $DyNi_2B_2C$ with $T_c < T_N$. It was not clear how $T_c(x)$ for the $(Ho_{1-x}Dy_x)Ni_2B_2C$ series would change as $T_c$ dropped below $T_N$. [107] We knew that there would be an answer to this question if we made the series and measured the properties. This is a clear known unknown. Similarly once we had i-RMgZn quasicrystals for R = Y, Gd – Er and appreciated that these were clear, potentially model, spin-glasses, then we could grow i-$(R_{1-x}R'_x)$MgZn samples and study the evolution of the freezing temperature with Heisenberg and non-Heisenberg moments. [76] These explorations of known unknowns is the essence of longer term NMP research on systems that have been mastered, at least in terms of growth.

A richer example of growths motivated by exploring known unknowns can be found in the Fe-based superconductor work surrounding the $AEFe_2As_2$ (AE = Ba, Sr, Ca) compounds. [13] Once we established that we could grow single crystals of the previously known $BaFe_2As_2$ and $SrFe_2As_2$ compounds there were a multitude of basic questions that could be asked by new growths. Given that not only substitution of K for Ba [83] but also substitution of Co for Fe [108] had been found to stabilize superconductivity a clear known unknown question was what did the T-x phase diagram of $Ba(Fe_{1-x}Co_x)_2As_2$ look like? (For the answer, refer to Fig. 2 above.) This motivated the systematic growth of scores of Co-substituted samples, and resulted in one of the iconic phase diagrams for the Fe-based superconductors. These samples not only generated great interest within our own NMP group, but also formed the basis of an extensive experimental, theoretical and computational collaboration that has led to our (Humanity's) current understanding of the interaction between this system's electronic, magnetic and structural degrees of freedom. [13] [109] [110] Given the richness of the $Ba(Fe_{1-x}Co_x)_2As_2$ T-x phase diagram, a related series of known unknown explorations emerged: studying the wider set of $Ba(Fe_{1-x}TM_x)_2As_2$ T-x phase diagrams (e.g. for TM = Co, Ni, Rh, Pd or Ru). [11] [13] [111] [112] [113] These studies entailed the growth of literally hundreds of different samples, but helped refined our understanding of how these competing phase transitions were tuned by substitution and interacted with each other.

A somewhat less certain known unknown (or at least a hoped for unknown) study was generated by the relatively obvious observation that whereas $BaFe_2As_2$ and $SrFe_2As_2$ were known members of this structural class, there were no reports of isostructural compounds with Ca or Mg (the next two elements up the AE column). As a result our first growths trying to address this question we were (i) able to discover $CaFe_2As_2$ [114] as well as (ii) perpetuate the non-discovery of $MgFe_2As_2$. $CaFe_2As_2$ was interesting regardless of its ground state; as an isostructural, isoelectronic analogue to $BaFe_2As_2$ and $SrFe_2As_2$ it offered insight into these materials no matter how it behaved. As it turned out $CaFe_2As_2$ has revealed itself to be the extreme member of this series in several ways. It is a very curious intermetallic, highly malleable and on the edge of a structural phase transition that involves a 10% collapse of the c-lattice parameter. [13] [115] It is phenomenally strain and pressure sensitive and can even be tuned by careful post growth annealing and quenching. [47] [48] It has even, recently been demonstrated as a superelastic and potential shape memory material. [116] The search for this particular known unknown led to one of the poster children for FeAs-based superconductivity in terms of competing ground states and the importance of magnetism for this form of superconductivity.

Sometimes a broad, or less than focused idea or hunch can lead to searches into the unknown. In the same manner that "Go west young man" broadly motivated explorations in the early years of the United States, a broad physical idea or desire can motivate exploratory growths. Based on our work in quasicrystals, spinglasses and heavy fermions we wanted to find and study dilute-rare-earth-bearing, ordered, crystalline compounds. The motivations were diverse: wanting to see if larger and larger unit cells had effects on properties (keeping in mind that a quasicrystal can be approximated as having a very large unit cell); wanting to see if fully ordered, but dilute, hybridizing rare earths could more readily map onto single ion Kondo models; wanting to see if larger and larger unit cell materials were harder or easier to nucleate and grow….As a result of these desires and motivations, we identified the, at the time, recently discovered (crystallographically) [117] series of $RT_2Zn_{20}$ compounds; materials with cubic unit cells that are over 14 Å on a side with 184 atoms per unit cell and yet only a single R-site (and also a single T-site as well).

The $RT_2Zn_{20}$ compounds were reported to form poorly, with potentially significant widths of formation, but there had been no systematic single crystalline work. In order to pursue these materials we needed to come to grips with the rather high vapor pressure of Zn. Fortunately Zn is not highly toxic (ZnO being used for sunscreen rather than rat poison), so our main concern was the loss of growths associated with the explosion of silica ampoules becoming over pressurized at high temperatures due to partial pressure of Zn. We found that whereas we could take Zn rich growths up to 1050 °C (roughly 4 atmospheres of Zn partial pressure) going to 1100 °C almost uniformly led to explosions. This is consistent with our experience that well sealed silica tubes with 1 mm wall thickness can withstand a few bars of pressure difference across them, but not too many. [118] We found that not only could we grow remarkably large (often crucible limited) single crystals (inset to Fig. 18a) of $RT_2Zn_{20}$ compounds, but they could be grown with little width of formation. More importantly they presented a plethora of unexpected properties, revealing a buffet of unknowns to be explored.

Given that R = Gd compounds can offer a measure of the magnetic ordering temperature for a rare-earth series (Gd having the largest dG value in Table 1), the first compounds we studied were the $GdT_2Zn_{20}$ (T = Fe, Ru, Os, Co, Rh, Ir) materials. [119] [120] Whereas $GdCo_2Zn_{20}$, $GdRh_2Zn_{20}$ and $GdIr_2Zn_{20}$ ordered antiferromagnetically near 4- 8 K, much to our surprise, the $GdT_2Zn_{20}$ compounds for T = Fe, Ru, Os ordered ferromagnetically with Curie temperatures ranging from 86 K for $GdFe_2Zn_{20}$ to 20 K for $GdRu_2Zn_{20}$ down to the more expected 4.3 K for $GdOs_2Zn_{20}$. This remarkable difference in ordering temperatures is not associated with the transition metals bearing local moments; they do not. Only Gd appears to be moment bearing in these six compounds, although changing the transition metal clearly has huge effects on the magnetic ordering.

As is often the case when confused, we made another set of compounds so as to be able to ask Nature (the Anthropomorphic Ideal rather than the journal) a clarifying question: Is there any difference between the non-moment bearing R = Y and Lu variants of these $RT_2Zn_{20}$ compounds? We found that $YFe_2Zn_{20}$ and $LuFe_2Zn_{20}$ are nearly, but not quite, Stoner ferromagnets. [119] Instead they are metals that are highly polarizable, with greatly enhanced electronic paramagnetism. The T = Ru analogues are somewhat enhanced and the T = Os analogues are not significantly enhanced. [119] [120] We were able to confirm this Stoner enhancement as origin of the large $GdFe_2Zn_{20}$ ferromagnetism by tracking the changes in $Y(Fe_{1-x}Co_x)_2Zn_{20}$ and $Gd(Fe_{1-x}Co_x)_2Zn_{20}$ as well as $YFe_2(Zn_{1-x}Al_x)_{20}$ and $GdFe_2(Zn_{1-x}Al_x)_{20}$ (Fig. 18a) and correlating them all to changes in the density of states with band filling. [119] [121]

Whereas our work on R = Y, Lu, Gd members of the $RT_2Zn_{20}$ series revealed clear d-shell correlated electron behavior, the $RT_2Zn_{20}$ family had an ever larger, correlated electron "unknown" to reveal: an unprecedented clustering of Yb-based heavy Fermions. Having mastered the growth of the R = Y, Gd, Lu members of the $RT_2Zn_{20}$ compounds, runs across the series, as discussed a few paragraphs above, were obvious known-unknown explorations. We started with the $RFe_2Zn_{20}$ series and when we got to $YbFe_2Zn_{20}$ we discovered very different and deviant behavior compared to the rest of the $RFe_2Zn_{20}$ compounds; Yb was very clearly hybridizing and was actually becoming very heavy (Fig. 18b). [122] Based on this discovery we examined the other five $YbT_2Zn_{20}$ (T = Ru, Os, Co, Rh, Ir) compounds and found that we had formally doubled the number of known Yb-based heavy Fermions (Fig. 18b). More importantly, these six new Yb-based heavy Fermions were isostructural with the Yb environment being essentially unchanged, given that the Yb site is surrounded by a 16-member nearest and next nearest neighbor, Zn polyhedron. This allowed for detailed study of the effects of changes in the Kondo temperature, $T_{K,}$ as compared to CEF splitting, and resulted in a clear experimental confirmation of the idea of a generalized Kadowaki-Woods ratio, i.e. demonstrated that the amount of entropy being transferred over to the heavy Fermi-liquid affects the value of the resistivity. [122]

The $RT_2Zn_{20}$ family was chosen for study based on somewhat broad, somewhat vague, somewhat intuitive criteria. Some of the compounds we studied were known structurally, some were unknown, but not unexpected. The result of this searching for known and unknown unknowns was the opening of a whole class of dilute, rare earth bearing materials that exhibited novel d-shell and f-shell physics.

The $RT_2Zn_{20}$ work represents a mixture of searching for known and unknown unknowns. Sometimes we wander even further from the light of the campfire and run around in the dark. Often explorations of the unknown unknowns are the most fun, even illicit (at least in terms of funding or justification). These searches are going back to the ideas of terra incognita and hic sunt dracones that we discussed near the beginning of this article. Sometimes regions of phase space simply beckon for our attention. Sometimes these searches yield little more than lost time and burnt fingers, at other times though they can provide profound and deeply rewarding surprises. Often it is just these sorts of "crazy ideas" that give birth to new research directions and efforts. As such, for a healthy NMP effort there should always be some degree of searching for unknown unknowns. In many cases these are "one-off" ideas or searches. As such they do not necessarily make good case studies or examples. On the other hand, there are three wider such searches that are perhaps illustrative: the "deep peritectic" search, the "we find solutions" search and, most recently, the "antagonistic pairs" search.

The "deep peritectic" search derives its name as well as its primary motivation from compounds like $A_2B$ in Fig. 9, compounds that decompose peritectically at a temperature very far below the liquidus temperature for their specific stoichiometry (that is why these are referred to as deep peritectics—an informal term that is probably considered improper in metallurgical circles, but does describe these materials well). The key idea or motivation for this search is that compounds that have very deep peritectic decomposition temperatures probably have not been grown in single crystal form unless they have been grown by solution growth. This means that most of these compounds have not been studied in single crystal form, or at all. This defines a large number of compounds, then, that have not been studied and that may well have interesting properties. An almost inherent bonus of trying to grow such compounds is that they have relatively low growth temperatures and, for many systems can involve relatively benign elements; this means that deep peritectic growths can be learning growths for new students.

The Pt-Sn binary phase diagram (Fig. 19a) is a good example of a system with a deep peritectic compound: $PtSn_4$. This compound is readily grown out of a $Pt_{04}Sn_{96}$ melt cooled from 600 to 300 °C over as little as 30 hours. [23] [123] The inset to Figure 19a shows a representative crystal, large, and often crucible limited. Although, at first glance, the low decanting temperature appeared to be only a good thing, we actually found that it posed an unexpected problem to us. When we first tried this growth we were still using a plug of silica wool in the catch crucible for separation of liquid from solid. We found that at such a relatively low decanting temperature, the surface tension of Sn did not allow it to pass through the very small holes that the wool presented; the Sn, along with all crystals basically sat on the surface of the silica wool, even with an effective acceleration of more than 100 x g. We found that increasing the hole size, by changing to a fritted crucible with well defined, relatively large holes (see Fig. 10 above), the Sn could pass through to the catch crucible allowing for clean separation of the crystal from the excess liquid. [22] [23]

Having grown $PtSn_4$ crystals, we needed to interrogate the compound so as to see what its properties were. We were rather surprised that this rapidly growing compound forms with fantastically high purity and very low electron scattering. [123] Figure 19b presents the temperature dependent resistivity of $PtSn_4$ taken in 0, 50, 100, and 140 kOe applied magnetic field. Whereas the residual resistivity ratio (RRR = R(300 K) / R(2 K)) of $PtSn_4$ is around 1000, the low temperature magnetoresistance (MR) increased so exceptionally rapidly that, for applied fields above 50 kOe the 2 K resistivity exceeds the room temperature resistivity value. $PtSn_4$ was the first example of what was subsequently called "titanic" or "exceptional" MR. Based on this exceptionally large MR, combined with the fact that $PtSn_4$ exfoliates and cleaves very well, we decided to study the details of its band structure and Fermi surface with angular resolved photoemission spectroscopy (ARPES). [124] We found clear signatures of topological band structure features (Dirac node arcs) that may be responsible for the phenomenal MR effects. (See inset to Fig. 19b.)

Another example of a delightfully successful, deep peritectic project can be found in our exploration of the Sc-Zn binary system. [24] Having mastered the use of Zn as a solvent during our $RT_2Zn_{20}$ project (as discussed above), [118] [119] [120] [121] [122] we were able to add Zn to our list of viable elements for general solution growth use. The Sc-Zn binary phase diagram, as it existed before we started this study (Fig 20a), showed $ScZn_{12}$ to be a promising target compound with a peritectic decomposition temperature (475 C) far below the liquidus temperature for the same melt composition (~750 C). For initial growths, to explore and master the phase diagram (which remember, is only a suggestion of reality), we put in growths with compositions ranging from $Sc_{02}Zn_{98}$ to $Sc_{04}Zn_{96}$. Instead of getting crystal of the cubic $ScZn_6$ and/or the tetragonal $ScZn_{12}$, we discovered an additional phase with a conspicuous morphology, as shown in the inset of Fig. 20b. [24] While on a hunt for possible interesting properties associated with $ScZn_{12}$ we discovered the second known, stable, binary quasicrystal: icosahedral $Sc_{12}Zn_{88}$ (i-ScZn).

We re-determined the Zn-rich portion of the Sc-Zn binary phase diagram (Fig. 20b), and found that i-ScZn has a very limited liquidus line. [24] This helps to explain how it was missed in earlier determinations of the phase diagram. The proximity of i-ScZn to the approximant phase, $ScZn_6$, combined with the low peritectic decomposition of i-ScZn led us to speculate that there should be other, yet to be found, examples of a binary quasicrystalline systems in similar proximity to other crystalline approximants. It was precisely this hypothesis that motivated our search for, and discovery of, the i-RCd phase, a search that led to the discovery of the only, R-bearing, binary QC system. [25] [102] [103] [104]

Since, as part of the deep peritectic project, we made the Pt-Sn and Sc-Zn growths with no expectations of finding specific phases or ground states, we cannot claim that the discoveries associated with $PrSn_4$ and i-ScZn were surprises, rather, we can claim them as validation of our premise that the deep peritectic search has indeed identified a promising cut through phase space that has yet-to-be-discovered treasures hiding along it. At some level, I feel a bit foolish, waving about a treasure map that I know to be reliable, but such is the nature of an article in ROPP. I will repeat this foolishness twice again as I review the remaining two, unknown unknowns searches.

The next example of searching for unknown unknowns is what I dub, “we find solutions”. As in many parts of my group’s research, this is a play on words, but in this case both meanings are true and salient. We, as a research group, find solutions, answers, to many problems associated with the growth of new materials. The CCS is an excellent solution to several problems associated with containment, decanting, and quantification. [23] In addition, as a group, we have been developing new melts, solutions, that we can use to grow known materials out of and search for new materials with. The “we find solutions” approach is based on the development of broadly versatile melts that will allow for the explorations of a new sets of phase spaces; new pathways into a poorly explored part of the forest… hic sunt leones.

We have been trying to develop workable solutions, often binaries, that can serve as the starting points for new growths and explorations of compounds with one or both of the elements used in the binary. For some binary pairs this is a trivial exercise, e.g. for Cu and Ge the eutectic near $Cu_{65}Ge_{35}$, or even the Ni-B binary shown in Fig. 14, [19] do not present any difficulty, the elements are not particular reactive or volatile, or, for the case of Cu-Ge, even toxic. For other binary pairs, though, life is not so easy.

Over the past decade we have been developing sets of chalcogen and pnictogen based melts that allow us to work with N, P, S, As, Se, and Te in solution, at temperatures that would not seem to be possible, based on their elemental properties. The motivation for wanting to work with these elements is multifold, ranging from interest in ground states as disparate as superconductivity, ferromagnetism, and wide-gap semiconductors.

Whereas Bi and Sb are relatively easily handled, without too large or worrisome vapor pressures, As, P, and N are each difficult to work with due to vapor pressure as well as varying degrees of toxicity. Se, Te and S are similarly problematic. It should be noted, that toxicity combined with high vapor pressures is a proverbial “double whammy” [125]; the high vapor pressure can lead to pressures large enough to rupture the silica containment ampoule (e.g. $P > 5$ atm) and once containment is lost the toxic material can readily spread if precautions are not taken. In order to mitigate this problem we have (and are) studying various binary melts containing these elements to identify which ones have reduced vapor pressures, which ones can be handled safely, and which ones can serve as versatile solutions for future growths.

One of our earlier campaigns was an attempt to work with sulfur-based melts so as to more readily grow and study mineral based or mineral inspired compounds that might manifest fragile magnetism and/or superconductivity. [43] [44] Upon heating, elemental S forms viscous, polyatomic/polymeric complexes, sublimes easily and boils below 450 °C. As such, S is not a promising element to use as a solvent. In order to use solution growth for S-based compounds, we surveyed possible solutions that would allow the incorporation of S. Broadly speaking we found two classes of solutions: (1) dilute S-bearing melts and (2) S-based eutectics.

Sulfur can be dissolved into relative low melting, elements such as Bi, Pb, Sb and Sn to form melts that can be heated to over 1000 °C. [43] [44] In each of these cases we approached the binary melt with some care, adding more and more S and heating to higher and higher temperatures to determine how S-rich and how hot we could go without encountering too much vapor pressure or S-loss. For example, in the case of Bi, we found we could bring $S_{40}Bi_{60}$ to 1000 °C without any indication of S condensation on the inside of the silica ampoule. [43] [44] Having determined a range of composition and temperature for use we found that not only can these melts be used to grow the obvious, readily accessible binaries (e.g. $Bi_2S_3$ or PbS), they could be used to growth ternary, S-bearing compounds such as $Ni_3Bi_2S_2$, $Co_3S_2Sn_2$, CoSSb, and even $Fe_2GeS_4$. [43] [44]

A more surprising, and potentially more versatile, approach was associated with binary, S –based eutectics with transition metals. In the cases of Co, Ni, Rh and Pd we found that the transition metal rich eutectics could be used to grow transition binary and ternary S-bearing compounds. These eutectics can have relatively large (~ 30%) amounts of sulfur in them and are liquid for temperatures above sulfur's boiling point. The reported eutectic temperatures, $T_{eu}$ associated with them are, for example: $Co_{60}S_{40}$, $T_{eu}$ ~870 °C; $Ni_{67}S_{33}$, $T_{eu}$ ~640 °C; $Rh_{63}S_{37}$, $T_{eu}$ ~ 925 °C; and $Pd_{72}S_{28}$, $T_{eu}$ ~625 °C. In each case we were able to create these liquids and hold them in an alumina crucible. We first tested them by decanting the liquid above the eutectic temperature; then we would grow either elemental transition metals or binary compounds. In the case of Pd, we grew $Pd_4S$ with a residual resistivity ratio in excess of 700. [43] [44] In the case of Co, we were able to grow elemental Co out of the Co-S melt, forming the crystals below their Curie temperature. [43] [44] In the case of Rh we were able to grow one of the few known superconducting minerals: miassite, $Rh_{17}S_{15}$. We could, of course, then use these transition metal-rich binary melts as the staging point for exploration for compounds; [43] [44] for example we were able to grow $Rh_9In_4S_4$, $Bi_2Rh_{3.5}S_2$ and $Bi_2Rh_3S_2$ and study superconductivity as well as it competition with structural phase transitions. [126] [127]

Moving beyond sulfur and widening our focus to include P, Se and Te, as part of our search for new transition metal based magnetic systems, both hard ferromagnets as well as fragile antiferromagnets, we have developed similar transition metal based eutectics with these volatile elements. Our basic modus operandi is to identify a promising TM-X eutectic, first establish that we can indeed form it and contain it at high temperatures without significant crucible attack, internal pressure, or evaporation of X. Then, once we feel that we indeed do, "own the eutectic", we use it to grow binary, ternary or higher compounds. Using this technique we have been able to study the magnetic anisotropy of $Fe_5B_2P$, [53] discover two new ferromagnetic compounds: MnPZr and MnPHf, [54] explore the physics of the $RPd_2P_2$ series, [52] and study the potential fragility of ferromagnetism in $Fe_3GeTe_2$. [128]

One other set of solutions that we have been developing is nitrogen bearing ones. Whereas oxide growth is generally possible due to the relative simplicity of working with solutions of low melting oxides, nitride growth is much more difficult. The motivation to study nitrogen based materials is strong, though. If we carry the progression of Bi, Sb, As and P-based materials (either for Fe-based materials or other systems) to the logical extreme, there is a strong motivation to see if N-based materials may show even narrower-band effects. In an attempt to see if we could develop N-based solutions for growth we studied the Li-N and Ca-N binary phase diagrams. [46] We found that we could readily contain and work with a $Li_{90}N_{10}$ melt (see Fig. 21a) and use it to grow crucible limited single crystals of $Li_3N$ (Fig. 21a inset) as well as $Li_2(Li_{1-x}TM_x)N$, for TM = Mn, Fe, Ni, Co (see Fig. 21a). [129] [130] [131] We used our Ta-3-cap crucibles (Fig. 11) and did not have any indication of over pressurization due to N gas. In a similar manner, the Ca-N binary phase diagram was proven to allow for the creation and use of a binary melt that did not suffer from any significant nitrogen overpressure. [46]

It should be noted, though, that not everything goes so smoothly every time a new solution is tried. To illustrate this point, Fig. 21b shows the binary phase diagram for In-N. At first glance, this phase diagram looks as promising, if not more so, than the one for Li-N shown in Fig. 21a. We tested this one as well, carefully adding a little InN to In, heating up to several hundred °C above the reported liquidus value and cooling down again. We found that (i) we did not recrystallize any InN and (ii) there was a significant gas over pressurization, even at room temperature. This indicates that Fig. 21b is, at best, inadequate. It is important to note, though, that whereas the Li-N binary phase diagram in Fig. 21a was experimentally obtained, the only reported In-N binary phase diagram in the ASM Alloy Phase Database (as of this writing), i.e. Fig. 21b, is a computationally based one. The morals of this last story are (a) caveat emptor, and (b) it is important to make sure that you can indeed work with a binary melt before you start adding other elements and making life more complicated.

The final, broad, unknown unknowns search is based on “antagonistic-pairs” of elements. This search is fundamentally different from the “deep-peritectic” or “we find solutions” searches in that both of these searches emphasized either exposed liquidus lines or eutectic points. In the “antagonistic-pairs” search the starting point is finding pairs of elements that are as immiscible as possible. A clear example of such immiscibility is the Co-Pb binary phase diagram shown in Fig. 22. It is manifestly clear that these two elements *really* do not want to mix. The heart of the “antagonistic-pairs” search is to ask the question, “What would happen if a third element could be introduced to stabilize a ternary compound?” It turns out the when two elements are highly immiscible, there are, statistically, not that many intermetallic (i.e. excluding oxides and halides) compounds existing, or at least reported, in the crystallographic data bases. [56] In addition, and most importantly, when actual ternary compounds do exist, they tend to have a structure that segregates the two antagonistic elements. For example, there are three reported La-Co-Pb ternaries, each manifesting clear segregation of the Co from the Pb. [56] The clearest of these is $La_6Co_{13}Pb$; there are dense slabs of Co separated from planes of Pb by La layers. $La_5CoPb_3$ has a very different stoichiometry and has one dimensional rows of Co separated from the Pb by sheaths of La. Finally there is $La_{12}Co_6Pb$; for this compound there is enough La to completely encapsulate each Pb atom and completely isolate it from the Co. These three structures each manifest a clear lowered dimensionality that results in the segregation of the antagonistic Pb and Co atoms: (i) two dimensional (Co-slabs in $La_6Co_{13}Pb$), (ii) one dimensional (Co-chains in $La_5CoPb_3$), and (iii) zero dimensional (encapsulated and isolated Pb atoms in $La_{12}Co_6Pb$).

Whereas the Co-Pb binary phase diagram is a clear example of such immiscibility, it is hardly unique. There are many such antagonistic pairs and, in many cases, the compounds formed by them are interesting, especially when the low dimensionality is associated with, for example, 3d-shell transition metals. To not make too fine of a point, the parent compound of many of the Fe-based superconductors, $BaFe_2As_2$, [11] [12] [13] [47] [48] [83] [84] [108] [109] [110] [111] [112] [113] [114] as well as one of the prototypical fragile magnetic systems, $LaCrGe_3$, [7] [132] [133] are also antagonistic pair ternary compounds. Fe-Ba (as well as Fe-Ca) are an antagonistic pair and, in $BaFe_2As_2$ they are segregated, in planes, from each other by As. La-Cr also form an antagonistic pair and in $LaCrGe_3$ the Cr forms chains that are separated from the La by sheaths of Ge. Both of these materials manifest fragile magnetic properties that may well be associated with their reduced dimensionality.

The antagonistic-pair idea offers a chance to find materials with reduced dimensionality for either or both of the pair-elements. Many of the ternary phase diagrams involving such pairs have been only poorly explored, at best. Given that these reduced dimensionalities reduce and simplify the types of structural motifs that may occur in potential ternary structures, these systems are amenable to computational searches for new materials as well. This search for unknown unknowns is the latest addition to our ongoing efforts. Again, I do feel a little silly outlining this so explicitly, but such is the nature of a review article.

## OUROBOROS

Now that we are near the end, we can see the beginning.

The ouroboros is the image of a tail-eating-snake that often symbolizes cyclicality, especially in the sense of something constantly re-creating itself like the phoenix which operate in cycles that begin anew as soon as they end. I cannot think of a better symbol for this last section of this overview of NMP, of for NMP itself. One of the difficulties in writing this paper, especially the latter part about motivations and ideas, has been that very rarely does a project or idea magically spring from the luminiferous aether, fully formed as Athena emerged from the forehead of Zeus. Instead, there are often very complex genealogies associated with ideas and research projects, mixing and blurring motivations, phase transitions and ground states.

Figure 23 is a schematic of the idea-phase-space and allows for the tracing of the genealogies and interconnectivity of a subset of some of the long term research projects discussed in this review. In this case we can trace a line between R-bearing quasicrystals to crystalline spin-glasses to d- and f-shell correlated electron physics and back to R-bearing quasicrystals again. As mentioned above, we became interested in i-RMgZn quasicrystals based on our own interest in R-magnetism and Alan I. Goldman's interest in quasicrystals. In the late 1990's we devised new growth techniques to let us grow single grain i-RMgZn [19] [71]. In the process of our i-RMgZn work we realized that several of the Al-based, non-R-bearing QC could also be readily grown in well faceted, free standing form from high temperature solutions [77] [78] [79]. Given that the i-RMgZn systems manifest very clear spinglass physics, we pursued this further by making non-magnetic site dilutions in the $(Y_{1-x}Tb_x)Ni_2Ge_2$ system [134]; $TbNi_2Ge_2$ being a member of the $RNi_2Ge_2$ family, that we had identified as being Ising-like in its local moment anisotropy (as part of a known-unknown run across the rare earth series). [33] With $(Y_{1-x}Tb_x)Ni_2Ge_2$ we found that we could stabilize an Ising-spinglass for $x < 0.25$. We returned to the $RNi_2Ge_2$ system when we wanted to see if the whole rare earth series offered enough diversity in local moment size and anisotropy to emulate a random moment route to a spinglass state. We grew what we called the "mélange" sample, in which we intentionally created a pseudo-ternary compound by mixing up to nine moment bearing rare earths on $RNi_2Ge_2$'s single R-site. [135] [136] These samples did not form low-temperature spinglass states, but rather ordered magnetically at more or less exactly the temperature very primitive de Gennes scaling would predict. Given recent interest in "high entropy" alloys and compounds, it seems that this particular growth may merit some further study.

Based on our QC and spinglass work, we started to wonder if the size of the unit cell might have some effect on how readily a local moment system could order. On one hand, strictly speaking a QC has a "unit cell" that is as large as the single grain in hand; perhaps more formally, the size of a QC unit cell exceeds the coherence volume probed by diffraction measurements that fail to detect any translational periodicity. On the other hand compounds such as $RNi_2Ge_2$ have relatively small unit cell volumes, e.g. tetragonal with a ~ 4.06 Å, c ~ 9.8 Å and volume ~ 207 $Å^3$. [33] Clearly larger unit cell compounds could be studied, so we searched for a system with (i) a very large unit cell, (ii) as dilute in rare earth as possible while still, (3) having only a single, fully occupied rare earth site. This is what led us to the cubic, $RT_2Zn_{20}$ family with a ~ 14.1 Å and volume ~ 2,800 $Å^3$. [117] [118] [119] [120] [121] [122] In order to grow these compounds we had to master the use of Zn as a solvent, particularly work with Zn's relatively high vapor pressure. [118]

Our group's access to Zn as a solvent opened up new binary phase diagrams for exploration, both in general, but also for the deep peritectic project type of growths. The exploration of the Zn-rich part of the Sc-Zn binary was motivated by the hope of (i) possible Sc-related correlated electron effects and (ii) finding a hidden surprise. The discovery of i-ScZn [24] was a better surprise than we could have imagined. Our analysis and redefinition of the Zn-rich part of the Sc-Zn binary phase diagram led us to speculate that other "hidden" binary QC systems may be lurking adjacent to crystalline approximants such as $ScZn_6$, and indeed, when we focused on the Cd-rich side of the $RCd_6$ approximant we found exactly what we had predicted: i-RCd. [25] [102] The development of the CCS, fritted-crucibles [23] made the redetermination of the Cd-rich side of the phase diagram much more reliable and the use of isotopically enriched $^{112}Cd$ and $^{114}Cd$ viable.

So, to summarize, when colleagues see an image of a binary, i-RCd quasicrystal on the cover of a glossy journal they should *not* think how wonderful it is to be lucky and just happen to discover this more or less by accident. The discovery of the i-RCd system has been decades in coming, stemming from a long standing interest in (1) quasicrystals, (2) spinglasses associated with 4f-local moment systems, (3) large unit-cell compounds, (4) the mastery of Zn as a viable solvent, (5) an exploratory growth effort associated with compounds with relatively low peritectic decompositions temperatures. The discovery of i-RCd quasicrystals had very, *very* little to do with luck.

Similar genealogies of ideas, innovations, discoveries and even temporary dead ends or failures can be made for almost all of the compounds we work on. Like the ouroboros, NMP research is constantly consuming and re-emerging from itself; what we learn from one set of growths and measurements leads to new ideas, new techniques and, of course, new insights that spawn dozens of ideas that, in a Linus Pauling-like manner, we need to prioritize. When working well, an active and thriving NMP effort can be like a nuclear reaction going critical, each event producing multiple new ideas that in themselves can produce multiple newer ideas, etc.

It may be that we need to explain or illustrate this richness more clearly, every now and then, so that our colleagues, both within our field as well as external to our field appreciate what is involved in designing, discovering and developing new materials with novel and/or desirable properties. Unfortunately, due to our own brevity in writing, often for articles with artificially constrained page limits, we leave out the context and history that led to the material du jour. When we do this, we run the risk of perpetuating this myth or perception that either we have more in common with a head chef at a Micheline 4 star restaurant than we do with other physicists, or we are just compulsive gamblers who every now and then get lucky. This is really, *really*, not the perception we want to engender or perpetuate.

As scientists we need to recall that explaining how we discovery what we do, how we decide to follow path 1 rather than path 2, why we decide to grow this rather than that, is perhaps more important than the mechanistic details of how the superconducting transition temperature was inferred from a specific criterion. The latter is rather straight forward and, although important, not particularly insightful. As we enter into an age that will be more and more dominated by a spectacular diversity of new material discoveries, providing our students and colleagues with a window into how we rationalize or motivate our searches for new materials is very important.

As part of this communication effort, we need to make it clearer to the broader public, and certainly to public and private funding agencies, that our science has, and *should* have, a large exploratory component to it. This is in contrast to the more firmly established image of "hypothesis driven" research that is drilled into student in middle school or high school. Although we all have some level of hypothesis for any given growth, it can be a rather vague one, such as, given what I know about these elements and/or this structure class, I hypothesize that there may be some novel electronic or magnetic transitions to be found if I can grow compound XYZ. Whereas some disparage this as, "just a fishing trip", [91] this is a very damaging and inaccurate statement. Not so much for the use of the fishing imagery, but for the use of the word "*just*". Skilled fishermen know where to find exactly what they want in what seems to be a large and featureless phase space. That is a highly admirable skill. The successful NMP research effort combines skills found in professional fishermen, [91] master chefs, [137] and trackers or explorers [138]. These are key parts of what we do and key parts of what we need the public and funding agencies to understand, appreciate and support. It is our task, *our duty*, to explain this as clearly and compellingly as possible.

I sincerely believe that the long-term prospects of our species can be dramatically improved by the discovery of new materials that will help alleviate the multiplicity of problems we face. I sincerely hope that this vital, NMP research is fostered and grows in multiple academic departments. It is inherently interdisciplinary and needs support from all of its contributing disciplines. I think that the future for this field as well as for our species is promising. We just need to broadly explore and widely harvest the multiplicity of materials and states that will help provide new options and capabilities to us.

Acknowledgements:

This article summarizes decades of work and experience. I specifically want to thank both Susan Anwar and Raquel Ribeiro, both of whom have given me the hope and strength to carry on over these many years. I would also like to thank Sergey L. Bud'ko who, for much of this time, has been my closest colleague and collaborator. Finally, I thank all my group members, past and present, as well as the multitude of collaborators whom have shared our enthusiasm for exploring known and unknown unknowns. For all of this time, my work has been supported by the U.S. Department of Energy, Office of Basic Energy Science, Division of Materials Sciences and Engineering. The research was performed at the Ames Laboratory. Ames Laboratory is operated for the U.S. Department of Energy by Iowa State University under Contract No. DE-AC02-07CH11358. Over the past five years, this research has also been supported by the Gordon and Betty Moore Foundation's EPiQS Initiative through Grant GBMF4411.

## Tables

Table 1: Filling of 4f-shell in trivalent lanthanides with S, L, J generated from Hund's rules. Magnetic parameters such as saturated moment, $\mu_{sat}$, and effective moment, $\mu_{eff}$, as well as de Gennes parmeter, dG, are readily and reliably determined as well. The Landé g-factor, $g_J$, was generated using: $g_J \sim 3/2 + [S(S+1) – L(L+1)] / 2J(J+1)$.

| Ion | $n_{4f}$ | S | L | J | $g_J$ | $\mu_{sat} = g_J J$ | $\mu_{eff} = g_J\sqrt{J(J+1)}$ | $dG = (g_J-1)^2 J(J+1)$ |
|---|---|---|---|---|---|---|---|---|
| La | 0 | 0 | 0 | 0 | 0 | 0 | 0 | 0 |
| Ce | 1 | 1/2 | 3 | 5/2 | 6/7 | 2.14 | 2.53 | 0.18 |
| Pr | 2 | 2/2 | 5 | 8/2 | 4/5 | 3.20 | 3.58 | 0.80 |
| Nd | 3 | 3/2 | 6 | 9/2 | 8/11 | 3.27 | 3.62 | 1.84 |
| Pm | 4 | 4/2 | 6 | 8/2 | 3/5 | 2.40 | 2.68 | 3.20 |
| Sm | 5 | 5/2 | 5 | 5/2 | 2/7 | 0.71 | 0.84 | 4.46 |
| Eu | 6 | 6/2 | 3 | 0 | -- | 0 | 0 | 0 |
| Gd | 7 | 7/2 | 0 | 7/2 | 2 | 7.00 | 7.94 | 15.75 |
| Tb | 8 | 6/2 | 3 | 12/2 | 3/2 | 9.00 | 9.70 | 10.50 |
| Dy | 9 | 5/2 | 5 | 15/2 | 4/3 | 10.00 | 10.60 | 7.08 |
| Ho | 10 | 4/2 | 6 | 16/2 | 5/4 | 10.00 | 10.60 | 4.50 |
| Er | 11 | 3/2 | 6 | 15/2 | 6/5 | 9.00 | 9.60 | 2.55 |
| Tm | 12 | 2/2 | 5 | 12/2 | 7/6 | 7.00 | 7.60 | 1.17 |
| Yb | 13 | 1/2 | 3 | 7/2 | 8/7 | 4.00 | 4.53 | 0.32 |
| Lu | 14 | 0 | 0 | 0 | 0 | 0 | 0 | 0 |

# Figures

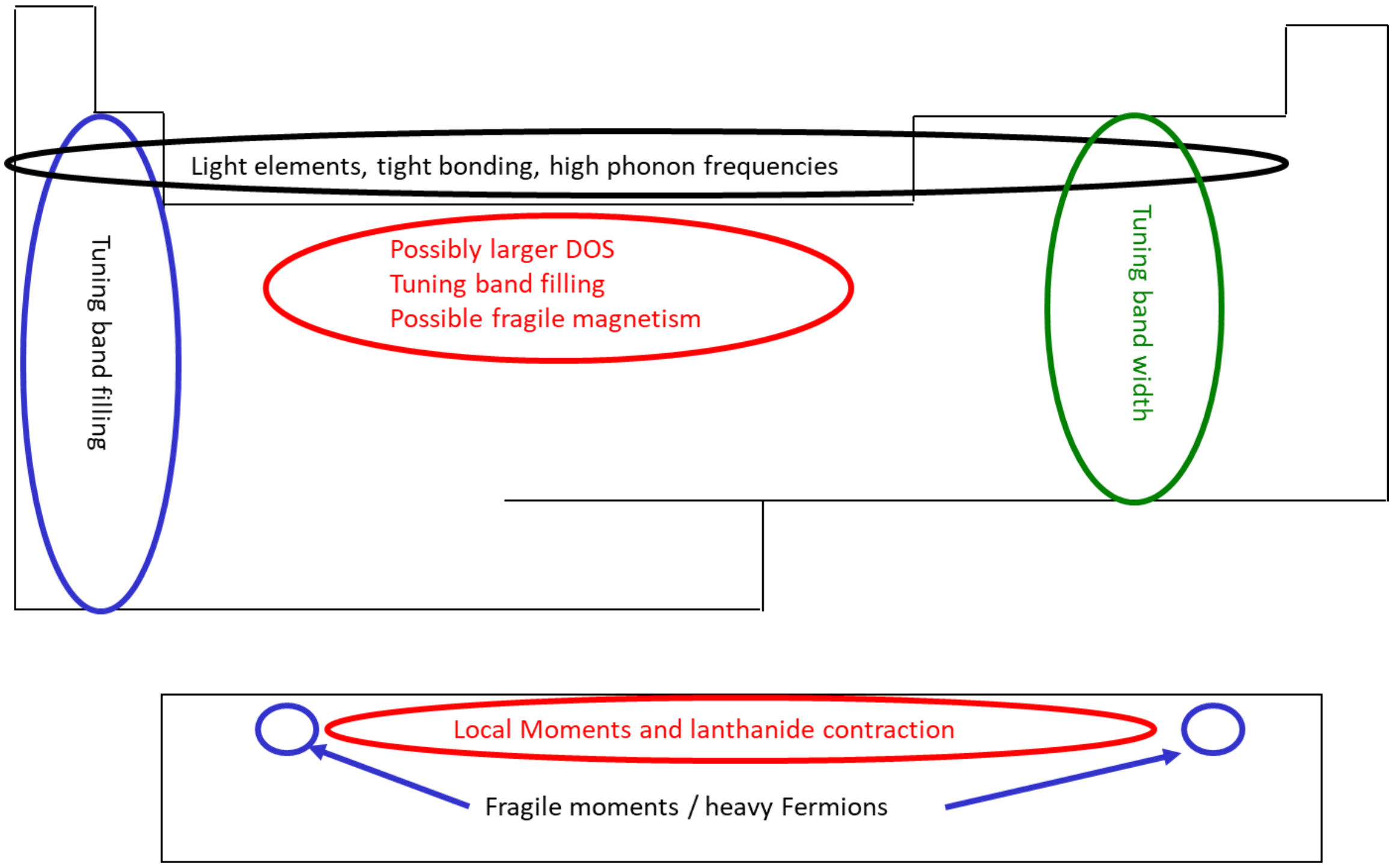

Figure 1: A schematic outline of the periodic table with various regions used for tuning terms in Hamiltonians used to describe superconductivity.

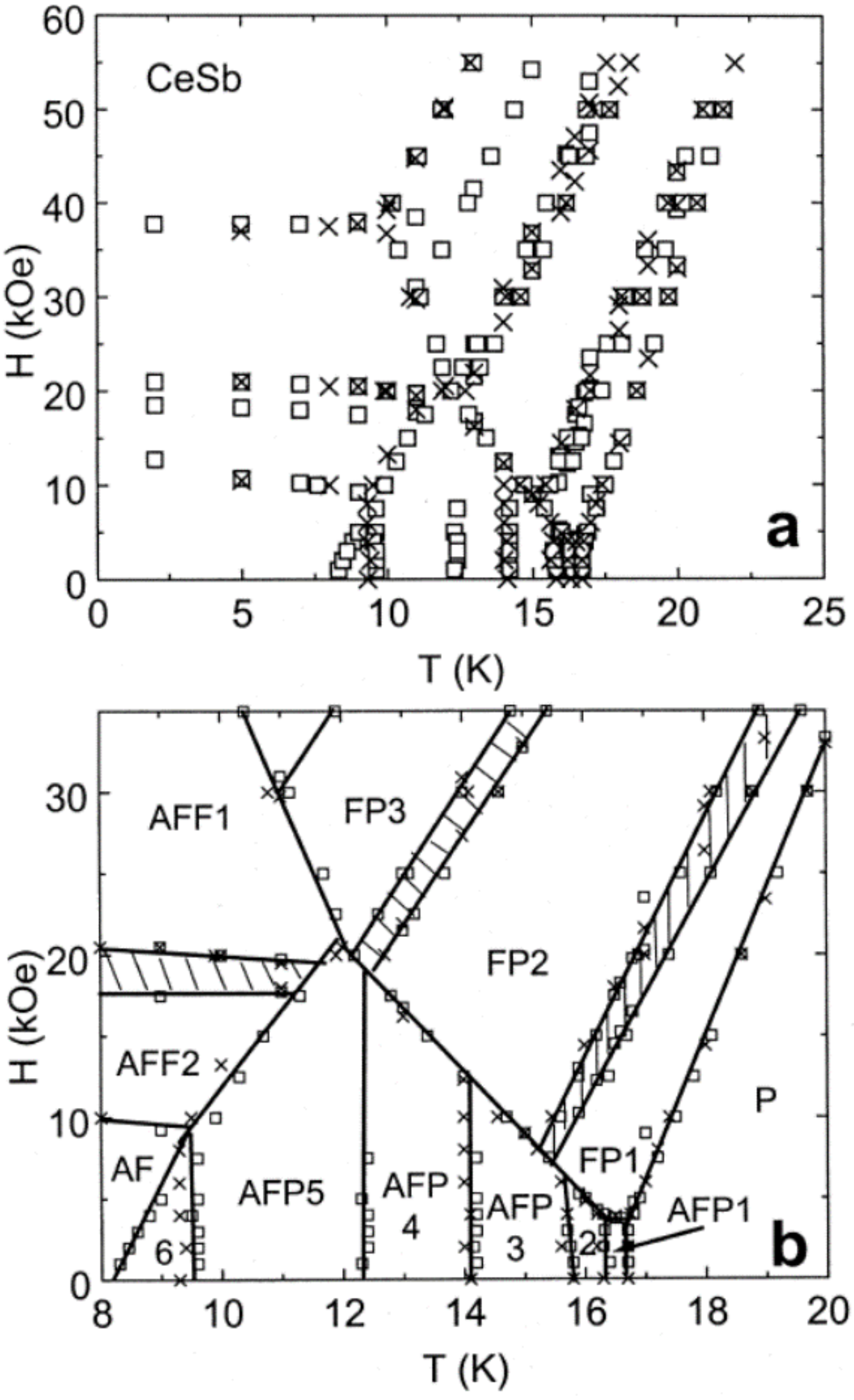

Figure 2: (a) CeSb H – T phase diagram for magnetic field applied along the (100) direction exemplifies the beauty and complexity of phases for a "simple binary" compound. (b) Enlarged plot with specific phases more clearly identified using notation from [8] and references there in; P - paramagnetic, AFP – mixed antiferromagnetic arrangement of planes and paramagnetic planes, FP – mixed ferromagnetic arrangement of planes and paramagnetic planes, AF - antiferromagnetic state, AFF - mixed ferromagnetic and antiferromagnetic arrangement of planes, shaded areas – new phases discovered in [8]. Reprinted from [8] with permission of Elsevier.

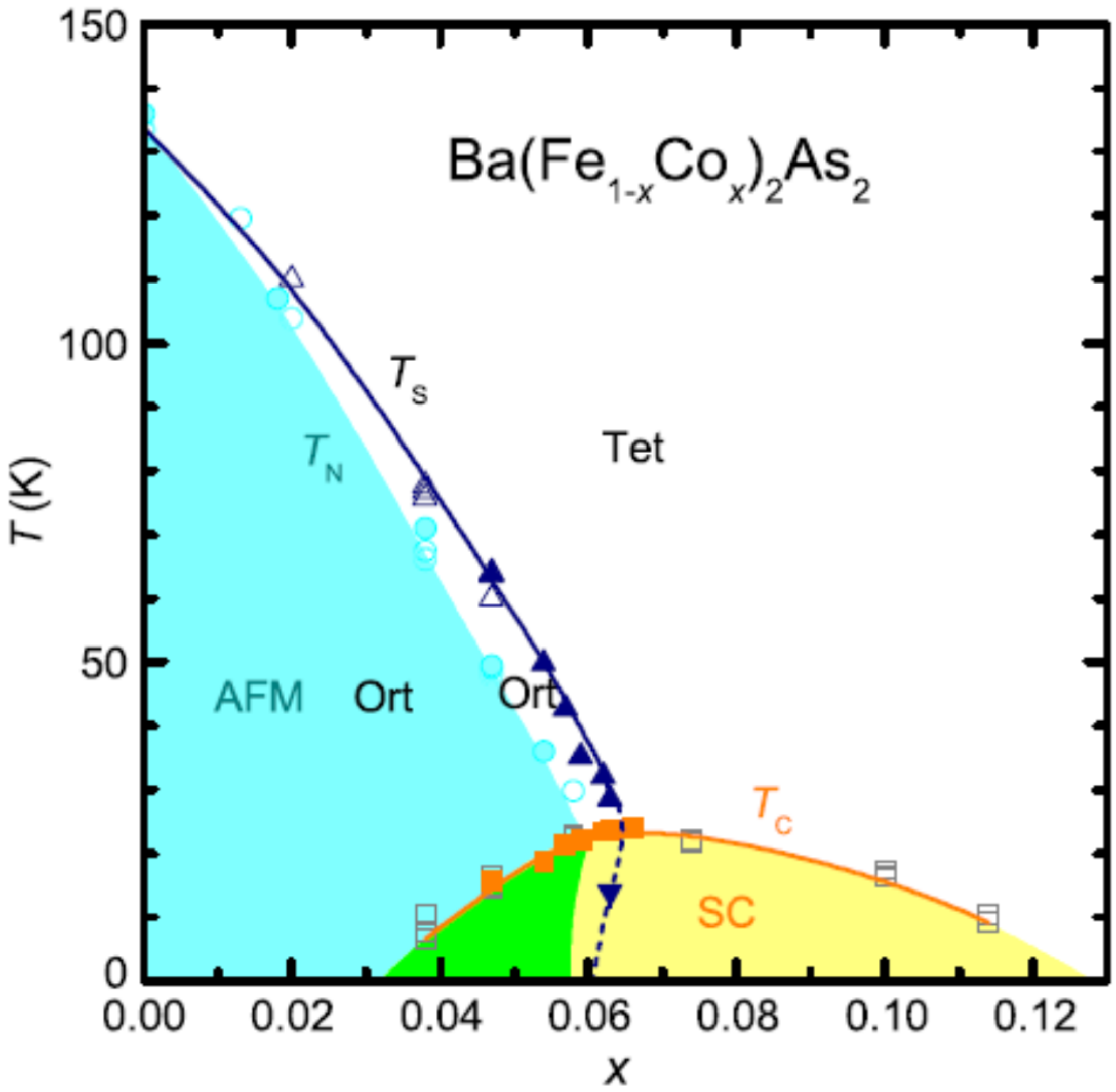

Figure 3: $Ba(Fe_{1-x}Co_x)_2As_2$ T – x phase diagram illustrating the interplay between electronic, magnetic, and structural degrees of freedom in Fe-based superconductors. [14] Structural ($T_S$), antiferromagnetic ($T_N$) and superconducting ($T_c$) phase lines associated with transitions to/from the orthorhombic (Ort) state, the antiferromagnetic (AFM) state and superconducting state (SC) respectively. These phase lines have been determined by diverse thermodynamic, structural and microscopic measurements. [13] Reprinted from [13,14] with permission from American Physical Society and Annual Reviews.

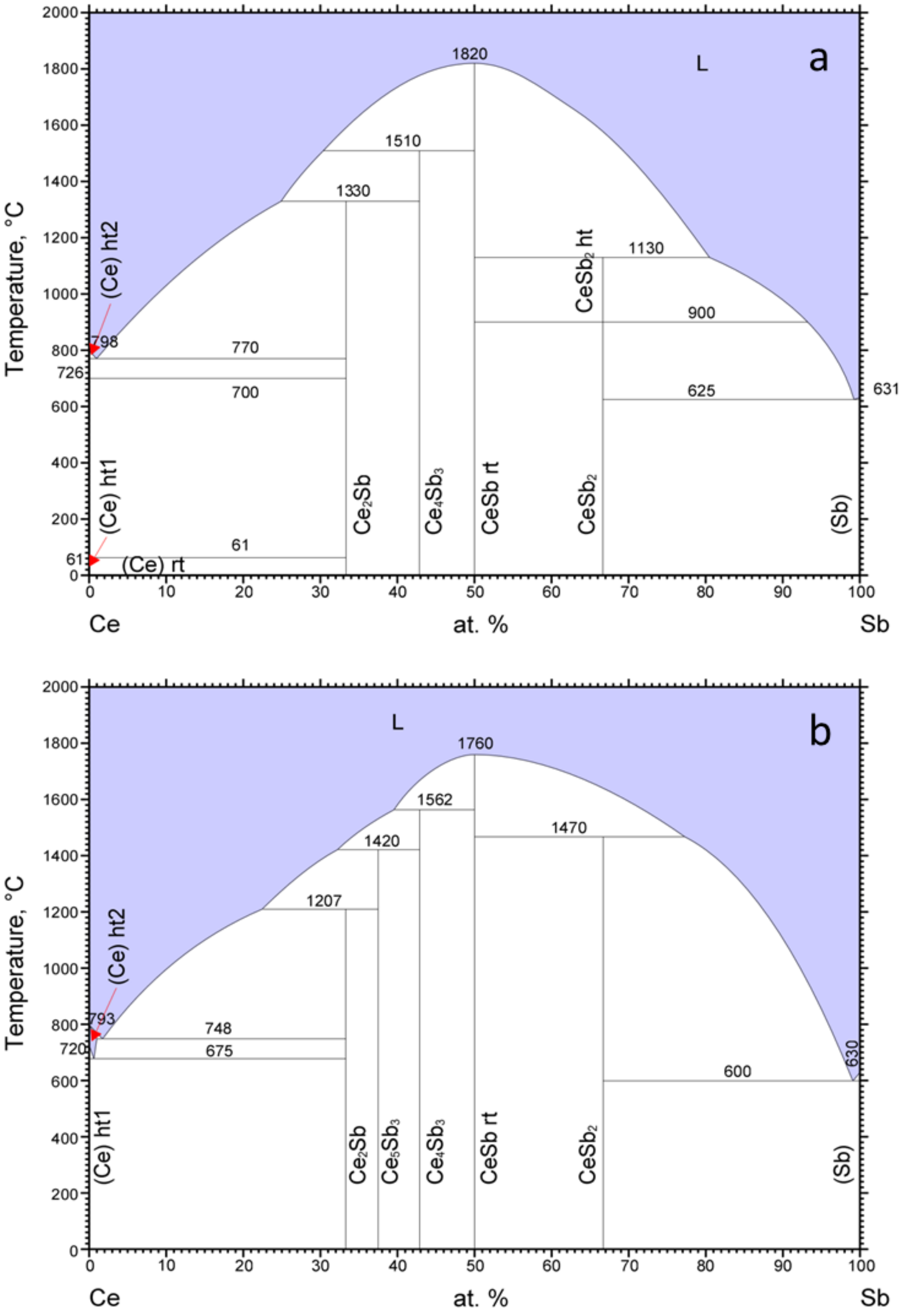

Figure 4: Two versions of the Ce – Sb binary phase diagram [16] (a) More recent one; ASM Diagram #1600390 (2001) based on differential thermal analysis, x-ray diffraction and metallography data. Reprinted with permission of ASM International. All rights reserved. www.asminternational.org. (b) Older one; ASM Diagram #902977 (1981) schematically drawn based on Nd-Sb data. Note that differences are both qualitative (extra compound such as $Ce_5Sb_3$ in older one and a high temperature phase, ht, of $CeSb_2$

in newer one) and quantitative (differences in temperatures). Reprinted with permission of ASM International. All rights reserved. www.asminternational.org.

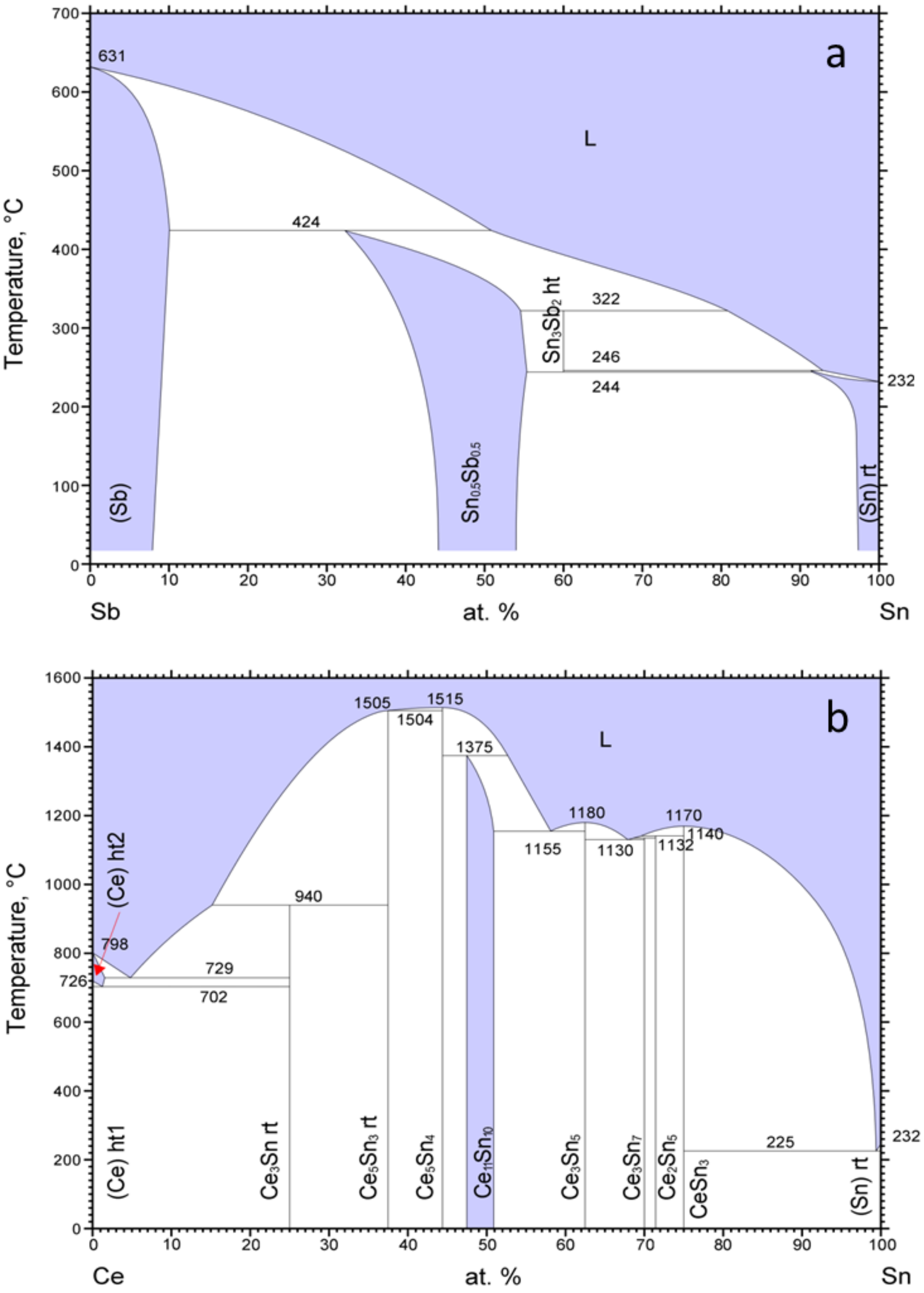

Figure 5: Binary phase diagrams [16] related to the growth of CeSb out of excess Sn. (a) Sb – Sn binary phase diagram; ASM Diagram #107140. (b) Ce – Sn binary phase diagram; ASM Diagram #2002056. These two phase diagrams, along with the Ce-Sb binary phase diagram from figure 4a allow for estimates of how to attempt the growth of CeSb using Sn as a solvent, or flux. Reprinted with permission of ASM International. All rights reserved. www.asminternational.org

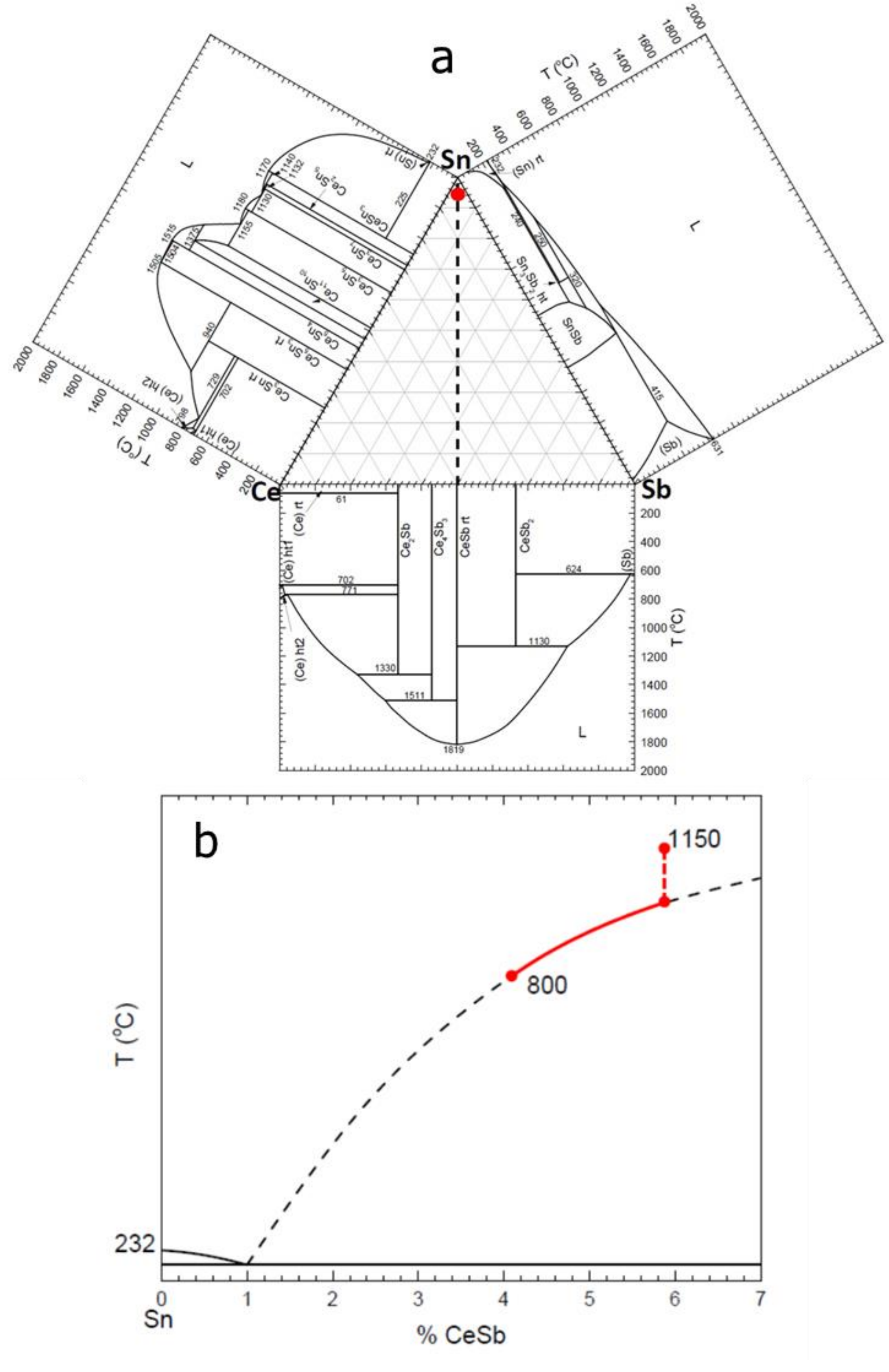

Figure 6: (a) Ce-Sb-Sn ternary, compositional phase diagram with binary phase diagrams alone three edges, red spot indicates region of pseudo binary cut. Each vertex of the equilateral, ternary phase diagram is the pure element. Any point in the ternary phase diagram is a unique ratio of $Ce_xSb_ySn_z$ with x + y + z = 1 (or 100%). (b) Schematic, pseudo binary CeSb – Sn phase diagram on the Sn rich side. The solid red segment of the liquidus line illustrates the temperature profile used to grow CeSb single crystals out of excess Sn.

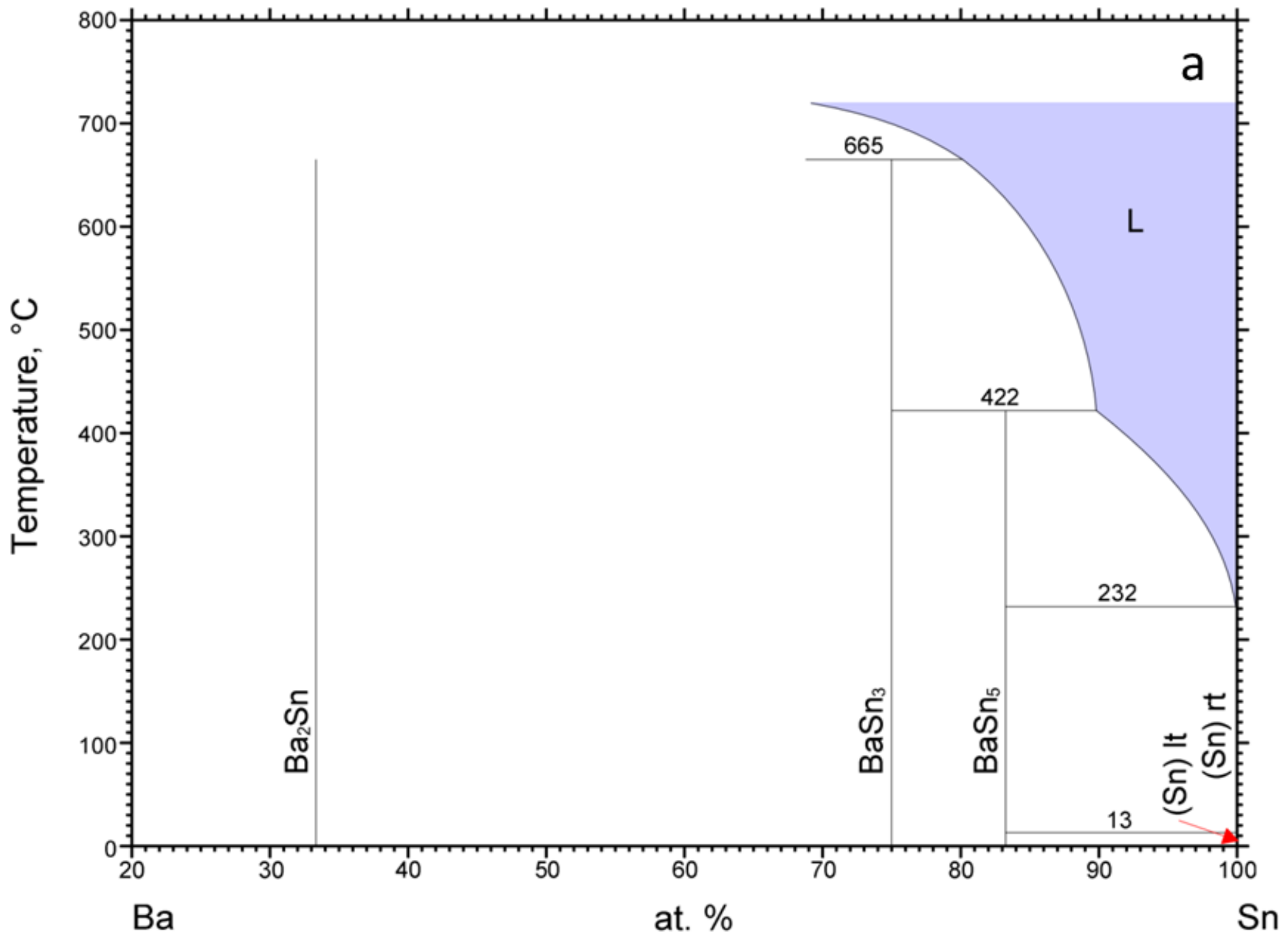

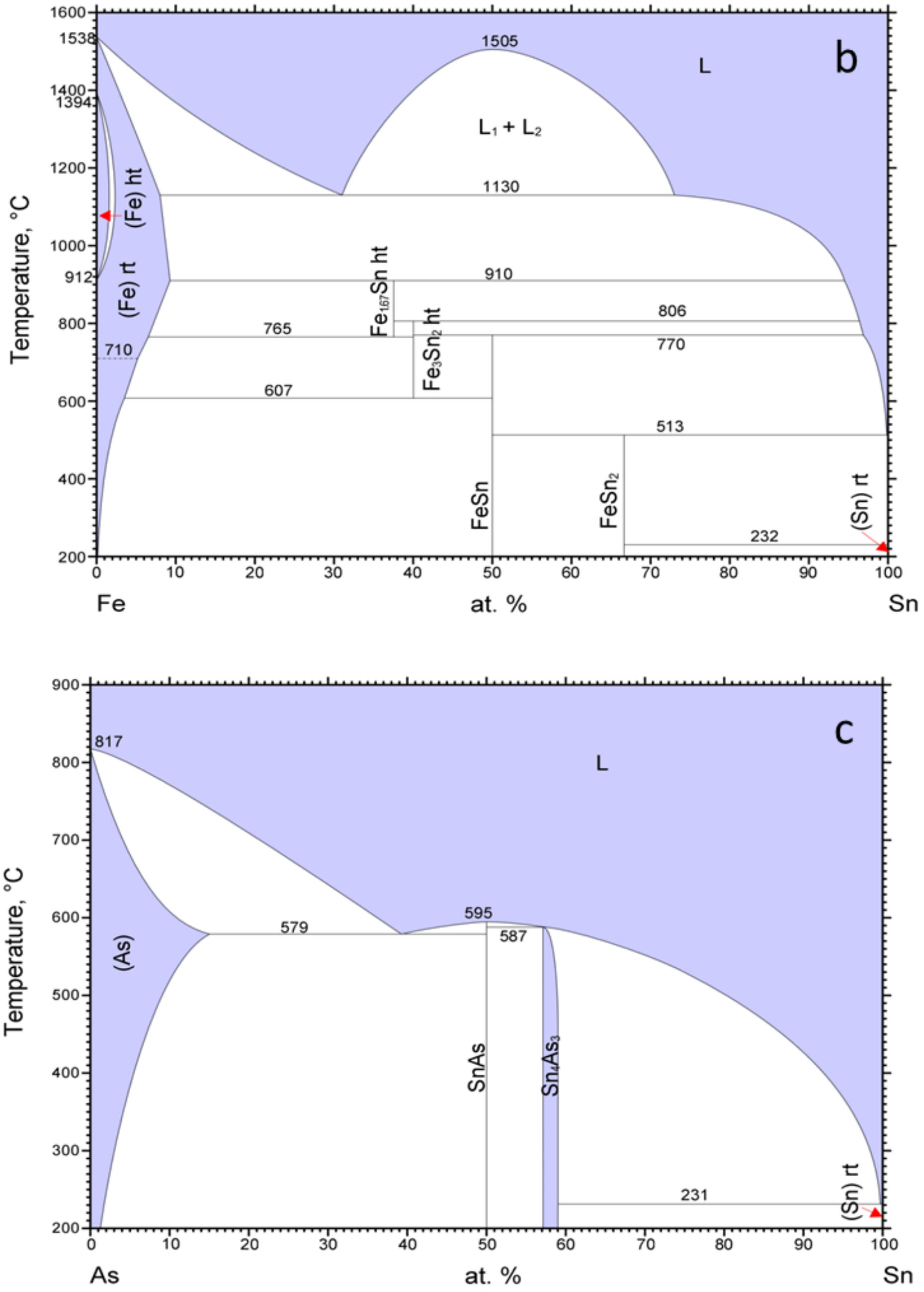

Figure 7: Binary phase diagrams [16] associated with growth of $BaFe_2As_2$ out of Sn: (a) Ba-Sn binary phase diagram; ASM Diagram #900375. (b) Fe-Sn binary phase diagram; ASM Diagram #901083. (c) As-Sn binary phase diagram; ASM Diagram #900186. Each element has adequate solubility in Sn suggesting that $BaFe_2As_2$ may be dissolved into, and even regrown out of, Sn.

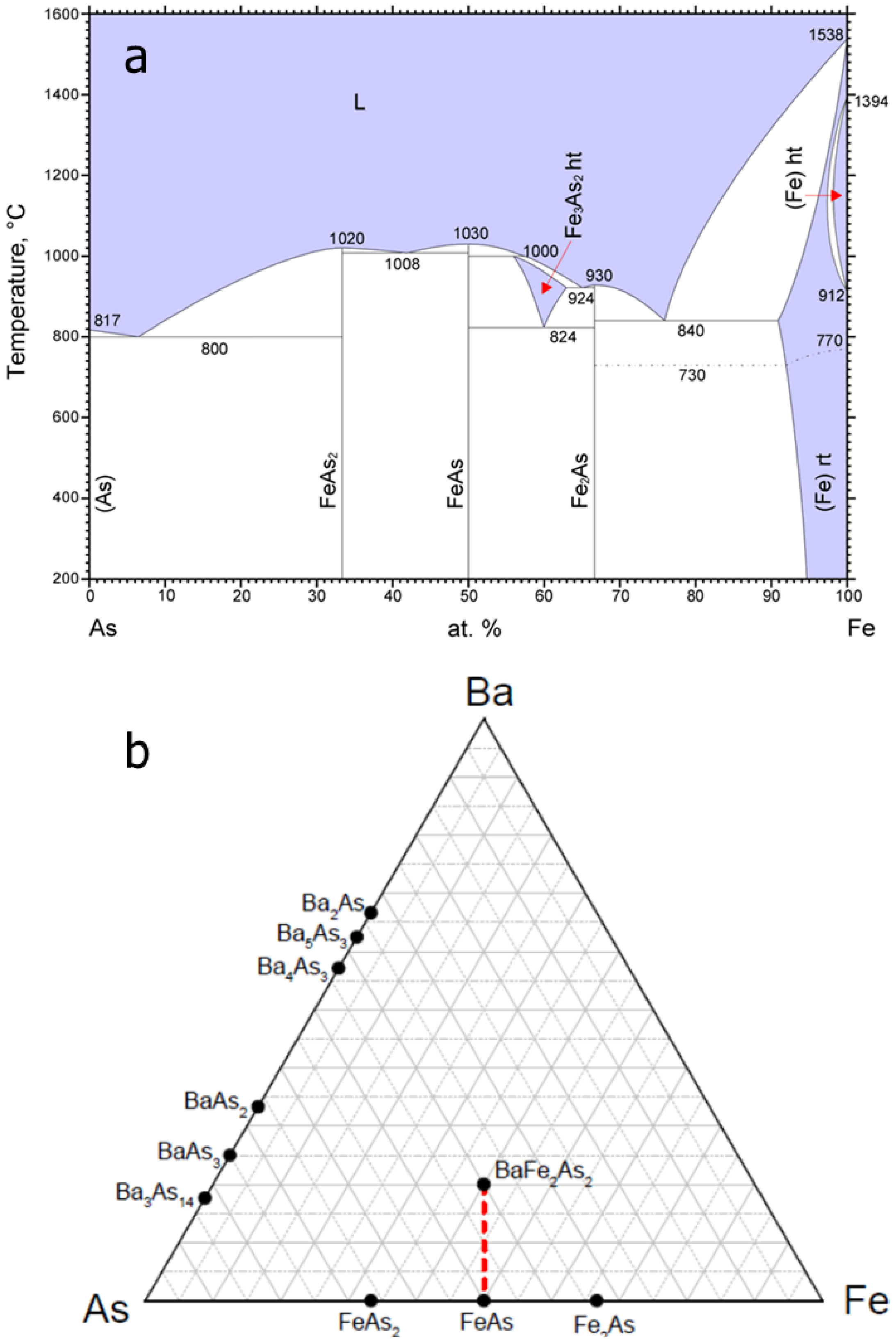

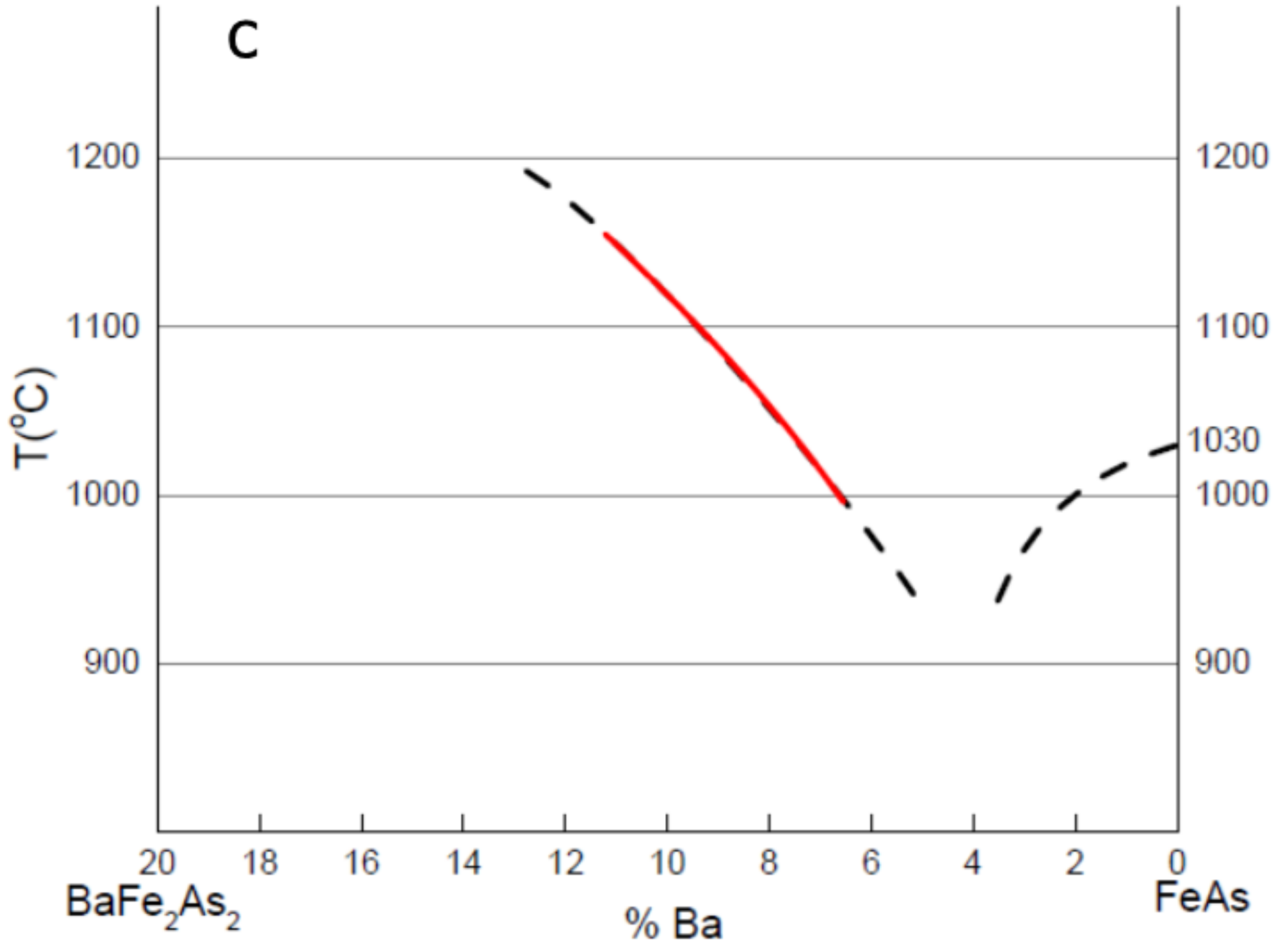

Fig. 8 (a) Fe-As binary phase diagram, [16] ASM Diagram #900163, showing that FeAs is a binary compound with a low enough melting point to act as a high temperature solution for the growth of $BaFe_2As_2$. Reprinted with permission of ASM International. All rights reserved. www.asminternational.org (b) Ba-Fe-As ternary, compositional, phase diagram showing known binary compounds as well as single, known ternary compound: $BaFe_2As_2$. Note: each vertex is a pure element and, for example, $BaFe_2As_2$ is located at the $Ba_{0.2}Fe_{0.4}As_{0.4}$ point. The red dashed line highlights the compositional extent of the pseudo-binary cut. (c) Pseudo-binary phase diagram between $BaFe_2As_2$ and FeAs; the red line is based on single crystal growths between 1180 and 1000 °C.

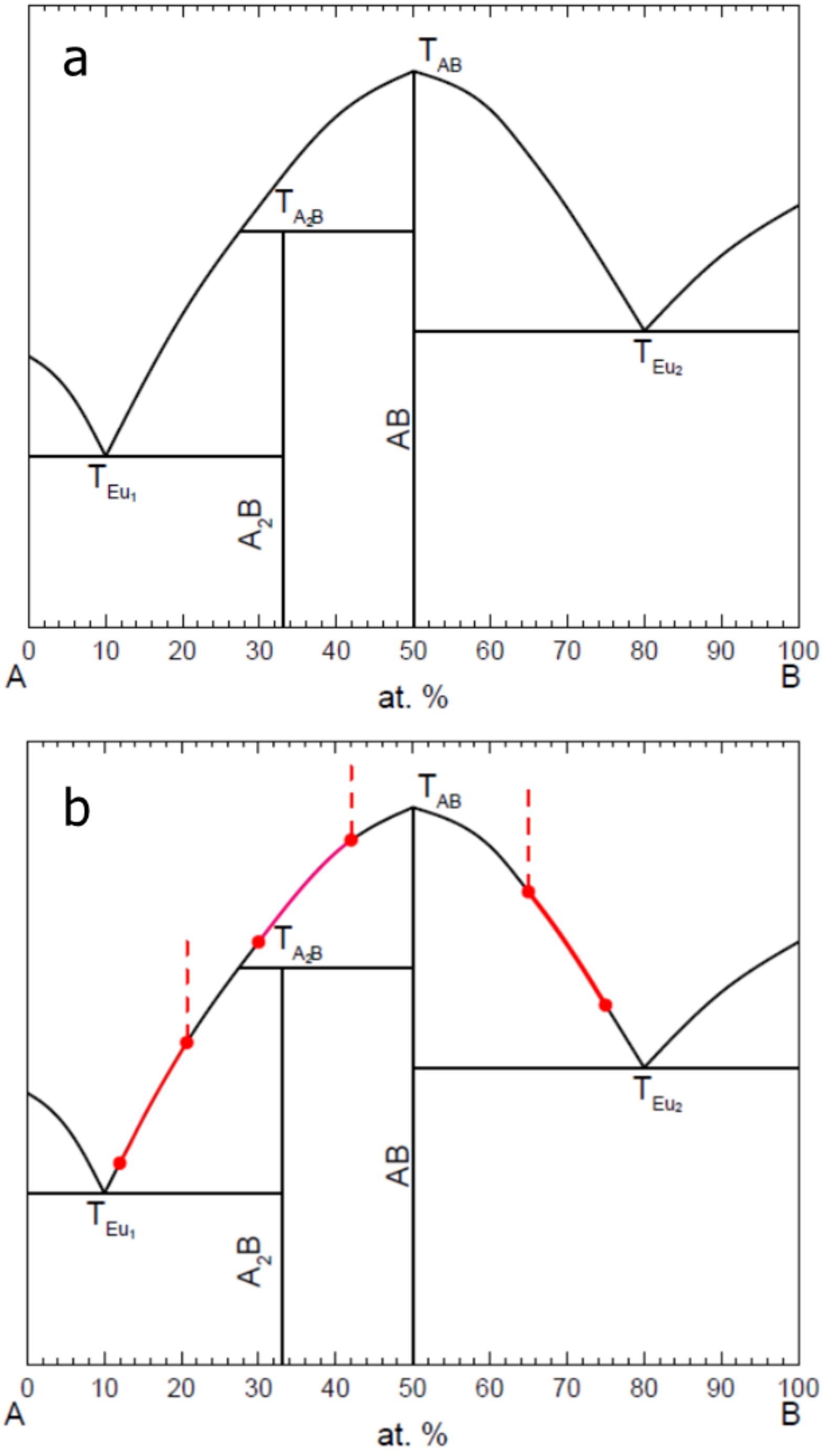

Fig. 9 (a) Generic A – B binary phase diagram with two binary compounds: congruently melting AB and incongruently melting $A_2B$. AB melts congruently at $T_{AB}$; $A_2B$ melts incongruently at $T_{A2B}$; $T_{Eu1}$ and $T_{Eu2}$ are the eutectic temperatures on the A-rich and B-rich sides of the diagram respectively. (b) Same A-B binary with possible growth ranges for growths of $A_2B$ out of excess A, AB out of excess A and AB out of excess B.

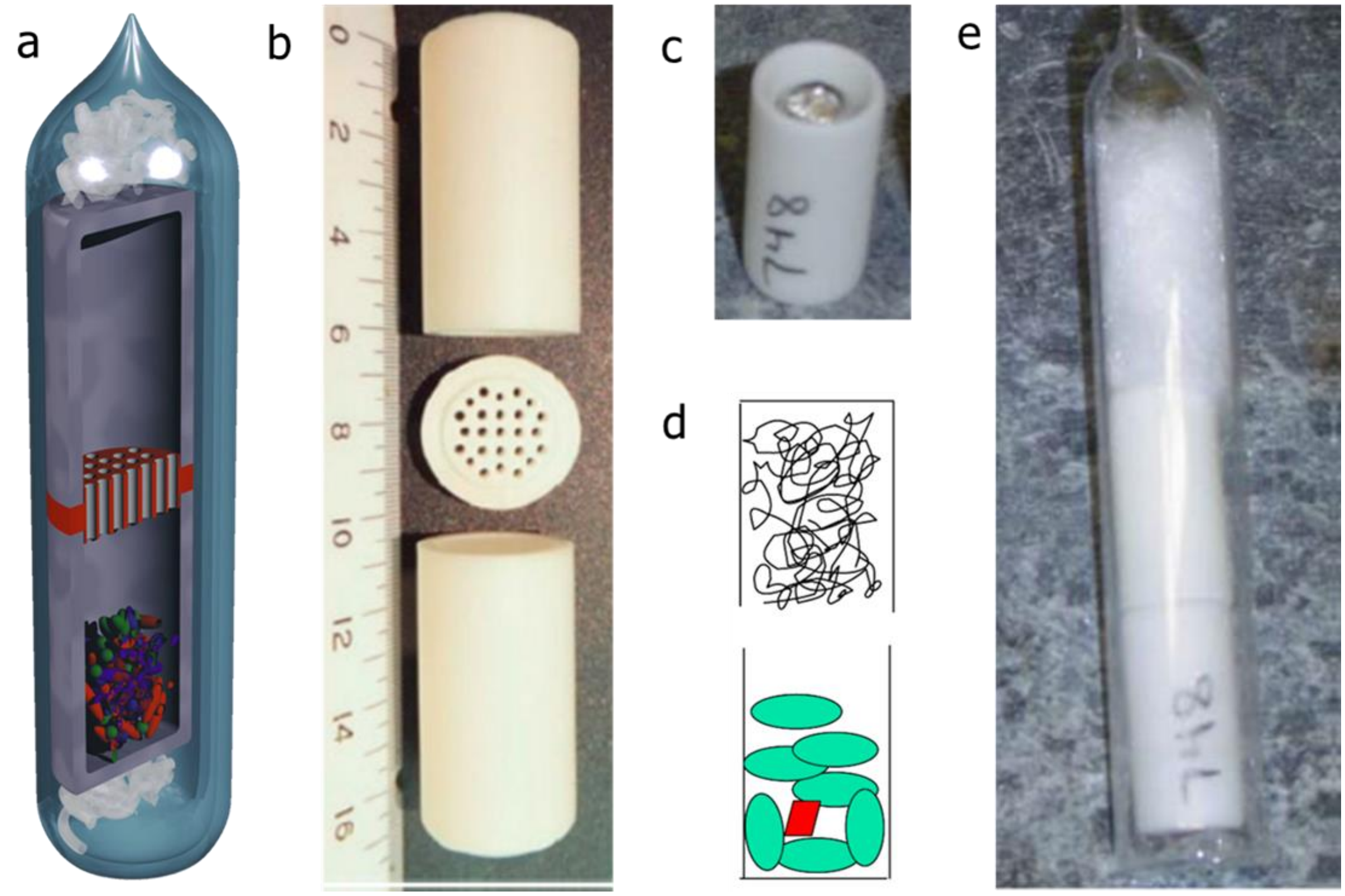

Fig. 10 (a) Schematic of crucibles (lower one with mixture of elements in it), frit, quartz wool sealed in silica tube; (b) fritted crucible set shown next to a cm scale; (c) photograph of crucible packed with elements, filled to 75-80% level; (d) schematic of packed growth crucibles with catch crucible packed with silica wool; (e) crucibles sealed into a silica ampoule with silica wool at top to act as cushioning during the decanting process.

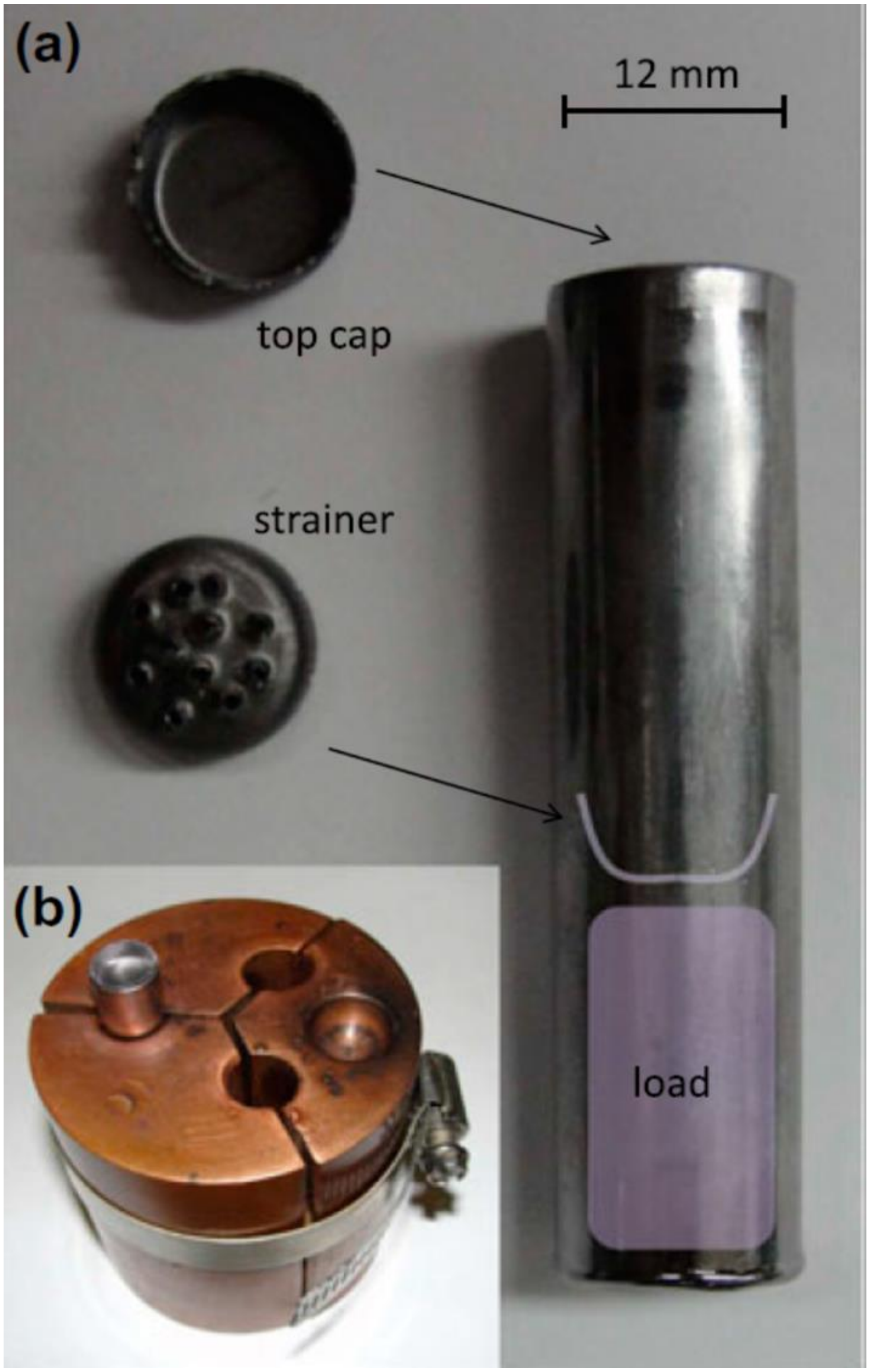

Figure 11: (a) Details of Ta-3-cap crucible assembly. Whereas the top and bottom caps are welded in place to seal the tube, the middle cap is perforated so as to create a metal frit and is held in place by pinching the outer tube slightly. (b) Picture of large thermal inertia welding jig we use to seal up to three such crucibles at a time. Note that the well at roughly 3:00 is for holding a button of Zr that is used to getter the atmosphere before welding. [46] Reprinted from [46] with permission from Taylor&Francis.

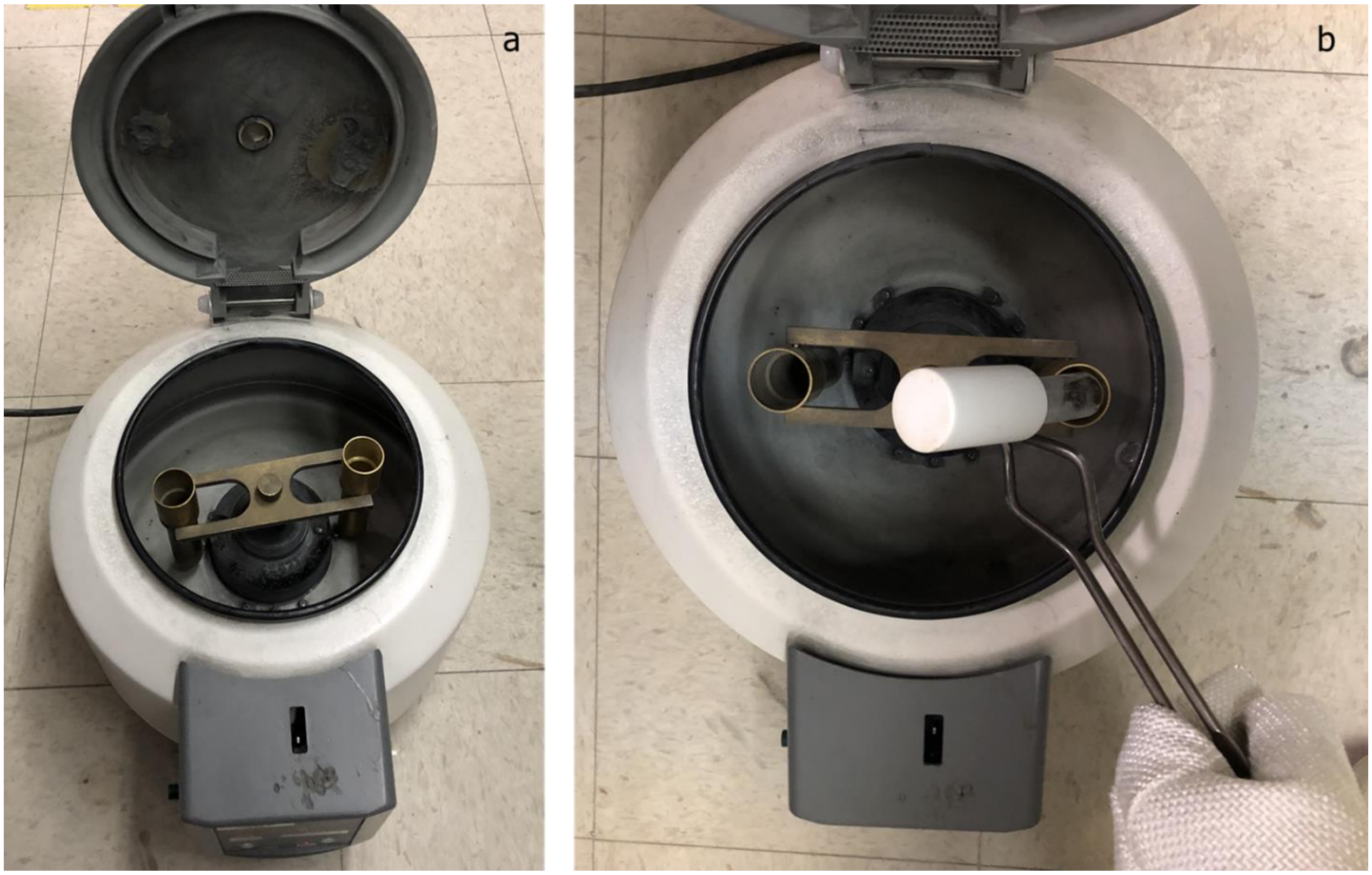

Fig. 12 (a) Centrifuge assembly with brass rotor and cups used for decanting ampoules as hot as 1200 °C, (b) a 50 ml crucible with growth ampoule being tipping into the right hand centrifuge cup.  Once the ampoule is in place the lid is shut and the centrifuge activated for 10-15 s of angular acceleration.  In the bottom right hand corner of the photograph a set of tongs and a high temperature glove can also be seen.

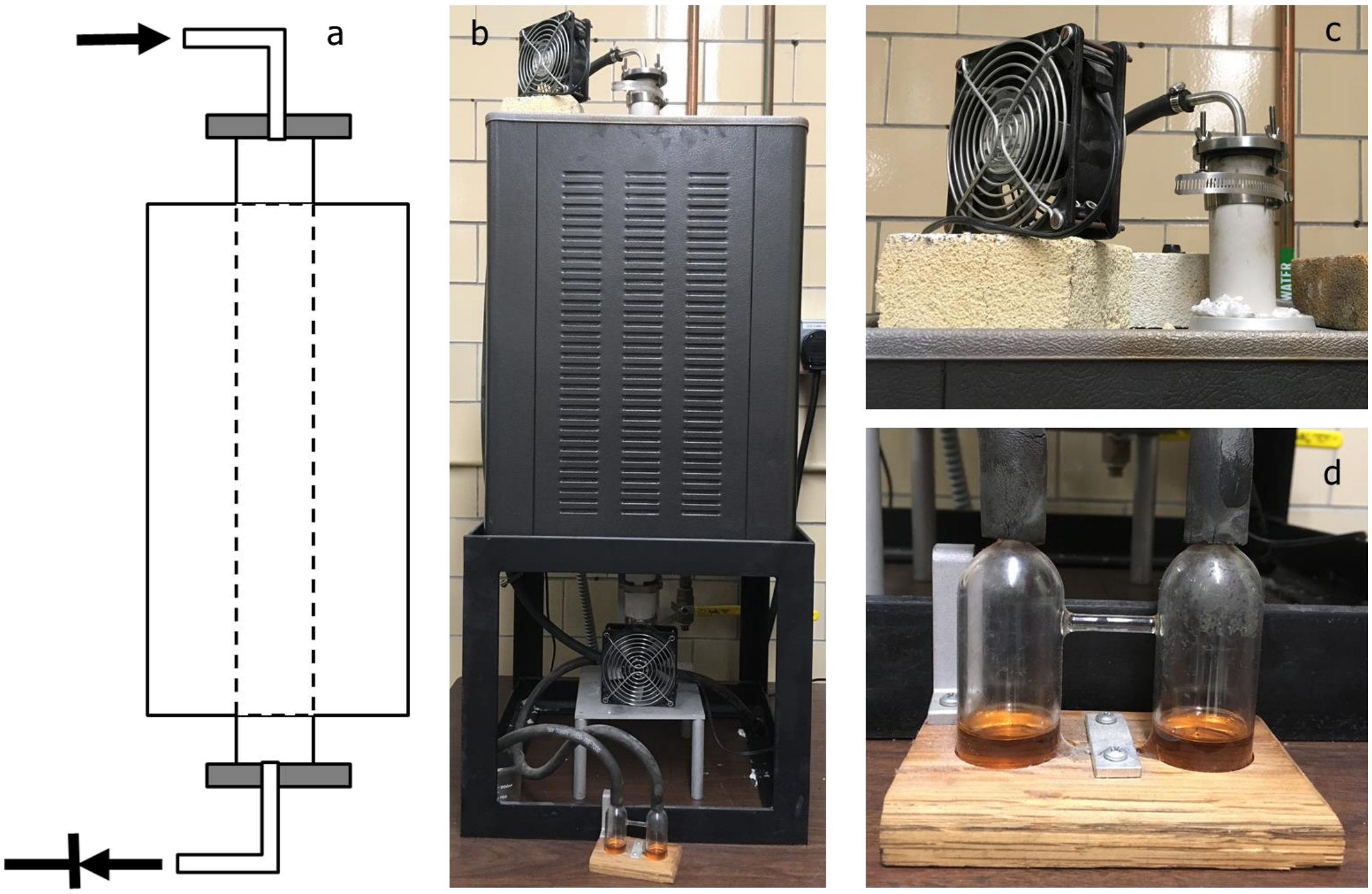

Fig. 13 (a) Schematic of a vertical tube furnace setup with gas flow controlled by bubbler acting as a gas-flow-diode. (b) Picture of vertical tube furnace with fans and bubbler. (c) Picture of mullite tube with gas feed-through cooled by muffin fan. (d) Bubbler made from two small flasks joined at their bases by hollow tube (under metal strap). Note that the solid glass rod connecting them near their tops is for structural stability, not gas flow.

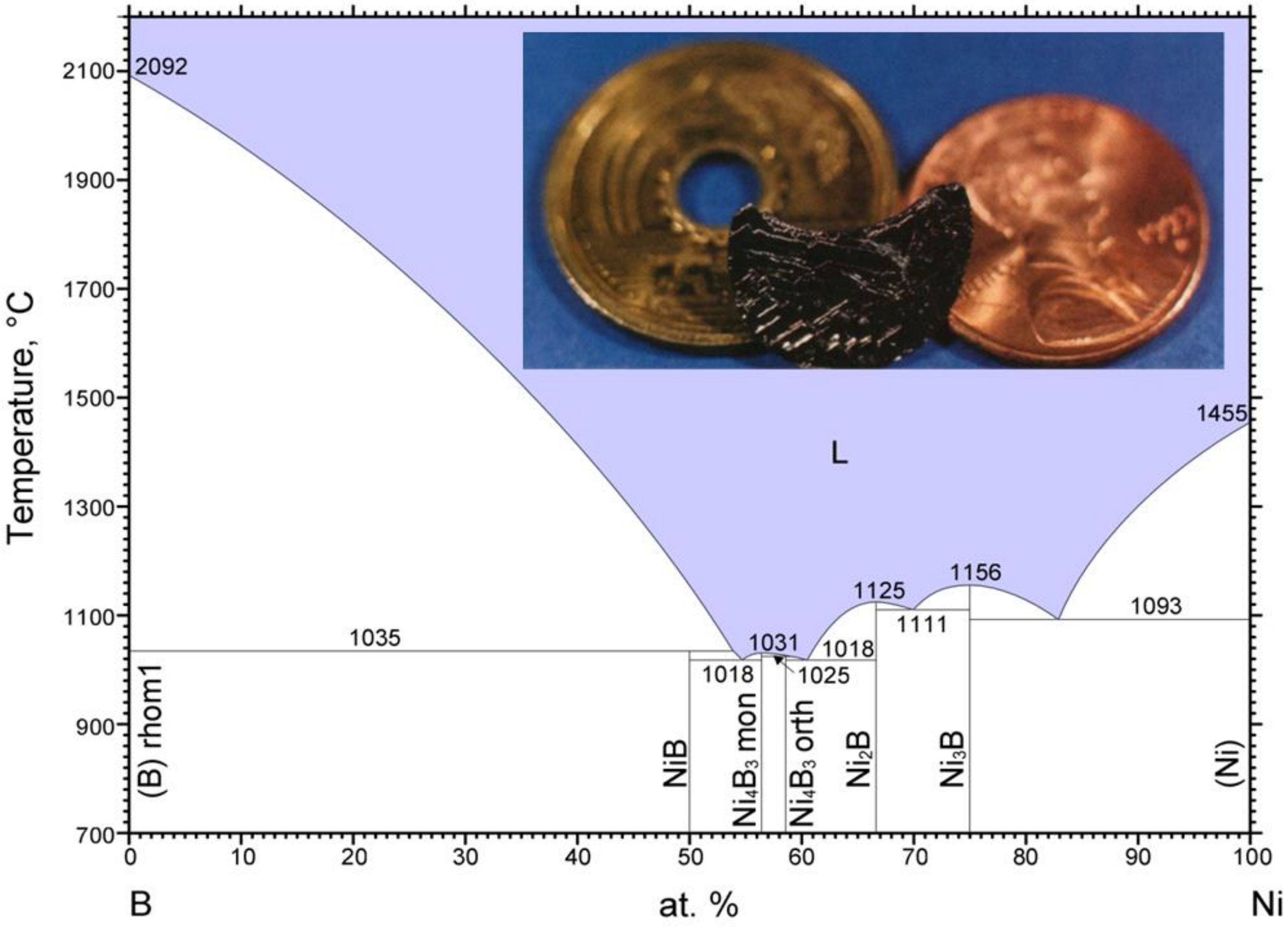

Figure 14: Ni-B binary phase diagram; [16] ASM Diagram #900303. Between roughly 50% and 85% Ni there are four eutectic regions giving rise to liquid existing below 1200 °C. This extended liquid region offered the possibility of growth of $RNi_2B_2C$ out of excess Ni-B binary melt. Reprinted with permission of ASM International. All rights reserved. www.asminternational.org; Inset: photograph of a $RNi_2B_2C$ single crystal grown out of excess $Ni_2B$ beside coins for scale. [60]

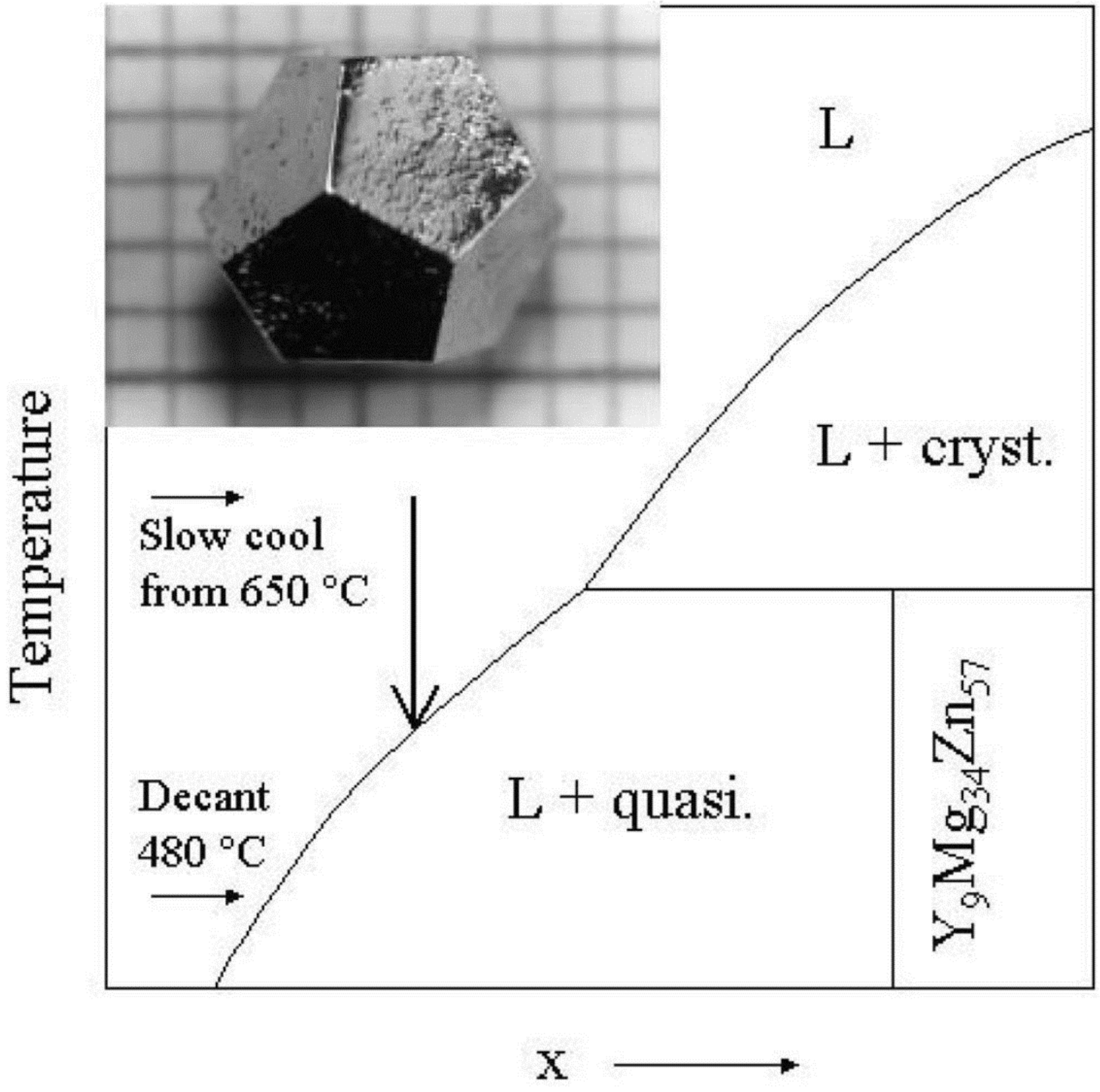

Figure 15: A pseudo-binary cut of Y-Mg-Zn ternary phase diagram with growth profile as well as inset picture of grown quasicrystal with pentagonal, dodecahedral morphology. [76] The existence of an accessible liquidus line for the formation of the quasicrystalline phase (quasi.) suggests that the growth of single grain quasicrystals is no different from the growth of their crystalline cousins. Reprinted from [76] with permission from Elsevier.

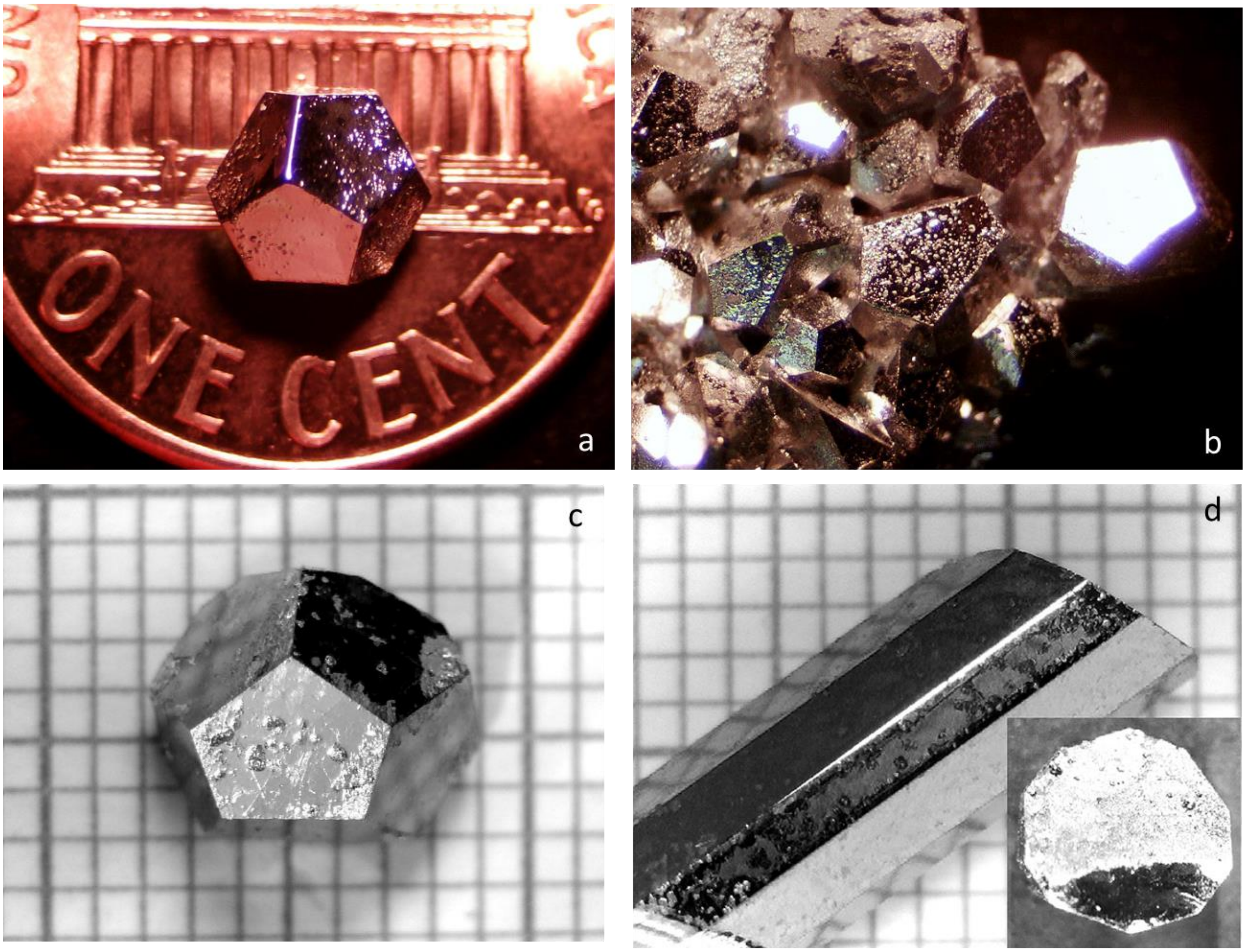

Figure 16: Images of single grains of icosahedral (i) and decagonal (d) quasicrystals: (a) i-HoMgZn quasicryatalline grain on penny; (b) i-RMgZn growth showing multiple grains; (c) i-AlPdMn (doped with with Ga) single grain shown over mm-grid; [77] Reprinted with permission from Taylor&Francis. (d) decagonal, d-AlNiCo single grain shown over mm-grid. [79] Reprinted with permission from Taylor&Francis.

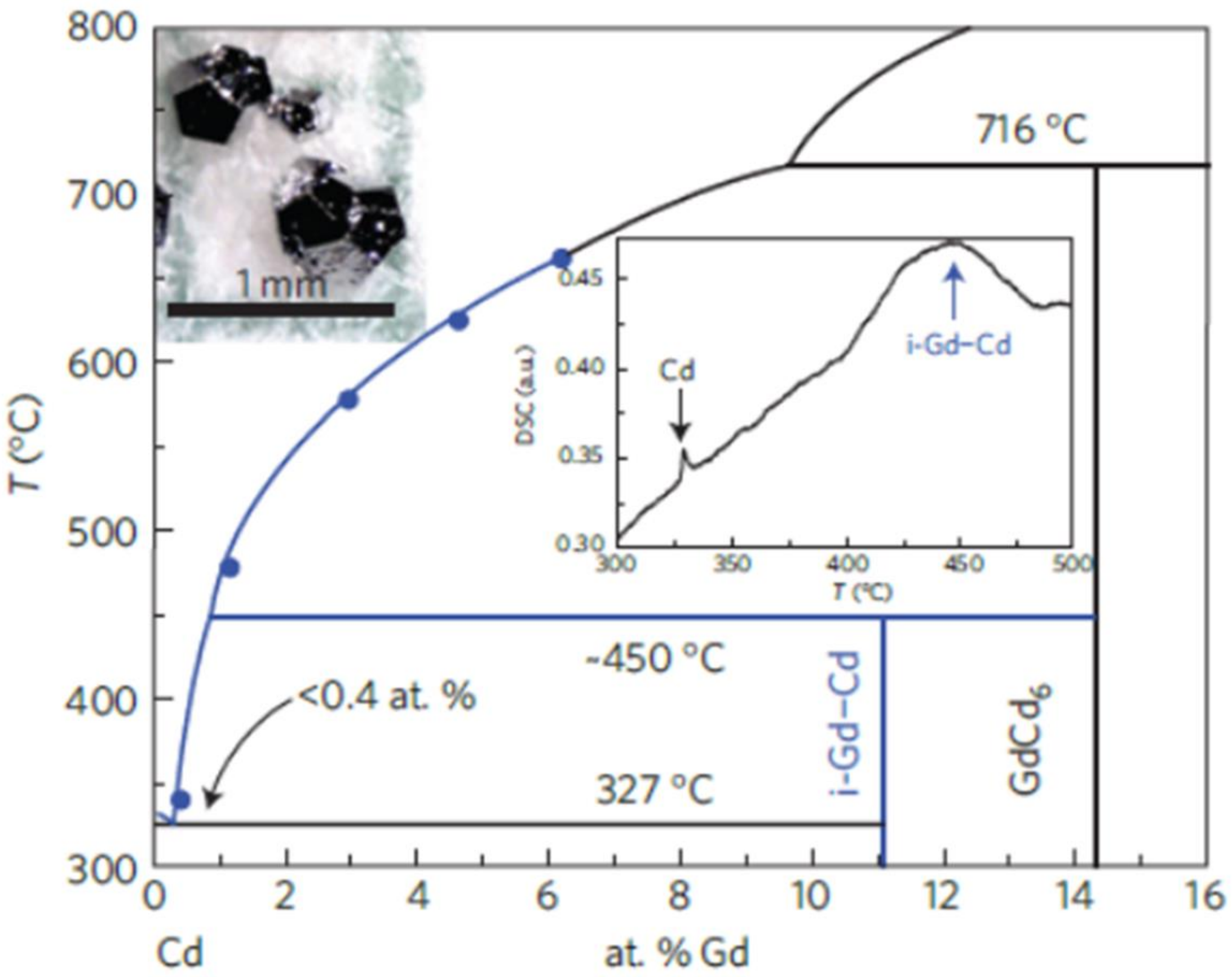

Figure 17: Updating of the Gd-Cd binary phase diagram and discovery of new binary, quasicrystalline phase: (a) Cd-rich part of Gd-Cd binary phase diagram. Black lines are based on existing phase diagram data (ASM Diagram #900581) and blue lines and data points are inferred from our growths and measurements. Inset shows single grains of i-GdCd quasicrystals. [25] Reprinted from [25] with permission from Springer Nature.

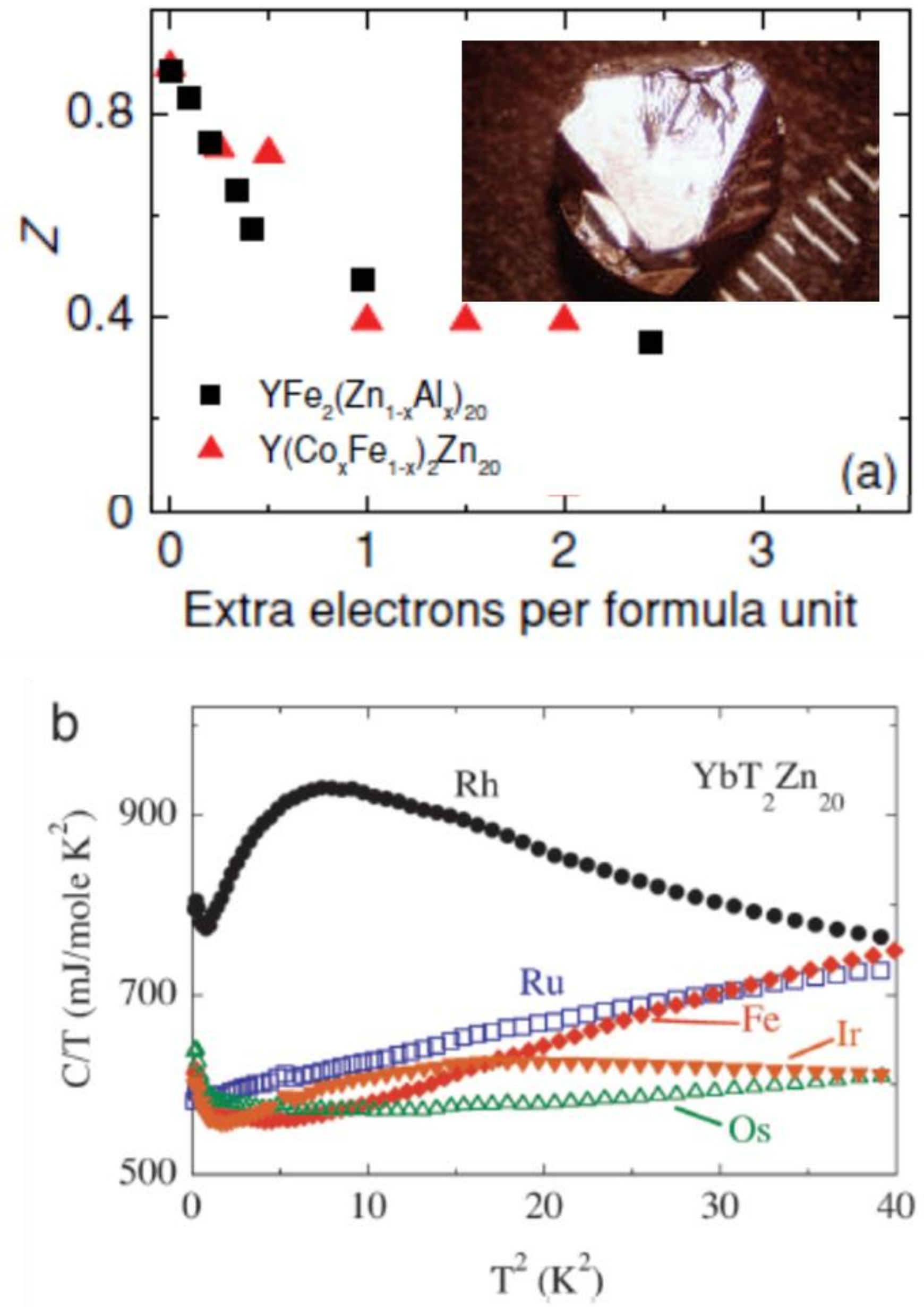

Figure 18: 3d-electron and 4f-electron related correlations in RT2Zn20 compounds (R = Y, Yb; T = Fe, Ru, Os, Rh, Ir. (a) The x-dependence of the Stoner enhancement factor, Z, for $YFe_2(Zn_{1-x}Al_x)_{20}$ and $Y(Fe_{1-x}Co_x)_2Zn_{20}$ both show similar behavior when plotted as extra electrons per formula unit suggesting band filling as key parameter. [121] Reprinted with permission from American Physical Society. Inset: single crystal of $YFe_2Zn_{20}$ next to a mm scale. [119] Reprinted with permission from Springer Nature. (b) Temperature dependent specific heat, C(T), divided by temperature, T, data for $YbT_2Zn_{20}$ compounds (T = Fe, Ru, Os, Rh, Ir). The discovery of six $YbT_2Zn_{20}$ (including T = Co as well) compounds doubled the number of known Yb-based heavy fermions. [122] Reprinted from own article.

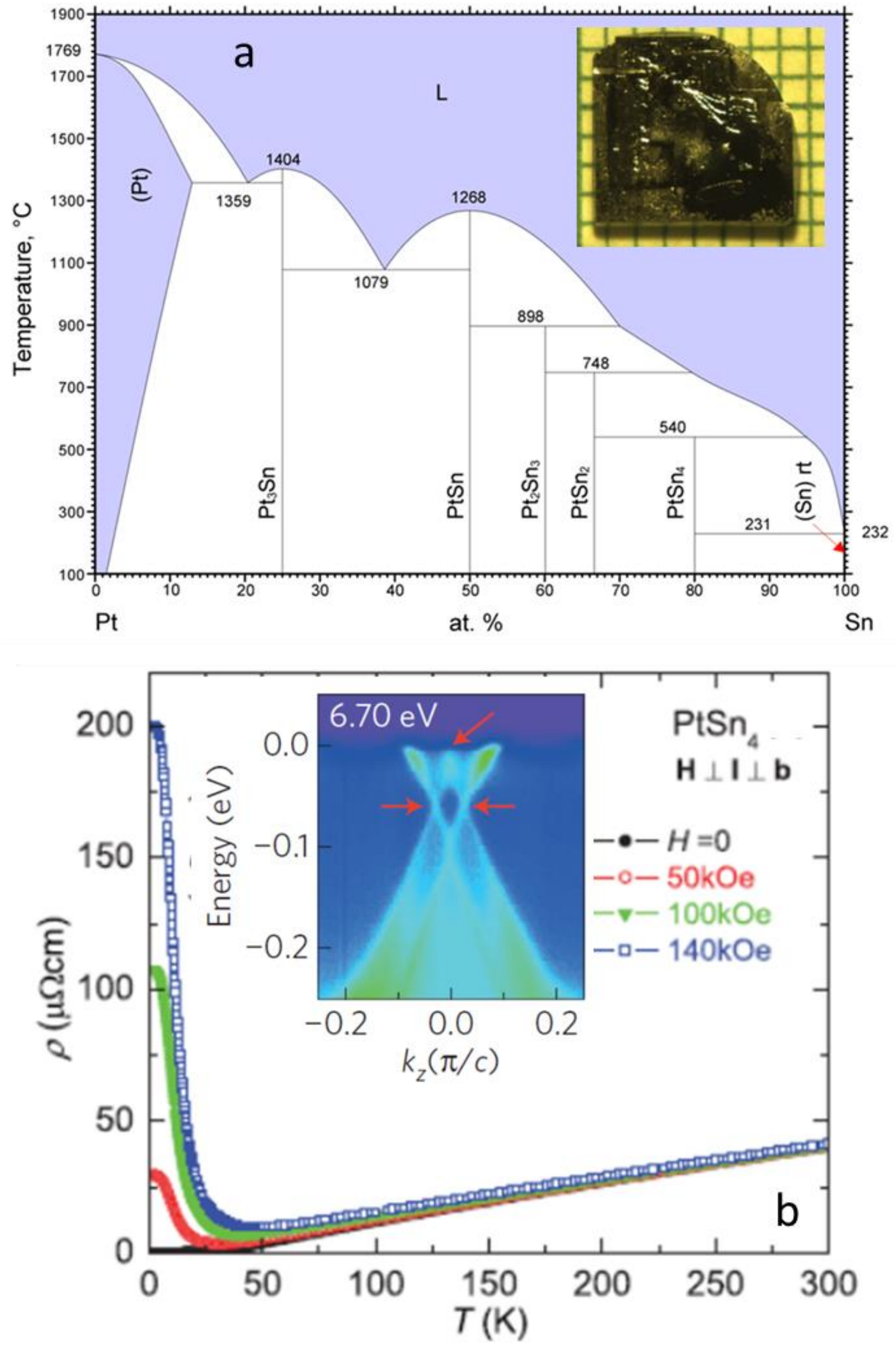

Figure 19: $PtSn_4$ was grown as part of an exploration of compounds that had relatively low peritectic decomposition temperatures. Its exceptionally large magnetoresistance suggested novel bandstructural features.(a) Pt-Sn binary phase diagram; [16] ASM Diagram #1600401. Inset: picture of single crystalline $PtSn_4$. [123] Reprinted with permission from American Physical Society. (b) resistivity versus temperature of $PtSn_4$ for applied magnetic fields of 0, 50, 100 and 140 kOe. [123] Inset: ARPES data from $PtSn_4$ showing Dirac Node Arc feature. [124] Reprinted with permission from Springer Nature.

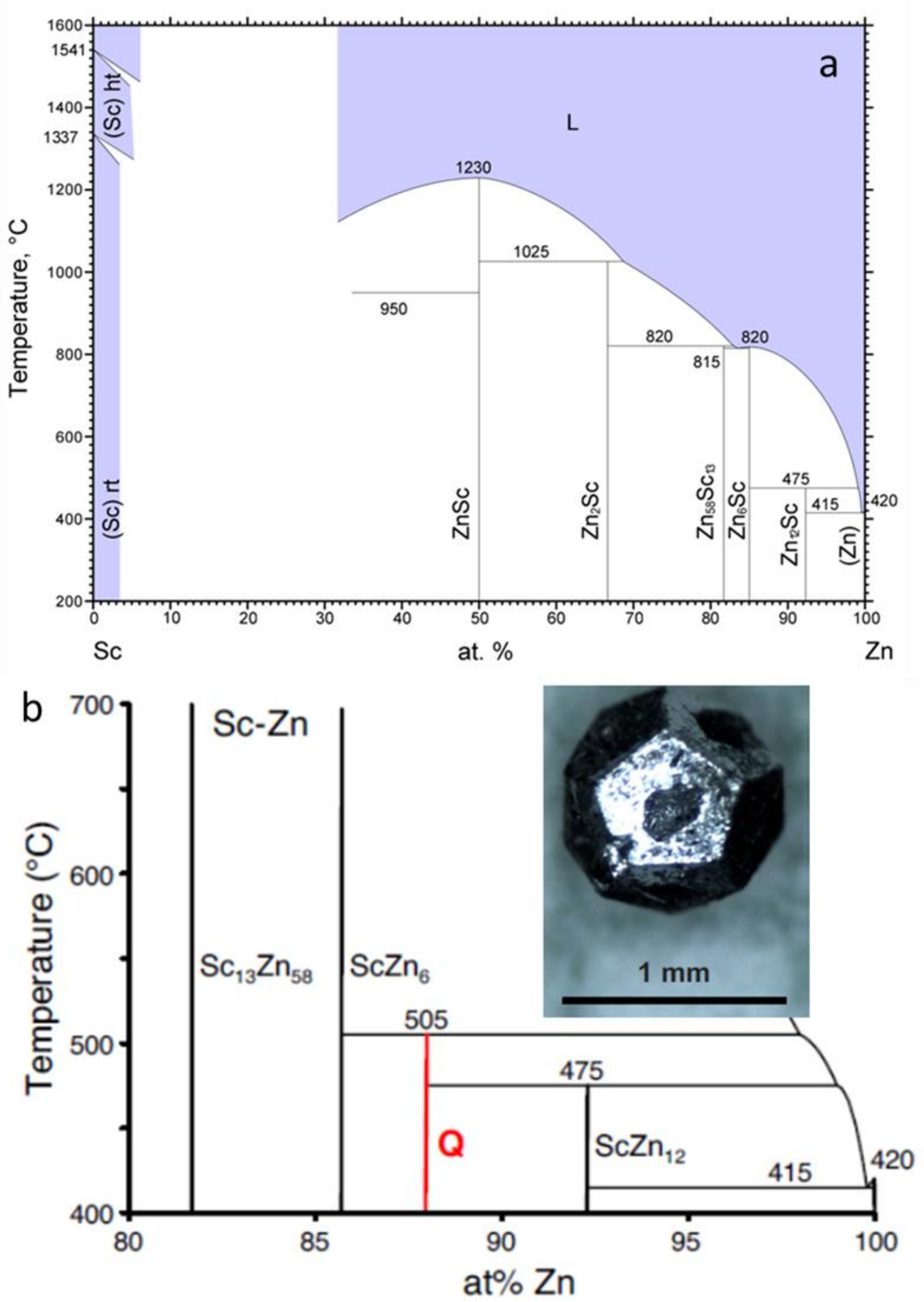

Figure 20: An icosahedral Sc-Zn binary quasicrysalline phase was discovered during a study of Zn-rich compounds with low peritectic decomposition temperatures. (a) Sc-Zn binary phase diagram, [16] ASM Diagram #1201640, missing quasicrystalline phase. (b) Updated version of Sc-Zn binary phase diagram with i-ScZn phase (Q-line, in red) and its peritectic temperature shown.  Inset shows single grain with mm-bar. [24] Reprinted with permission from American Physical Society.

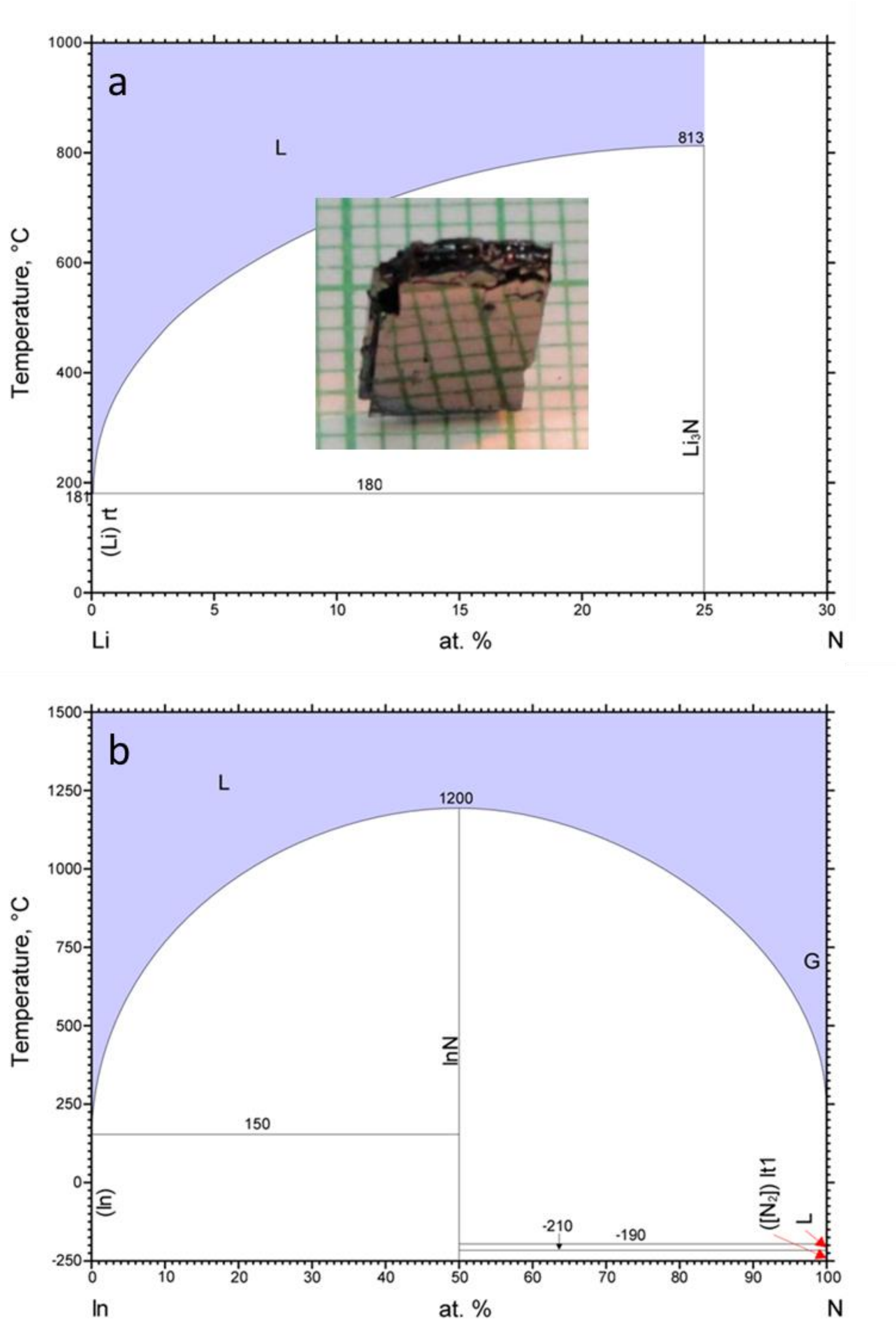

Figure 21: Solution growth of N-based compounds requires a solution that contains N.  In the case of the Li, a Li-rich Li-N solution can be used to grow N-bearing compounds.  (a) Li-N binary phase diagram; [16] ASM Diagram #901511. Reprinted with permission of ASM International. All rights reserved. www.asminternational.org; Inset: $Li_3N$ single crystal on mm-grid with mm-grid reflections on facing surface. In the case of In, although the calculated phase diagram implies that an In-rich solution of In and N may exist, there is limited (if any) solubility of N in the In. (b) In-N binary phase diagram; [16] ASM Diagram #907517. Reprinted with permission of ASM International. All rights reserved. www.asminternational.org.

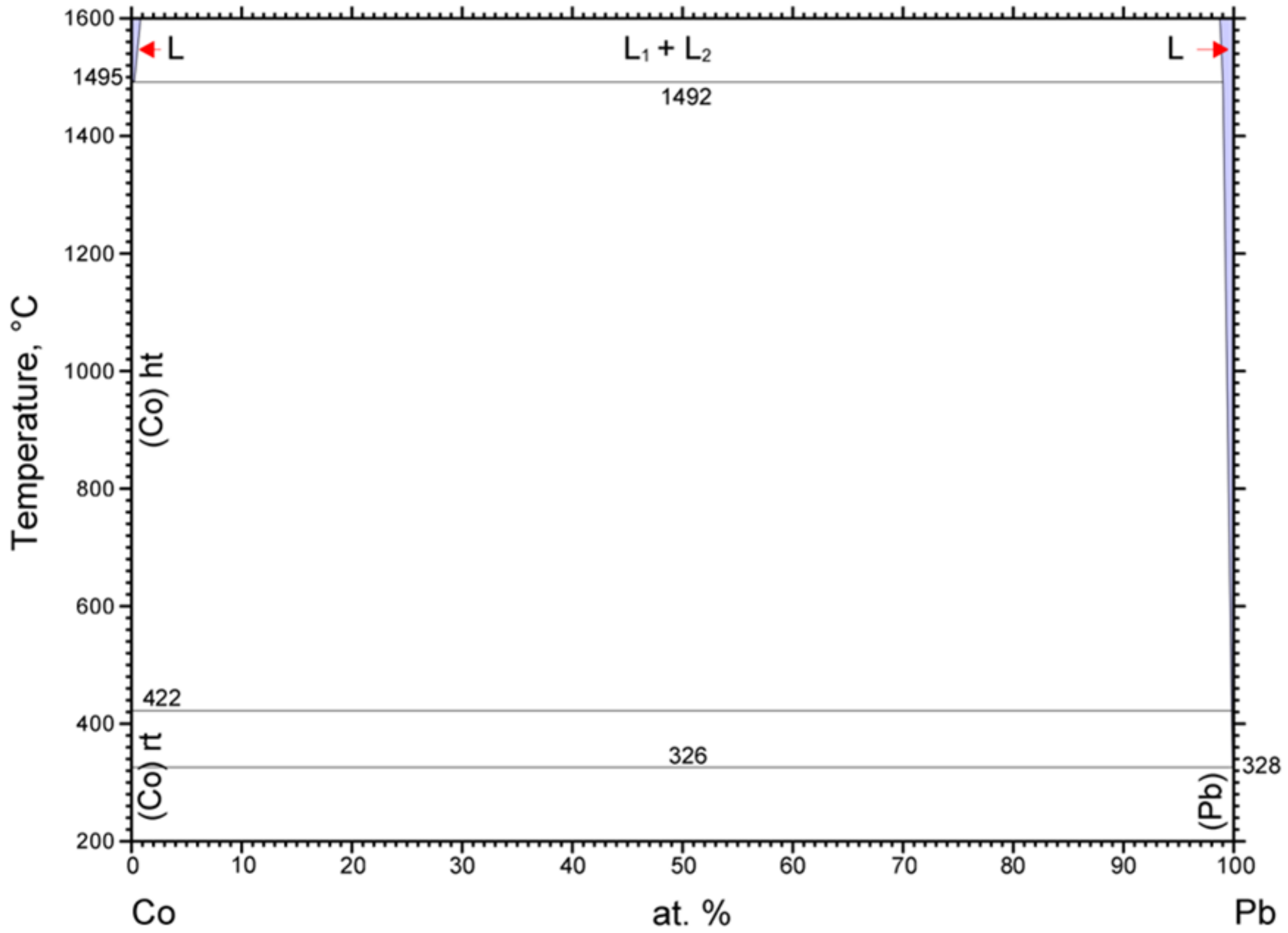

Figure 22: Co-Pb binary phase diagram; [16] ASM Diagram #900732. Co and Pb are highly immiscible; they simply do not like to mix or have atoms near each other. Reprinted with permission of ASM International. All rights reserved. www.asminternational.org

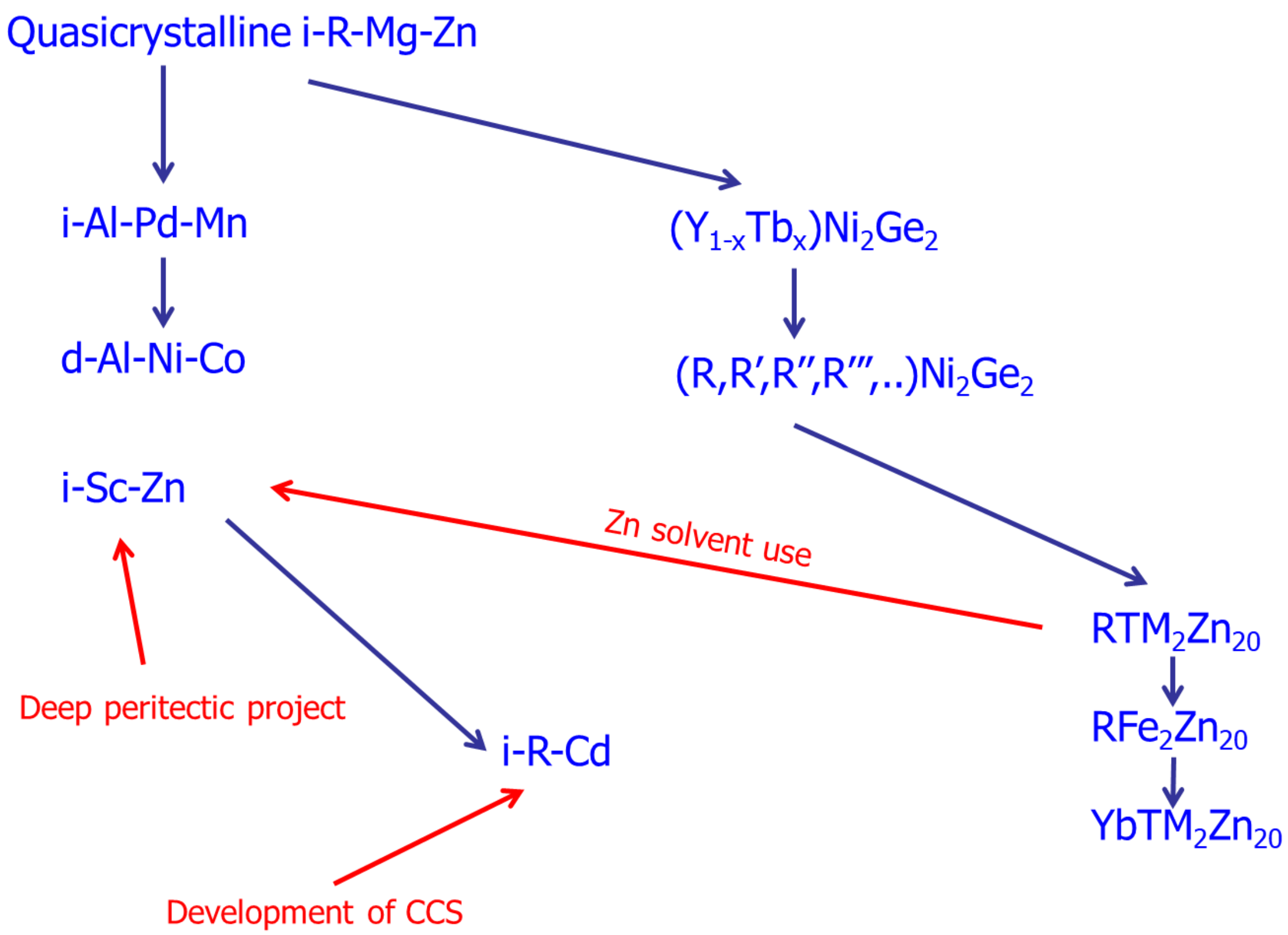

Figure 23. One example of a family tree of ideas; in this case how work on RMgZn quasicrystals led studies of spinglasses, heavy fermions and back to new families of binary quasicrystals. The influence of new experimental capabilities (such as working with the high vapor pressure of Zn or the development of new crucibles) and new research projects (such as the exploration of compounds with low peritectic decomposition temperatures) are shown in red.